\PassOptionsToPackage{dvipsnames}{xcolor}
 \documentclass[a4paper,12pt,dvipsnames]{article}
\usepackage{jheppub_modified_green}
\usepackage{caption}

\usepackage{placeins}
\usepackage{bbold}

\usepackage{multirow} 
\usepackage{simpler-wick} 
\usepackage{epsfig}
\usepackage[makeroom]{cancel}
\usepackage{amsfonts}
\usepackage{amssymb,esint}
\usepackage{enumitem}
\usepackage{comment}
\usepackage[normalem]{ulem}
\usepackage{amsthm}
\usepackage{subfig}
\usepackage{bm}
\usepackage{hyperref}
\usepackage{mathrsfs}
\usepackage{bbm}
\usepackage{cancel}
\usepackage{xcolor}
\usepackage{relsize}
\usepackage{float, slashed, graphicx, amssymb, amsmath}
\usepackage{shuffle}
\usepackage{pgfplots}
\usepackage{tikz-feynman}
\usepackage{silence}
\usepackage{orcidlink}
\pgfplotsset{compat=1.18}
\tikzfeynmanset{compat=1.1.0}
\tikzfeynmanset{warn luatex=false}
\tikzfeynmanset{graviton/.style={circle, draw=green!60, fill=green!5, very thick, minimum size=7mm}}
\tikzfeynmanset{codot/.style={/tikz/shape=circle,/tikz/fill=white,/tikz/minimum size=0.1cm,/tikz/inner sep=1.8pt}}
\tikzfeynmanset{myblob/.style={/tikz/shape=rectangle,/tikz/fill=red,/tikz/minimum size=0.2cm,/tikz/inner sep=1.8pt} }
\tikzfeynmanset{ghc/.style={/tikz/shape=circle,/tikz/fill=white,/tikz/minimum size=0.05cm,} }
\tikzfeynmanset{HV/.style=
{/tikz/shape=circle,
/tikz/fill={rgb:black,1;white,2},
 /tikz/minimum size=0.01cm,/tikz/inner sep=3.0pt
 } }
\tikzfeynmanset{GR/.style={/tikz/shape=ellipse,/tikz/fill={rgb:black,1;white,2},/tikz/minimum size=0.3cm,} }
\tikzfeynmanset{GR2/.style={/tikz/shape=ellipse,/tikz/fill={rgb:black,1;white,2},/tikz/minimum width = 1.2cm, 
    /tikz/minimum height = 3.4cm} }
\tikzset{box/.pic={\filldraw[fill=black]  (0,0) circle (2.5pt); \filldraw [fill=black] (0.5,0) circle (2.5pt); \draw [line width=5pt] (0,0) -- (0.5,0);}}
 \tikzfeynmanset{myblob2/.style=
{/tikz/shape=rectangle,
/tikz/fill=black,
 /tikz/minimum width=0.1cm,/tikz/inner sep=1.8pt
 } }
 \tikzfeynmanset{sb/.style=
{/tikz/shape=circle,
/tikz/fill=black,
 /tikz/minimum size=0.3cm,/tikz/inner sep=1.pt
 } }
 \tikzfeynmanset{myblob/.style=
{/tikz/shape=ellipse,
/tikz/fill=red,
 /tikz/minimum width=0.5cm,
 } }

\usetikzlibrary{arrows.meta} 
\usetikzlibrary{calc}
\usetikzlibrary{decorations.pathmorphing}
\usetikzlibrary{decorations.pathreplacing}
\usetikzlibrary{decorations.markings}
\usetikzlibrary{decorations.text}
\tikzset{
   vector2/.style={decorate, decoration={snake, amplitude=1pt, segment length=6pt}, draw,double},
   vector/.style={decorate, decoration={snake, amplitude=1pt, segment length=6pt}, draw},
	provector/.style={decorate, decoration={snake,amplitude=2.5pt}, draw},
	antivector/.style={decorate, decoration={snake,amplitude=-2.5pt}, draw},
    fermion/.style={draw=black, postaction={decorate},
        decoration={markings,mark=at position .55 with {\arrow[draw=black]{>}}}},
    fermionbar/.style={draw=black, postaction={decorate},
        decoration={markings,mark=at position .55 with {\arrow[draw=black]{<}}}},
    fermionnoarrow/.style={draw=black},
    gluon/.style={decorate, draw=black,
        decoration={coil,amplitude=4pt, segment length=5pt}},
    scalar/.style={dashed,draw=black, postaction={decorate},
        decoration={markings,mark=at position .55 with {\arrow[draw=black]{>}}}},
    scalarbar/.style={dashed,draw=black, postaction={decorate},
        decoration={markings,mark=at position .55 with {\arrow[draw=black]{<}}}},
    scalarnoarrow/.style={dashed,draw=black},
    electron/.style={draw=black, postaction={decorate},
        decoration={markings,mark=at position .55 with {\arrow[draw=black]{>}}}},
	bigvector/.style={decorate, decoration={snake,amplitude=4pt}, draw},
massive/.style={},
	R/.style={blue},
	A/.style={green},
	cut/.style={postaction={decorate, decoration={markings, mark=at position 0.5 with {
					\draw[opacity=1,red] (0,#1*-2pt) -- (0,#1*2pt);
	}}}},
	cut/.default=1,
	Rnew/.style={postaction={decorate, decoration={markings, mark=at position #1 with {
					\arrow[xshift=3pt]{Latex}
	}}}},
    Rnew/.default=0.5,
	Anew/.style={postaction={decorate, decoration={markings, mark=at position #1 with {
					\arrowreversed[xshift=-3pt]{Latex}
	}}}},
    Anew/.default=0.5,
    thetaLineOld/.style={postaction={decorate, decoration={markings,
			mark=at position 0 with {\node (s) at (0,1pt*#1){};},
			mark=at position 0.999  with {\node (e) at (0,1pt*#1){};},
			mark=at position 1 with {\draw[red] (s.center) -- (e.center);}
		},
	}},
	thetaLineOld/.default=1,
	bothLines/.style={decoration={markings,
				mark=at position 0 with {\node (s1) at (0,0.5pt*#1){};\node (s2) at (0,-0.5pt*#1){};},
				mark=at position 0.999  with {\node (e1) at (0,0.5pt*#1){};\node (e2) at (0,-0.5pt*#1){};},
				mark=at position 1 with {\draw[red] (s1.center) -- (e1.center); \draw[black] (s2.center)-- (e2.center);} 
			},
	},
	bothLines/.default=1,
    causArrow/.style={postaction={decorate}, decoration={markings, mark=at position #1 with{\arrow[xshift=3pt,scale=0.7]{Latex}}}},
    causArrow/.default=0.5,
    causArrowR/.style={postaction={decorate}, decoration={markings, mark=at position #1 with{\arrow[xshift=-3pt,scale=-0.7]{Latex}}}},
    causArrowR/.default=0.5,
    noncausArrow/.style={},
    thetaLine/.style={thin, red, causArrow=#1},
    thetaLine/.default=0.5
}
\tikzset{cross/.style={cross out, draw, 
         minimum size=2*(#1-\pgflinewidth), 
         inner sep=0pt, outer sep=0pt}}

\tikzstyle{block} = [draw, rectangle, 
    minimum height=3em, minimum width=6em]

\usepackage[T1]{fontenc} 

\def\sc#1{\overline{#1}}
\makeatletter
\newcommand \UPlus {\mathop {\operator@font \uplus }\limits }
\makeatother
\makeatletter
\newcommand \Bigcup {\mathop {\operator@font \bigcup }\limits }
\makeatother
\def\LabelNote#1{}
\def\Label#1{\label{#1}%
\smash{\hbox to\phipt{\raise1ex\hbox{\tiny[#1]}\hss}}}

\definecolor{bananayellow}{rgb}{1.0, 0.88, 0.21}
\definecolor{amber}{rgb}{1.0, 0.75, 0.0}

\newcommand{\cA}{\mathcal{A}}
\newcommand{\cI}{\mathcal{I}}
\newcommand{\cM}{\mathcal{M}}

\newcommand{\cC}{\mathcal{C}}

\newcommand{\cH}{\mathcal{H}}

\newcommand{\cO}{\mathcal{O}}

\newcommand{\tb}{\hat b}
\newcommand{\eps}{\epsilon}
\newcommand{\vareps}{\varepsilon}

\newcommand{\ta}{\tilde{a}}

\def\nn{\nonumber}

\newcommand{\red}{\color{red}}

\def\bra#1{\langle #1|}
\def\ket#1{|#1 \rangle}

\def\braket#1{\langle #1 \rangle}

\def\spa#1.#2{\left\langle#1\,#2\right\rangle}
\def\spb#1.#2{\left[#1\,#2\right]}

\def\be{\begin{equation}}
\def\ee{\end{equation}}
\def\bea{\begin{eqnarray}}
\def\eea{\end{eqnarray}}  
\allowdisplaybreaks

\newcommand{\NOwick}{\text{normal-ordered contractions}}

\newcommand{\cT}{\mathcal{T}}

\usepackage{amssymb,amsmath}
\usepackage{mathtools} 
\usepackage{cancel} 
\usepackage{graphicx} 

\usepackage{hyperref}
\hypersetup{
	colorlinks=true,
	linktoc=page,
	citecolor=americanrose,
	linkcolor=cadmiumgreen,
	urlcolor=blue} 
\definecolor{americanrose}{rgb}{1.0, 0.01, 0.24}
\definecolor{cadmiumgreen}{rgb}{0.0, 0.42, 0.24}

\usepackage[colorinlistoftodos]{todonotes}

\newcommand{\Cdot}{{\cdot}} 
\makeatletter
\newcommand*{\bigcdot}{}
\DeclareRobustCommand*{\bigcdot}{%
  \mathbin{\mathpalette\bigcdot@{}}%
}
\newcommand*{\bigcdot@scalefactor}{.6}
\newcommand*{\bigcdot@widthfactor}{1.25}
\newcommand*{\bigcdot@}[2]{%
  \sbox0{$#1\vcenter{}$}
  \sbox2{$#1\cdot\m@th$}%
  \hbox to \bigcdot@widthfactor\wd2{%
    \hfil
    \raise\ht0\hbox{%
      \scalebox{\bigcdot@scalefactor}{%
        \lower\ht0\hbox{$#1\bullet\m@th$}%
      }%
    }%
    \hfil
  }%
}
\makeatother
\def\nn{\nonumber}

\newcommand{\sgn}{\text{sgn}}

\title{The Magnus expansion in relativistic quantum field theory}
\author{Andreas Brandhuber,\,\orcidlink{0000-0002-4203-8811}$\mbox{}^{a}$}
\author{Graham R.~Brown,\,\orcidlink{0000-0002-2178-0267}$\mbox{}^{b}$}
\author{Paolo Pichini,$\, \orcidlink{0000-0001-7539-740X}\mbox{}^{a}$}
\author{\\Gabriele Travaglini$\,\orcidlink{0000-0002-6699-3960}\mbox{}^{a}$}
\author{and Pablo Vives Matasan$\,\orcidlink{0009-0009-8404-0301}\mbox{}^{a}$}
\affiliation{$\mbox{}^{a}$Centre for Theoretical Physics, Department of Physics and Astronomy, \\
Queen Mary University of London, Mile End Road, London E1 4NS, United Kingdom}
\affiliation{$\mbox{}^{b}$Higgs Centre for Theoretical Physics, School of Physics and Astronomy, \\
The University of Edinburgh, Edinburgh EH9 3JZ, Scotland, United Kingdom}
\emailAdd{a.brandhuber@qmul.ac.uk}
\emailAdd{graham.brown@ed.ac.uk}
\emailAdd{p.pichini@qmul.ac.uk}
\emailAdd{g.travaglini@qmul.ac.uk}
\emailAdd{p.vivesmatasan@qmul.ac.uk}
\begin{document}
\begin{flushright}
	QMUL-PH-25-27
\end{flushright}

\abstract{We investigate the Magnus expansion of the $N$-operator in relativistic quantum field theory, which is related to the $S$-matrix via $S = e^{iN}$. We develop direct methods to compute matrix elements of the $N$-operator, which we refer to as Magnus amplitudes, bypassing scattering amplitudes entirely.
At tree level, Magnus amplitudes are expressed in terms of retarded and advanced propagators, with each diagram weighted by factors that we identify as Murua coefficients. 
At loop level this structure is augmented by the Hadamard cut function, and we establish remarkable relations between loop- and tree-level Magnus amplitudes. Among these, we find that $n$-point one-loop Magnus amplitudes are entirely determined by phase-space integrals of forward limits of $(n{+}2)$-point tree-level amplitudes, and hence related to Murua coefficients, and we generalise this to a class of higher-loop contributions. Furthermore, in the case of heavy  particles interacting via massless mediators, we conjecture that Magnus diagrams that contribute to the classical limit are always given by forward limits of trees, and we show this explicitly in a one-loop example.
We derive these results studying theories of scalar fields with cubic interactions, but our methods are applicable to general theories as well as to integral functions appearing in gravitational-wave computations. Given that Magnus amplitudes are free of hyper-classical terms, and the known relations between Magnus amplitudes and the radial action, our results lay the groundwork for systematic and efficient calculations of classical observables from quantum field theory.

}

\vspace{-2.6cm}

\maketitle

\flushbottom
 \tableofcontents
\newpage 

\section{Introduction}

The $S$-matrix is the cornerstone of quantum field theory calculations, encoding all physical scattering amplitudes from which  observable quantities can be computed. Since the development of perturbation theory, physicists have predominantly relied on the Dyson series \cite{Dyson:1949bp} to compute $S$-matrix elements -- a time-honoured approach that expands the time evolution operator as an infinite sum of time-ordered products of interaction Hamiltonians. However, an alternative formulation exists that has remained relatively unexplored in the context of relativistic quantum field theory: the Magnus expansion
\cite{Magnus:1954zz}.

The Magnus expansion  takes a  different approach to the time evolution operator  by writing it in exponential form,  $U(t,t_0) = e^{\frac{i}{\hbar}N(t,t_0)}$, and then finding $N(t, t_0)$ as a perturbative series -- the Magnus expansion. 
Applied to the 
 $S$-matrix  this yields 
{$S {=} e^{\frac{i}{\hbar} N}$}, with $N$ determined order by order through the Magnus expansion. There are two important advantages of this approach. First, by writing the evolution operator or the $S$-matrix as exponentials, unitarity is manifest at every order, since 
$N(t, t_0)$ and $N$  are hermitian  at any finite order by construction.%
\footnote{The $S$-matrix computed using standard Feynman rules is only unitary when all orders of perturbation theory are included. In contrast, the Magnus expansion naturally incorporates resummations, ensuring that the $S$-matrix remains unitary at every finite order in perturbation theory.}  
Second, the exponential representation is particularly well suited when the interaction Hamiltonian is not small, such as in collision processes relevant for chemical physics \cite{Pechukas1966}. 

The $S$-matrix can be seen as the limit of a sequence of infinitesimal steps of the form $e^{- \frac{i}{\hbar} H_I (t) \Delta t}$,  where $H_I(t)$ is the interaction Hamiltonian. In this sense, the evolution operator in its exponential form itself can be viewed as  a ``continuous'' generalisation of the Baker-Campbell-Hausdorff formula \cite{Magnus:1954zz, Pechukas1966}. It is therefore not surprising that 
 the $N$-operator  is constructed from nested commutators of the interaction Hamiltonian, in stark contrast to the time-ordered products appearing in the Dyson series. 
For example,  at second  order in perturbation theory, the Dyson series  for the $S$-matrix   contains the time-ordered product of two Hamiltonians, $T\{ H_I(t_1)H_I(t_2)\}$, while the Magnus expansion  for the $N$-operator  features terms such as 
$\theta(t_1-t_2)[H_I(t_1),H_I(t_2)]$, which in turn give rise to causal propagators. 

This reorganisation of the perturbative series has profound consequences for its mathematical structure and physical implications.
In particular, at the $n^{\rm th}$ order the relevant term 
scales as $[H_I(t_1), [H_I(t_2), [\ldots, [H_I(t_{n-1}), H_I(t_n)]\ldots]]] / \hbar^{n} \!\sim\! \mathcal{O}(1/\hbar)$ if we assume that $[H_I(t_i), H_I(t_{i+1})] \sim \hbar$, and hence such a contribution is classical.
Crucially, this structure ensures that no hyper-classical terms%
\footnote{These are terms that diverge faster than $1/\hbar$ in the classical $\hbar\to 0$ limit which can appear in $S$-matrix elements, but cancel out when one computes classical observables.} 
appear at any order  in the expansion of the $N$-operator (and of its matrix elements),  as discussed in \cite{Damgaard:2021ipf,Damgaard:2023ttc,Kim:2024svw}, and 
taking the classical limit becomes a simple task.

While Magnus's original work focused on linear differential equations, its application to quantum mechanics was recognised early on \cite{osti_4703875,Pechukas1966}. The expansion was quickly adopted in atomic and molecular physics, and became especially valuable in nuclear magnetic resonance spectroscopy, where it provided the foundation for Average Hamiltonian Theory in analysing time-dependent spin dynamics \cite{EVANS196872,Haeberlen1968,Klarsfeld1989}.
In condensed matter physics, the Magnus expansion became an important tool for understanding driven quantum systems, particularly in the context of Floquet theory and periodically driven Hamiltonians 
\cite{Casas2001,Eckardt2015,Bukov2015,Kuwahara2016}. Mathematicians have also long been fascinated by the Magnus expansion, initially motivated by  its connection to the Baker-Campbell-Hausdorff formula and Lie group theory mentioned earlier, see e.g.~\cite{Ebrahimi_Fard_2025}, with more recent applications extending to the computation of Feynman integrals%
\footnote{A comprehensive treatment of the Magnus expansion and its applications, with extensive references, is given in \cite{Blanes2009}, while a more pedagogical exposition is provided in \cite{Blanes_2010}.} \cite{Argeri:2014qva}. 

Recent work \cite{Kim:2024svw,Kim:2025olv}  has demonstrated that the Magnus expansion provides a natural framework for computing the radial action in worldline quantum field theory (WQFT) \cite{Mogull:2020sak} 
-- a quantity that encodes classical scattering data without quantum corrections, and which  is also free of hyper-classical contributions.  
This development represents a significant advance in extracting classical physics from quantum  calculations.
The key technical innovation in the approach of \cite{Kim:2024svw}  is  a  representation of the Magnus expansion in terms of ``directed trees'', diagrams where retarded and advanced propagators are shown as arrows. Each tree diagram~$\tau$ is weighted by a combinatorial factor $\omega(\tau)$ that can be computed algorithmically  using an extension  of a formula by Murua \cite{Murua_2006}, thus
providing diagrammatic rules for matrix elements at tree level in the WQFT. For this reason, we refer to $\omega(\tau)$ as Murua coefficients in the rest of this work.

The radial action $I_r(J) = \int\!dr\, p_r$ is a central object in classical gravitational scattering, as 
it encodes the complete information about classical trajectories, both for bound and unbound orbits.  In particular, the scattering angle can then be obtained  as $\chi = -\partial I_r/\partial J$,
and other observables such as the periastron advance for bound orbits can also be derived from it \cite{Kalin:2019rwq}.
The connection between scattering amplitudes and the radial action was introduced in \cite{Bern:2021dqo}, schematically written as 
$M({\vec{q}}) = \int_J \big( e^{I_r(J) }-1\big)$, where $M$ is the four-point (elastic) amplitude and ${\vec{q}}$ is the transferred momentum. 
It is also   believed  that the radial action describing the classical dynamics of spinless black holes or neutron stars 
can be extracted  from the  four-point matrix element of the $N$-operator \cite{Damgaard:2021ipf}. 
In that paper,  the scattering angle of two spinless black holes was computed to the third Post-Minkowskian (PM) order  precisely from the four-point $N$-matrix element (see also \cite{Brandhuber:2021eyq} for a derivation from the closely related HEFT expansion). The use of the $N$ operator was further explored in \cite{Damgaard:2023ttc}, where the momentum kick  (and hence  the scattering angle) was obtained at fourth PM order.  The relation between the radial action and other,   well-known exponential representations of  classical scattering (e.g.~the eikonal, see the review \cite{DiVecchia:2023frv} and references therein) was explored recently in \cite{Kim:2025gis, Kim:2025sey}.
The $N$-operator is also a key ingredient in the recently developed Dirac bracket formalism \cite{Gonzo:2024zxo,Alessio:2025flu,Kim:2024grz, Kim:2024svw, Kim:2025hpn}, where observables are derived by taking iterated brackets of $N$-matrix elements. Indeed, at least in the conservative sector of two-to-two scattering, iterated quantum commutators like those described above become Dirac brackets in the classical limit.

The calculations of \cite{Damgaard:2021ipf,Damgaard:2023ttc} were performed by relating $N$-matrix elements (which we will 
refer to as Magnus amplitudes) to $S$-matrix elements,  that is ordinary scattering amplitudes, which are computed with the usual Feynman $i\varepsilon$ prescription. 
 As mentioned earlier, $N$-matrix elements are free of hyper-classical terms;  furthermore  the $N$-matrix is hermitian and this implies that $N$-matrix elements of real scalars are real. 
However, by computing them via scattering amplitudes, one is forced to compute such hyper-classical terms, even though ultimately they cancel. The HEFT approach of \cite{Brandhuber:2021eyq} partially improves on this by working directly with classical quantities and thus 
avoiding the computation and the need to explicitly cancel hyper-classical terms, at least up to two loops.  Yet this approach leads to matrix elements which contain an imaginary, divergent part and hence are not immediately  (though  very closely)  related to Magnus amplitudes and the radial action (up to two loops, the radial action can be identified with the real part of the four-point HEFT matrix element). 

The preceding  discussion motivates  the main goal of this paper: the direct computation of Magnus amplitudes    completely bypassing scattering amplitudes, in a  relativistic quantum field theory setting. While ultimately, in a gravitational-wave context, one may be interested in just computing classical terms, we will keep our discussion general. Unlike \cite{Kim:2024svw}, we  work in a quantum field theory, specifically that   of a real scalar field with a  $\phi^3$ coupling. This is   a toy model that nevertheless captures  all the essential features of a general calculation, and the main lessons can then  be  exported to other theories. Furthermore, many integral functions in a theory with a single mass (or no mass) can be represented within $\phi^3$ theory. Therefore one can imagine performing a calculation in a general quantum field theory, reducing it to  a set of integrals that correspond  to diagrams in our $\phi^3$ theory,  which can then be evaluated with the prescriptions developed in this paper.

Generally speaking, these prescriptions correspond to replacing Feynman propagators with advanced, retarded or cut propagators, and weighting each diagram with specific factors. Our calculations will reveal that, both at tree level and one loop (and partially at higher loops), these factors are precisely the  Murua coefficients. Once this is established, one may  compute them either directly or, much more economically,   by applying   Murua's formula. Importantly, as we shall see, all the $\theta$-functions that appear in (and are a key feature  of)  the Magnus expansion 
(see e.g.~\eqref{eq:CompactMagnusExprs})  recombine to yield advanced and retarded propagators, as expected from Lorentz invariance.
 
We will also find that  $N$-matrix elements at different loop orders are tightly interconnected  (in a way illustrated in Figure~\ref{fig:MagnusCutStructureFL}). At one loop this relation takes on an especially simple and beautiful form: the $n$-point one-loop Magnus amplitudes are completely  determined by the phase-space integral of forward limits of the corresponding $(n+2)$-point tree-level Magnus amplitudes.  Interestingly, this statement can be generalised to higher loops, as we will study  in the following sections.%
\footnote{The idea of obtaining loops from integrated trees appears in several contexts. The Feynman tree theorem provides an early example \cite{Feynman:1963ax,Feynman:1972mt}, with modern applications related to the tree  theorem appearing in \cite{Brandhuber:2005kd,Caron-Huot:2010fvq}, in the loop–tree duality approach of \cite{Catani:2008xa,Bierenbaum:2010cy}, and in the $Q$-cuts of \cite{Baadsgaard:2015twa}.}

The relations described above were found using an improved set of Wick contractions specialised to the Magnus expansion. We have implemented these Wick contractions in code, obtaining explicit results for (unintegrated) Magnus amplitudes in $\phi^3$ theory up to six vertices and four loops, and seven vertices and two loops. These results are provided in the companion repository \href{https://github.com/QMULAmplitudes/Magnus-Expansion-in-QFT}{{\it Magnus expansion in QFT}}.

Up to this point the  discussion has concerned the full quantum theory, but for applications to gravity and the radial action it is important to focus on the  classical limit. Based on a simple example, we  put forward a  conjecture directly relating particular cut diagrams in the field theory to diagrams in the WQFT. 

The Magnus expansion offers then a powerful framework for studying perturbative quantum field theory, making unitarity manifest and providing direct access to classical observables. By developing explicit prescriptions and illustrating them in a simple yet representative scalar model, we aim to clarify the underlying structure of Magnus amplitudes at loop level and its connection to Murua's work, at the same time  establishing practical tools for future applications, particularly to gravitational physics.

The rest of the paper is organised as follows. In the next section we give a brief review of the Dyson and Magnus expansions, contrasting the main characteristics of the two: while the first involves time-ordered products of the Hamiltonian, the second involves nested commutators of the same quantity. We will also present a new closed formula for the $n^{\rm th}$ order term in the Magnus expansion, as an alternative to the known Chen-Strichartz formula. Section~\ref{sec:functions-review} summarises the cast of characters that will appear in what will follow: commutation functions, propagators and the Hadamard function. We then begin to explore Magnus amplitudes perturbatively, first at tree level 
in Section~\ref{sec:treelevel} and then at one loop in Section~\ref{sec:oneloop}. We will compute Magnus amplitudes explicitly, via Wick contractions, up to six (three) points at tree level (one loop), from which we will learn two key lessons: that each diagram is weighted with particular factors, which we will later associate to  
Murua coefficients; and that one-loop coefficients are related to tree-level ones. After gaining  this insight, in Section~\ref{sec:redoMurua} we show how at tree level these coefficients can be obtained  in a much simpler way  from the Murua formula. We also establish a tree/one-loop correspondence: as mentioned earlier each one-loop diagram contains one cut propagator,  and its coefficient is half the Murua coefficient of the tree-level diagram obtained by deleting  the cut line (see \eqref{eq:oneLooprelationship}). 
This fact reflects a  general correspondence between loops and trees, which we explore in Section~\ref{sec:treestoloops}.
We establish this relation using a set of simplified Wick-contraction rules that we have developed, and summarise it in Figure~\ref{fig:MagnusCutStructureFL}.
 The main lesson is that Magnus amplitudes with $L+\ell$ loops and $C+\ell$ cuts can be obtained from $\ell$ forward limits of Magnus amplitudes with $L$ loops and $C$ cuts. In particular, this means all amplitudes with $\ell$ loops and $\ell$ cuts can be obtained from $\ell$ forward limits of tree-level amplitudes; we believe these contain all classical contributions and present evidence for this in Section~\ref{sec:classlim}, where we also show in a simple example the cancellation of hyper-classical terms. 
We conclude in Section~\ref{sec:bye} by summarising our results and outlining possible directions for future work.
A few Appendices complete the paper. In particular Appendix~\ref{App:summaryrules} summarises all the main diagrammatic properties of the Magnus expansion, while the remaining appendices 
expand on some technical aspects of our work.

\section{Dyson and Magnus: old and new}
\subsection{Review of the Dyson and Magnus expansions}
\label{sec:review}

We begin by recalling  that  the evolution of a state $|t\rangle$  in the interaction picture is governed by the equation (setting $\hbar=1$ from now on)
\begin{equation}
\label{schro}
    i\frac{\partial}{\partial t}\ket{t} = H_I(t)\ket{t}\, . 
\end{equation}
Here the Hamiltonian is $H\coloneq H_0 + V$, where 
 $H_I(t)$ is the interaction Hamiltonian time-evolved with the free Hamiltonian $H_0$, that is $H_I(t)\coloneq e^{i H_0 t} V e^{- i H_0 t}$. By writing the states in terms of the time-evolution operator, $\ket{t} = U(t,t_0)\ket{t_0}$, we can recast \eqref{schro} as an operator differential equation,
\begin{equation}\label{eqn:TimeEvolOpEqn}
    i\frac{\partial}{\partial t}U(t,t_0) = H_I(t) U(t,t_0)\, .
\end{equation}
This can be solved iteratively to yield the Dyson series \cite{Dyson:1949bp},
\begin{align}
\label{eq:U}
    U(t,t_0) &= 1 + (-i) \int_{t_0}^tdt_1\ H_I(t_1) + (-i)^2\int_{t_0}^t\int_{t_0}^{t_1}dt_1 dt_2\ H_I(t_1) H_I(t_2)+\cdots \nonumber\\
    &= 1 + (-i)\int_{[t_0,t]}dt_1\ H_I(t_1)+\frac{(-i)^2}{2!}\int_{[t_0,t]^2}dt_1 dt_2\ T\big \{H_I(t_1) H_I(t_2)\big \} +\cdots \nonumber\\
    &= 1 + \sum_{n=1}^\infty \frac{(-i)^n}{n!}\int_{[t_0,t]^n}dt_1\cdots dt_n\ T\big \{H_I(t_1)\cdots H_I(t_n)\big \}\, .
\end{align}
Here $T$ denotes the usual Dyson time-ordering prescription
 \begin{align}
    T\!\left\{ H_I(t_1)\cdots H_I(t_n)\right\}
\!=\! \sum_{\pi \in S_n}
\theta\big(t_{\pi(1)} - t_{\pi(2)}\big)\cdots
\theta\big(t_{\pi(n-1)} - t_{\pi(n)}\big)
H_I\big(t_{\pi(1)}\big)\cdots H_I\big(t_{\pi(n)}\big),
\end{align}
where the sum is over all permutations of $\{1,\dots,n\}$.
Evaluating \eqref{eq:U} from $t_0=-\infty$ to $t= \infty$ computes the $S$-matrix 
\begin{align}
    S := U(+\infty,-\infty)\, , 
\end{align}
which is then expanded as
\begin{align}
    S &= 1  + \sum_{n=1}^\infty \frac{(-i)^n}{n!}\int dx_1\cdots dx_n\ T\big \{\cH_I(x_1)\cdots \cH_I(x_n)\big \}\, , 
\end{align}
 with $H_I(t)\coloneq \int\!d^3x \, \cH_I(x)$.

An alternative way to solve  \eqref{eqn:TimeEvolOpEqn} is to instead write the time evolution operator itself in exponential form \cite{Magnus:1954zz},
\begin{equation}
    U(t,t_0):=e^{iN(t_,t_0)}\, ,
\end{equation}
and  solve for the operator $N(t,t_0)$. The derivation of this solution is more involved, with a general form given by Magnus \cite{Magnus:1954zz}. The first few terms of this Magnus series are
\begin{subequations}
\begin{align}
    iN^{(1)}(t,t_0) &= -i\int_{[t_0,t]}dt_1\ H_1 \, , \\
    iN^{(2)}(t,t_0)&=\frac{(-i)^2}{2}\int_{[t_0,t]^2} dt_1 dt_2\ \theta_{12} \ [H_1,H_2]\, ,   \\
   i N^{(3)}(t,t_0)&=\frac{(-i)^3}{6}\int_{[t_0,t]^3} dt_1 dt_2dt_3\ \theta_{12}\theta_{23}\,  \left([H_1,[H_2,H_3]]+[H_3,[H_2,H_1]]\right)\\
   i N^{(4)}(t,t_0)& =\frac{(-i)^4}{12}\int_{[t_0,t]^4} dt_1 dt_2 dt_3 dt_4\ \theta_{12}\theta_{23}\theta_{34} \big( [H_1,[H_2,[H_3,H_4]]] \\
    &+[H_4,[H_3,[H_1,H_2]]]+[H_2,[H_3,[H_4,H_1]]] + [H_1,[H_4,[H_3,H_2]]] \big)\, , \nn
\end{align}
\end{subequations}
where we have written $H_i:=H_I(t_i)$ and $\theta_{ij}:=\theta(t_i-t_j)$ for compactness. 
We can write similar formulae for the $S$-matrix. Defining \begin{align}
\label{SdefN}
    S\coloneq e^{i N} = e^{iN(+\infty,-\infty)}\, ,
\end{align} 
 we have, for  the first few orders,%
 \footnote{We note that there are other formalisms where nested commutators make an appearance, e.g.~the computation of in-in correlators \cite{Schwinger:1960qe,Keldysh:1964ud}, cosmological correlators \cite{Weinberg:2005vy}, and the direct computation of probabilities in scattering processes \cite{Dickinson:2025kuy}. }
\begin{subequations}\label{eq:CompactMagnusExprs}
\begin{align}
    iN^{(1)} &= -i\int\!d^4x_1\ \cH_1\, ,  \\
    iN^{(2)}&=\frac{(-i)^2}{2}\int\!d^4x_1 d^4x_2\ \theta_{12} \ [\cH_1,\cH_2]\, ,   \\
   i N^{(3)}&=\frac{(-i)^3}{6}\int\!d^4x_1 d^4x_2 d^4x_3\ \theta_{12}\theta_{23}\,  \left([\cH_1,[\cH_2,\cH_3]]+[\cH_3,[\cH_2,\cH_1]]\right) \,\\
   \label{N4}
    i N^{(4)}& =\frac{(-i)^4}{12}\int d^4x_1 d^4x_2 d^4x_3 d^4x_4\ \theta_{12}\theta_{23}\theta_{34} \big( [\cH_1,[\cH_2,[\cH_3,\cH_4]]] \\
    &+[\cH_4,[\cH_3,[\cH_1,\cH_2]]]+[\cH_2,[\cH_3,[\cH_4,\cH_1]]] + [\cH_1,[\cH_4,[\cH_3,\cH_2]]] \big)\, . \nn
\end{align}
\end{subequations}
 At low perturbative orders, expressions involving a small number of nested commutators are known and have been reproduced in the formulae above. However, there is no general formula which provides  compact expressions for any $n$, though they can in principle be produced algorithmically%
 \footnote{For this reason there is an interest in working with truncations of the Magnus expansion, see for instance \cite{Apel:2025upy} for recent work in this direction. }
 \cite{Prato:1997zz}. Relaxing the requirement to have as few terms as possible, there do exist general closed form expressions for any $n$. One of these, which we will make use of later, is known as the Chen-Strichartz (CS) formula \cite{STRICHARTZ1987320}, and  reads
\begin{align}\label{eqn:ChenStrichartzFormula}
    iN=\sum_{n\geq1}\sum_{\sigma\in S_n}\frac{(-1)^{d_\sigma}}{n^2\binom{n-1}{d_\sigma}}\int&\! d^4x_1\cdots d^4x_n\ \theta_{12}\theta_{23}\cdots \theta_{n-1\, n}\\ \nn
&\Big[\cH_{\sigma(1)},\big[\cH_{\sigma(2)},\cdots[\cH_{\sigma(n-1)},\cH_{\sigma(n)}]\cdots\big]\Big]\, .
\end{align}
Here the natural number $d_\sigma = |D_\sigma|$ is the descent number  of the permutation $\sigma$, and $D_\sigma$ its descent set:%
\begin{equation}\label{eq:DescentSetDef}
    D_\sigma = \big\{i\ | \ \sigma(i)>\sigma(i+1), 1\leq i <n\big\}\, .
\end{equation}
The descent set labels at which points in the permutation two consecutive numbers are descending, and the descent number just counts the number of such occurrences. For example, for the set $\{ 1,2,3,4\}$, 
\begin{equation}
    D_{4213} = \{1,2\}\, ,\qquad D_{3124} = \{1\}\, ,\qquad D_{4321} = \{1,2,3\}\, , \qquad 
    D_{2413} = \{2\}\, , 
\end{equation}
 with
 \begin{equation}
    d_{4213} = 2\, ,\qquad d_{3124} = 1\, ,  \qquad d_{4321} = 3\, , \qquad 
    d_{2413} = 1\  .
\end{equation}
The CS formula contains $n!$ terms.  There are quick ways to reduce the number of terms contributing to a certain order in the Magnus expansion, such as flipping the innermost commutator on half of the terms, but this spoils the structure of \eqref{eqn:ChenStrichartzFormula} as doing so  mixes permutations of different descent number. %
In Section~\ref{sec:newformula}, we will be able to present an alternative formula for the Magnus expansion where the sum is not over permutations but over ``descent sets'',  which 
contains only $2^{n-1}$ terms.

We also notice that, due to Jacobi relations, the nested commutators form an overcomplete basis. In particular one can recast \eqref{eqn:ChenStrichartzFormula} as \cite{arnal2017generalformulamagnusexpansion,Arnal:2020xpt}
\begin{align}
\label{CS-2}
    iN=\sum_{n\geq1}\sum_{\sigma\in S_{n-1}}\frac{(-1)^{d_\sigma}}{n \binom{n-1}{d_\sigma}}\int&\! d^4x_1\cdots d^4x_n\ \theta_{12}\theta_{23}\cdots \theta_{n-1\, n}\\ \nn
&\Big[\cH_{\sigma(1)},\big[\cH_{\sigma(2)},\cdots[\cH_{\sigma(n-1)},\cH_n]\cdots\big]\Big]\, , 
\end{align}
where one Hamiltonian is inserted at a fixed point $n$, which is reminiscent of the work of \cite{Weinberg:2005vy}.

\subsection{A new formula for the Magnus expansion}
\label{sec:newformula}
In this section we present an alternative formula \eqref{eq:ReformulationCS} for the Magnus expansion, requiring the reorganisation of the sum over permutations present in the Chen-Strichartz formula \eqref{eqn:ChenStrichartzFormula}.  Our formula  contains $2^{n-1}$ terms at order $n$ in perturbation theory, hence  it presents a computational advantage compared to the $n!$ scaling of the CS formula.

Consider the term in the CS formula \eqref{eqn:ChenStrichartzFormula} at a fixed order $n$ in perturbation theory, $iN^{(n)}$. We can group all the permutations in $S_n$ appearing in the sum by their descent set, defined in \eqref{eq:DescentSetDef}. Suppose we label the set of all permutations with descent set $D$ by $S_n^D$. Consider similarly dividing the sum in $iN^{(n)}$ into chunks of constant descent set,
\begin{align}
    iN^{(n)}=&\sum_{D} \frac{(-1)^{|D|}}{n^2\binom{n-1}{|D|}}\sum_{\sigma\in S^D_n}\int\! d^4x_1\cdots d^4x_n\ \theta_{12}\theta_{23}\cdots \theta_{n-1\, n}\nn\\
    &\hspace{85pt}\Big[\cH_{\sigma(1)},\big[\cH_{\sigma(2)},\cdots[\cH_{\sigma(n-1)},\cH_{\sigma(n)}]\cdots\big]\Big]\ \nn \\
    \coloneq&\sum_{D} \frac{(-1)^{|D|}}{n^2\binom{n-1}{|D|}} \mathrm{CS}^{(n)}_D\, .
\end{align}
For the case of $n=4$, the grouping of permutations is shown in Table~\ref{tab:PermutationsByDescent}.
\begin{table}[t]
\centering
\begin{tabular}{l|lll}
\hline
$d$                  & $D_\sigma$     & $\sigma$ & $\Theta_D$ \\
\hline\hline
0                 & $\{\}$     & 1234 & $\theta_{12}\theta_{23}\theta_{34}$ \\
\hline
\multirow{3}{*}{1} & $\{1\}$     & 2134, 3124, 4123 & $\theta_{21}\theta_{23}\theta_{34}$ \\
                     & $\{2\}$     & 1324, 1423, 2314, 2413, 3412 & $\theta_{12}\theta_{32}\theta_{34}$ \\
                     & $\{3\}$     & 1243, 1342, 2341 & $\theta_{12}\theta_{23}\theta_{43}$ \\
\hline
\multirow{3}{*}{2} & $\{1,2\}$   & 3214, 4213, 4312 & $\theta_{21}\theta_{32}\theta_{34}$ \\
                     & $\{1,3\}$   & 2143, 3142, 3241, 4132, 4231 & $\theta_{21}\theta_{23}\theta_{43}$ \\
                     & $\{2,3\}$   & 1432, 2431, 3421 & $\theta_{12}\theta_{32}\theta_{43}$ \\
\hline
3                  & $\{1,2,3\}$ & 4321 & $\theta_{21}\theta_{32}\theta_{43}$\\
\hline
\end{tabular}
\caption{Permutations $\sigma\in S_4$ grouped by descent set $D_\sigma$. The final column shows the corresponding value of $\Theta_D$ as given by \eqref{eq:DescentSetThetaIdentity}.}
\label{tab:PermutationsByDescent}
\end{table}
For concreteness, focus on the set of permutations with $D=\{1\}$ and $n=4$. Reading from the table, these are $2134, 3124, 4123$. Suppressing the integrations,  
\begin{equation}
    \mathrm{CS}^{\{1\}}_4=\theta_{12}\theta_{23}\theta_{34} \left([\cH_2,[\cH_1,[\cH_3,\cH_4]]] + [\cH_3,[\cH_1,[\cH_2,\cH_4]]]+[\cH_4,[\cH_1,[\cH_2,\cH_3]]]\right).
\end{equation}
Since the variables $x_1,\dots, x_4$ are integrated over, we are allowed to relabel them as we please. In particular, we can relabel them in such a way that at the end the term is proportional to the $1234$ nested commutator,
\begin{align}
    \mathrm{CS}^{\{1\}}_4 &\rightarrow \left(\theta_{21}\theta_{13}\theta_{34}+\theta_{23}\theta_{31}\theta_{14}+\theta_{23}\theta_{34}\theta_{41}\right) [\cH_1,[\cH_2,[\cH_3,\cH_4]]]\nn\\ &=\theta_{21}\theta_{23}\theta_{34}[\cH_1,[\cH_2,[\cH_3,\cH_4]]]\, ,
\end{align}
where we have repeatedly used the $\theta$-function identities
\begin{equation}\label{eq:ThetaIds}
    \theta_{xa}\theta_{ab} + \theta_{xb}\theta_{ba} = \theta_{xa}\theta_{xb} \qquad \theta_{ab}\theta_{bx} + \theta_{ba}\theta_{ax} = \theta_{ax}\theta_{bx}\, .
\end{equation}
This can be done for every descent set, and the results are tabulated in the final column of Table~\ref{tab:PermutationsByDescent} above. Interestingly, we see that the final $\theta$-functions correspond to the original ordered $\theta_{12}\theta_{23}\theta_{34}$, with some of the $\theta$-functions flipped $\theta_{ij}\to\theta_{ji}$. The location of these flips exactly coincides with the values of the descent set, 
in the sense that for $D=\{1,2\}$, the first and second  $\theta$-functions are flipped, and so on.

We see this pattern appearing for any $n$, with explicit checks up to and including $n=8$. In general, if we sum over the set of permutations of $n$ objects with descent set $D=\{i_1,\dots,i_k\}$ (and consequently with ascent set $A=\{j_1,\dots,j_{(n-1)-k}\}$, such that $A\cup D=\{1,\dots, n{-}1\}$), then we find that
\begin{equation}\label{eq:DescentSetThetaIdentity}
    \Theta_D\coloneq \sum_{\sigma\in S^D_n}\sigma^{-1}(\theta_{12}\cdots\theta_{n{-}1\ n}) = \left(\prod_{i\in D}\theta_{i{+}1\ i}\right)\left(\prod_{j\in A}\theta_{j\ j{+}1}\right)\, .
\end{equation}
A proof of the identity above is given  in Appendix~\ref{sec:ProofNewMagnusFormula}. The value of $\Theta_D$ is tabulated in the last column of Table~\ref{tab:PermutationsByDescent}. This means that we can rewrite the CS formula in the following new  form:
\begin{equation}\label{eq:ReformulationCS}
    \boxed{
    iN^{(n)}=\int\! d^4 x_1\cdots d^4 x_n\Big[\cH_{1},\big[\cH_{2},\cdots[\cH_{n-1},\cH_n]\cdots\big]\Big]\sum_{D\subseteq \{1,\dots,n-1\}} \frac{(-1)^{|D|}}{n^2\binom{n-1}{|D|}} \Theta_{D}  }\ .
\end{equation}
We now wish to comment on some features of this formula which, 
to the best of our knowledge,  has not been presented in the literature before.

{\bf 1.}~First, note that in the new formula only one particular ordering of nested commutators of  Hamiltonians appears, namely $\{1, 2, \ldots, n\}$. 

{\bf 2.}~The sum to be performed is not over permutations, but over descent sets. Each descent set has a unique product of $\theta$-functions associated to it, given in 
\eqref{eq:DescentSetThetaIdentity}. 

{\bf 3.}~We also note that \eqref{eq:ReformulationCS} is a sum over $2^{n{-}1}$ terms, which presents a computational improvement over the $n!$ terms in the original CS formula (or the $(n{-}1)!$ terms of the linearly independent version \eqref{CS-2}).

While for small values of $n$, up to $n=4$, the compact formulae in \eqref{eq:CompactMagnusExprs} are shorter than \eqref{eq:ReformulationCS}, starting at $n=5$ the expression found by \cite{Prato:1997zz} contains 22 terms, compared to the 16 coming from our new formula.

{\bf 4.}~Lastly, we can reduce the number of terms further by rewriting the sum in \eqref{eq:ReformulationCS} in a different way. If we replace each $\theta_{i+1\,i}=1-\theta_{i\, i+1}$ and collect like terms, we can rewrite the sum in 
\eqref{eq:ReformulationCS} to obtain the following expression \cite{mielnik1970combinatorial}
\begin{align}\label{eq:AscentMagnusFormula}
    iN^{(n)}&=\frac{(-1)^{n-1}}{n}\int\! d^4 x_1\cdots d^4 x_n\Big[\cH_{1},\big[\cH_{2},\cdots[\cH_{n-1},\cH_n]\cdots\big]\Big]\sum_{A\subseteq\{1,\dots,n{-}1\}}\frac{(-1)^{|A|}}{n-|A|} \tilde{\Theta}_A\ ,
\end{align}
where we have defined
\begin{equation}
    \tilde{\Theta}_A\coloneq\prod_{j\in A}\theta_{j\, j{+}1}\, .
\end{equation}
We have shown that our result  \eqref{eq:ReformulationCS}
agrees with \eqref{eq:AscentMagnusFormula} up to $n=15$. For example, for the $n=4$ case, the sum inside \eqref{eq:AscentMagnusFormula} reads
\begin{equation}
    -\frac{1}{4}\left(\frac{1}{4}-\frac{1}{3}(\theta_{12}+\theta_{23}+\theta_{34})+\frac{1}{2}(\theta_{12}\theta_{23}+\theta_{12}\theta_{34}+\theta_{23}\theta_{34})-\theta_{12}\theta_{23}\theta_{34}\right)\, . 
\end{equation}
While the sum produces $2^{n-1}$ terms, \cite{SAENZ2002357} argues that due to the symmetries of the nested commutators in the integrand the number of non-zero terms when using \eqref{eq:AscentMagnusFormula} is $3\cdot2^{n-3}$ (in the case of $n=4$ the vanishing terms are given by the term with no $\theta$-functions and $\theta_{12}$). 

We wish to conclude this section with some intriguing observations on these new formulae.
Looking at the new formula \eqref{eq:ReformulationCS}, one particular term in the sum corresponding to $D=\{1,\dots,n-1\}$ produces the combination of a chain of $\theta$-functions $\theta_{21}\theta_{32}\cdots\theta_{n\, n-1}$ where position $x_n$ is future-most. This is reminiscent of a causal response function, which also has nested commutators and a flow of time towards one point. In particular \cite{Lehmann:1957zz},
\begin{equation}
R(\cH_n;\cH_{n-1},\dots,\cH_1) {=} {\sum_{\sigma\in S_{n-1}}} \theta_{n\sigma(n-1)}\cdots\theta_{\sigma(3)\sigma(2)}\theta_{\sigma(2)\sigma(1)}[\dots[\cH_n,\cH_{\sigma(n-1)}],\dots,\cH_{\sigma(1)}] \, .
\end{equation}
We can easily undo this sum over permutations by considering the \emph{integrated} causal response function
\begin{align}
    \int\! d^4x_1 &\cdots d^4 x_n\  R(\cH_n;\cH_{n-1},\dots,\cH_1) \nn\\
    &= (-1)^{n-1}(n-1)!\int\! d^4 x_1\cdots d^4 x_n\ \theta_{21}\cdots \theta_{n\, n-1} [\cH_1,\dots,[\cH_{n-1},\cH_n]\dots]\, .
\end{align}
This causal response function is commonly computed by using the Schwinger-Keldysh formalism, using a contour with a single time-fold
\begin{align}
R(\cH_n;\cH_{n-1},\dots,\cH_1) 
&=\begin{tikzpicture}[baseline={([yshift=-0.5ex]current bounding box.center)}]
%
    \draw (0,0.5) -- ++(6,0) arc(90:-90:0.5) -- ++(-6,0);
    \draw[decorate,decoration={markings, mark=at position 0.5 with {\arrow[scale=1.5,xshift=3pt]{Latex}}}] (0,0.5) -- ++(1.75,0);
    \draw[decorate,decoration={markings, mark=at position 0.5 with {\arrow[scale=1.5,xshift=3pt]{Latex}}}] (1.75,-0.5) -- ++(-1.75,0);
    \node at (0.6,0.75){I};
    \node at (0.6,-0.75){II};
%
    \draw[fill=black] (5.75,0.5) circle(2pt) node[above]{$n$};
    \draw[fill=black] (4.75,0.5) circle(2pt) node[above]{$n-1$};
    \node at (3.75,0){$\cdots$};
    \draw[fill=black] (2.75,0.5) circle(2pt) node[above]{$2$};
    \draw[fill=black] (1.75,0.5) circle(2pt) node[above]{$1$};
%
    \draw[fill=black] (4.75,-0.5) circle(2pt) node[below]{\phantom{1}}; 
    \draw[fill=black] (2.75,-0.5) circle(2pt);
    \draw[fill=black] (1.75,-0.5) circle(2pt);
%
    \draw[dashed] (4.75,0.5) -- (4.75,-0.5);
    \draw[dashed] (2.75,0.5) -- (2.75,-0.5);
    \draw[dashed] (1.75,0.5) -- (1.75,-0.5);
%
\end{tikzpicture}\ ,
\end{align}
where the dashed line indicates a difference between the Hamiltonian on fold I and fold II. One might wonder whether the other terms in the expression for the $N$-operator can be interpreted in this contour language. Extending the basic Schwinger-Keldysh contour as done in \cite{Haehl:2017qfl}, we can indeed express each term in \eqref{eq:AscentMagnusFormula} as a particular generalised contour. For example 
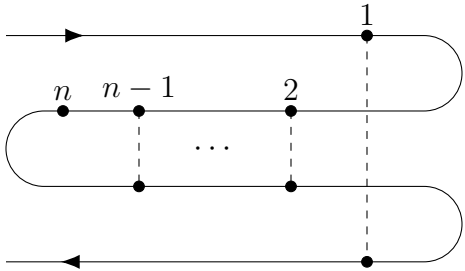
\begin{gather}
\begin{tikzpicture}[baseline={([yshift=-0.5ex]current bounding box.center)},xscale=-1]
%
    \draw (5.5,1.5) -- ++(-5.5,0) arc(90:270:0.5) -- ++(5,0) arc(90:-90:0.5) -- ++(-5,0) arc(90:270:0.5) -- ++(5.5,0);
    \draw[decorate,decoration={markings, mark=at position 0.5 with {\arrow[scale=1.5,xshift=3pt]{Latex}}}] (5.5,1.5) -- ++(-1.75,0);
    \draw[decorate,decoration={markings, mark=at position 0.5 with {\arrow[scale=1.5,xshift=3pt]{Latex}}}] (3.75,-1.5) -- ++(1.75,0);
%
    \draw[fill=black] (4.75,0.5) circle(2pt) node[above]{$n$};
    \draw[fill=black] (3.75,0.5) circle(2pt) node[above]{$n-1$};
    \node at (2.75,0){$\cdots$};
    \draw[fill=black] (1.75,0.5) circle(2pt) node[above]{$2$};
    \draw[fill=black] (0.75,1.5) circle(2pt) node[above]{$1$};
%
    \draw[fill=black] (3.75,-0.5) circle(2pt);
    \draw[fill=black] (1.75,-0.5) circle(2pt);
    \draw[fill=black] (0.75,-1.5) circle(2pt) node[below]{\phantom{$n$}};
%
    \draw[dashed] (3.75,0.5) -- (3.75,-0.5);
    \draw[dashed] (1.75,0.5) -- (1.75,-0.5);
    \draw[dashed] (0.75,1.5) -- (0.75,-1.5);
\end{tikzpicture}
\notag \\
\rightarrow (-1)^{n-1}(n-2)!\ \theta_{23}\cdots\theta_{n{-}2\,n{-}1}\theta_{n{-}1\, n}[\cH_1,\dots,[\cH_{n-1},\cH_n]\dots]
\end{gather}
omitting the integrations. Note that this is no longer an integrated response function, as the $\theta_{12}$ term is missing; this is what necessitates an additional time-fold. This shift in perspective demonstrates that each term of \eqref{eq:AscentMagnusFormula} is Lorentz invariant%
\footnote{We have also verified this experimentally for \eqref{eq:ReformulationCS}.
}
despite containing $\theta$-functions, as each contour can be viewed as an analytic continuation
\cite{Osterwalder:1973dx, Osterwalder:1974tc} of Euclidean correlator (Schwinger functions) \cite{Schwinger:1958mma}.


\section{Field commutators and propagators}

\label{sec:functions-review}

Compared  to the Dyson series, the Magnus series forgoes the time ordering in favour of imposing a fixed ordering among the $t_i$ variables. Specifically, it features nested commutators of  Hamiltonians instead of their time-ordered  products.
As a consequence, when evaluating matrix elements of the $N$-operator we need to use a version of Wick's theorem for products of fields, rather than time-ordered products.
The main relation, taking as an example the case of a single real scalar field $\phi$, is
\begin{align}
    \phi (x_i) \phi (x_j) = :\!\phi (x_i) \phi (x_j)\!: + \, \wick{\c1\phi(x_i) \c1\phi(x_j)}\, , 
\end{align}
where $:\!\phi (x_i) \phi (x_j)\!:$ is referred to as the normal-ordered product of $\phi (x_i)$ and $ \phi (x_j)$, and 
the Wick contraction, or pairing, of the two fields is%
\footnote{An  expression for the  function $i \Delta^{(+)} (x)$ will be given in \eqref{explexpredpm} below.}
\begin{align}
\label{firstW}
\wick{\c1\phi(x_i) \c1\phi(x_j)} = 
\langle 0 | \phi (x_i) \phi (x_j) | 0\rangle \coloneq i \Delta^{(+)} (x_i - x_j) =
i\Delta^{(+)}_{ij}\, .
\end{align}
Here we have introduced the abbreviated notation $\Delta_{ij}^{(\pm)}=\Delta^{(\pm)}(x_i-x_j)$.
In the case of many operators this relation extends to
\cite{Bogolyubov:1959bfo}
\begin{align}\label{eq:Wick}
\begin{split}
    \phi (x_1) \cdots \phi (x_n) &= :\!\phi (x_1) \cdots \phi (x_n)\!: + :\!\wick{\c1\phi(x_1) \c1\phi(x_2)}\cdots \phi (x_n)\!: + \cdots \\ &
    + 
    :\!\wick{\c1\phi(x_1) \cdots \c1\phi(x_n)}\!: + 
    :\!\wick{\c1\phi(x_1) \c1\phi(x_2) \c2 \phi (x_3) \c2 \phi (x_4) \cdots }\phi(x_n)\!: + \cdots 
    \, .  
    \end{split}
\end{align}

In what follows, we consider products of normal-ordered Hamiltonians. As in the standard application of Wick's theorem to time-ordered products, no contractions occur between fields within the same normal-ordered block.

In the following, we  introduce a number of functions we will need which describe commutators and anticommutators  of two fields, as well as various Green's functions. For the sake of completeness we now review  these quantities in some detail.

\subsection{Wightman commutation functions}

We start with  the standard mode expansion of a real, massive scalar field: 
\begin{align}
\begin{split}
\label{fieldexp}
\phi(x) & = \phi^{(+)}(x)+\phi^{(-)}(x) =  \int \frac{d^3p}{(2 \pi)^3 2 E(\vec{p})} \Big[  a(\vec{p}) e^{-i p \cdot x} + a^\dagger(\vec{p}) e^{i p \cdot x} \Big]  \, , \, \end{split}
\end{align}
where the positive/negative frequency parts $\phi^{(\pm)}(x)$ by definition contain the lowering/raising operators,  
with
\begin{align}
    [a(\vec{p}),a^\dagger(\vec{q})] = (2 \pi)^3 2 E(\vec{p}) \delta^{(3)}(\vec{p}-\vec{q}) \, .
\end{align}
The Wightman commutation function $\Delta^{(+)}(x) $ that has appeared in \eqref{firstW} and the related one $\Delta^{(-)}(x) $ are defined as%
\footnote{Our conventions are almost identical to those of Appendix C of Bjorken and Drell (BaD) \cite{Bjorken:1965zz}. We depart from theirs in that $\Delta^{(-)}_{\rm US} (x) = - \Delta^{(-)}_{\rm BaD}(x)$. 
}
\begin{align}
i\Delta^{(\pm)}_{ij}&=i\Delta^{(\pm)}(x_i-x_j)\coloneqq [ \phi^{(\pm)} (x_i), \phi^{(\mp)} (x_j)]  
\, ,
\end{align}
and satisfy 
\begin{align}
\Delta^{(\pm)} (-x) = - \Delta^{(\mp)} (x) \, . 
\end{align}
Explicit expressions for $\Delta^{(\pm)}(x)$ are  
\begin{align}
\begin{split}
    \label{explexpredpm}
    i \Delta ^{(\pm)} (x) &=[\phi^{(\pm)}(x)\, , \, \phi^{(\mp)}(0) ]\\ &= 
    \pm
    \int\!\frac{d^3k}{(2\pi)^3 (2 E(\vec{k})) }
\, e^{\mp i E(\vec{k}) x_0 + i \vec{k} \cdot \vec{x}} 
\\ & =
    \pm
   \left. \int\!\frac{d^3k}{(2\pi)^3 (2 E(\vec{k})) }
\, e^{\mp i k\cdot x} \right|_{k_0=E(\vec{k})} 
\\
& = \pm \int\!\frac{d^4k}{(2\pi)^4}\, e^{- i k\cdot x}\, (2\pi) \delta(k^2-m^2) \theta (\pm k_0) 
\  , 
\end{split}
  \end{align}
  where $E(\vec{k})\coloneq \sqrt{{\vec{k}}^{\, 2} + m^2}$ is the relativistic energy, 
so that in  momentum space the  Wightman functions are 
\begin{align}
i \Delta^{(\pm)} (k) & = \pm  (2\pi) \delta(k^2-m^2) \theta (\pm k_0)\,  .
\end{align}

\subsection{Pauli-Jordan and Hadamard functions}
\label{app:comm-fun}

Next we introduce the commutator of two fields (also known as Pauli-Jordan function) as well as the anticommutator (sometimes called Hadamard function):
\begin{align}
i \Delta_{ij} = i \Delta (x_i-x_j) &\coloneqq [ \phi (x_i), \phi (x_j)] =  i \Delta^{(+)}_{ij} + i \Delta^{(-)}_{ij}\, \label{eq:CommutatorFunction}  , 
\end{align}
and
\begin{align}
\Delta_{ij}^{(1)}=\Delta^{(1)}(x_i-x_j) &\coloneqq \langle 0 | \big \{ \phi (x_i), \phi(x_j)\big \}| 0\rangle 
=i \Delta^{(+)}_{ij} - i \Delta^{(-)}_{ij}\, \label{eq:AnticommutatorFunction} , 
\end{align}
with 
\begin{align}\label{eqn:DeltaDelta1Properties}
\Delta_{ji} = - \Delta_{ij} \, , \qquad  \Delta^{(1)}_{ji} =  \Delta^{(1)}_{ij} \, . 
\end{align}
Note that $\Delta(x)$ and $\Delta^{(1)}(x)$ are odd and even solutions to the homogeneous Klein-Gordon equation, 
\begin{align}
\begin{split}
    (\Box + m^2) \Delta(x) = (\Box + m^2) \Delta^{(1)}(x)=0\, . 
\end{split}\end{align}
We also quote the representations: 
\begin{align}
i \Delta (x) &= \int\!\frac{d^4k}{(2\pi)^4}\, e^{- i k\cdot x}\, (2\pi) \delta(k^2-m^2) \, \text{sgn} ( k_0) \, , 
\\
\label{Hadamard}
\Delta^{(1)} (x) &=\int\!\frac{d^4k}{(2\pi)^4}\, e^{- i k\cdot x}\, (2\pi) \delta(k^2-m^2) \,  , 
\end{align}
so that in momentum space the  Pauli-Jordan and Hadamard  functions are\footnote{As standard, we use the same symbol $\Delta$ for propagators in position and momentum space.}
\begin{align}
%
i \Delta (k) & = (2\pi) \delta(k^2-m^2) \, \text{sgn} (k_0)\, , \\
\label{Hadamard-mom}
 \Delta^{(1)} (k) & =  (2\pi) \delta(k^2-m^2) \, ,  
\end{align}
where $\text{sgn} (k_0) \coloneqq \theta (k_0) - \theta ( - k_0)$ is the  sign of $k_0$. 
 Note that  $\Delta^{(1)} (k)$  is proportional to an on-shell delta function with no restriction on the sign of the energy.  It is equal to twice a cut-propagator $\pi\,  \delta (k^2 - m^2)$. Finally, $\Delta(x)$ (unlike $\Delta^{(1)}(x)$) vanishes outside the lightcone, \begin{align}
     \Delta(x) = 0 \quad \text{for}\quad x^2 <0\, . 
 \end{align}

\subsection{Propagators}
\label{sec:propagators}

The  relevant Green functions are  
\begin{align}
i \Delta^F (x)  &\coloneqq i \Delta^{(+)}(x) \theta(x_0)  -  i \Delta^{(-)}(x) \theta(-x_0)\, ,  & \text{\tt Feynman}
\\
i \Delta^R (x)  &\coloneqq i \Delta(x) \theta(x_0)  \, , & \text{\tt Retarded}
\\
i \Delta^A (x)  & \coloneqq -i \Delta(x) \theta(-x_0)  \, , & \text{\tt Advanced}
\\
i \Delta^{\bar{F}} (x)  &\coloneqq i \Delta^{(-)}(x) \theta(x_0)  -  i \Delta^{(+)}(x) \theta (-x_0)\, . 
&\text{\tt Anti-Feynman}
\end{align}
Here $i \Delta^{\bar{F}} (x)$ is the Green function obtained with anti time ordering, picking a contour that is the parity-reflected of the Feynman contour. Since $\Delta (x) =0$ 
for $x^2<0$, i.e.~outside the 
lightcone, we see that the retarded and  advanced propagators have support within the future and past 
lightcone, respectively.
%
Note also the relations: 
\begin{align}
\label{advplusret}
i\Delta^F (x) + i\Delta^{\bar F} (x) &= i\Delta^R (x) + i\Delta^A (x)\, , \\
\label{FminusFbar}
i\Delta^F (x) - i\Delta^{\bar F} (x) &= \Delta^{(1)} (x) \, , 
\end{align}
as well as 
\begin{align}
\label{relretadv}
    \Delta^R(-x) = \Delta^A(x)\, . 
\end{align}
The  connection between propagators and (anti-)time-ordered products is 
\begin{align}
i \Delta^F (x) &= \langle 0 | T \big(\phi (x) \phi (0)\big) | 0\rangle \, , 
\\
 i\Delta^{\bar{F}} (x) &= -\langle 0 | \bar{T} \big(\phi (x) \phi (0)\big) | 0\rangle \, , 
\end{align}
with
\begin{align}
T\big( A(x) B(0) \big) &= \theta(x_0) A(x) B(0) + \theta(-x_0) B(0) A(x)\, , 
\\
\bar{T}\big( A(x) B(0) \big) &= \theta(x_0) B(0) A(x)+\theta(-x_0) A(x) B(0)  
\, . 
\end{align}
\noindent
In momentum space, the propagators  are%
\begin{align}
i\Delta^F(k) & = \frac{i}{(k_0-E(\vec{k}) + i \vareps)(k_0+E(\vec{k}) - i \vareps)}= \frac{i}{k^2 - m^2 + i \varepsilon}\, , \\
\label{RRR}
i\Delta^R(k) & = \frac{i}{(k_0 + i \varepsilon)^2 - E^2(\vec{k}) } = \frac{i}{k^2 - m^2 + i \vareps \, \text{sgn}(k_0)  } = \begin{tikzpicture}[baseline={([yshift=-2.8ex]current bounding box.center)},thick]
        \draw[Rnew] (0,0) -- (2,0);
        \draw[-Latex] (0.3, 0.3) -- (1.7, 0.3) node[midway, above]{$k$};
    \end{tikzpicture}
\, , \\
\label{AAA}
i\Delta^A(k) & = \frac{i}{(k_0 - i \varepsilon)^2 - E^2(\vec{k}) } = \frac{i}{k^2 - m^2 - i\vareps \, \text{sgn}(k_0)  } =
\begin{tikzpicture}[baseline={([yshift=-2.8ex]current bounding box.center)},thick]
        \draw[Anew] (0,0) -- (2,0);
        \draw[-Latex] (0.3, 0.3) -- (1.7, 0.3) node[midway, above]{$k$};
    \end{tikzpicture}
\, , 
\\
i\Delta^{\bar{F}}(k) & = \frac{i}{(k_0-E(\vec{k}) - i \vareps)(k_0+E(\vec{k}) + i \vareps)}= \frac{i}{k^2 - m^2 - i \varepsilon}\, . 
\end{align}
Note that we have introduced two arrows in 
\eqref{RRR} and \eqref{AAA}:
the arrow on the line carries the causality flow while the external arrow indicates the momentum flow. The two need not be the same.

\section{Tree-level amplitudes  in the 
\texorpdfstring{$\cH_I=\frac{:\phi^3:}{3!}$}{phicubed}
  theory}
  \label{sec:treelevel}

To describe the main ideas around the Magnus expansion, we focus  on the theory of a real scalar field  $\phi$ of mass $m$ with an  interaction Hamiltonian
\begin{align}
\label{phicubed}
\mathcal{H}_I = \frac{:\!\phi^3\!:}{3!}\, . 
\end{align}
We omit the coupling factor in the interaction above to simplify later expressions.
The examples we will present below will suffice to illustrate the key features of the matrix elements of the 
$N$-operator introduced in \eqref{SdefN}, which we will  refer to as Magnus amplitudes.  This includes the emergence of retarded and advanced propagators, and the characteristic weights of the various diagrams. The latter will be related to certain Murua coefficients, as we will discuss in Section~\ref{sec:redoMurua}.

\subsection{The three-point amplitude}
\label{3ptphicubed}

As a first simple example, we illustrate the three-point matrix element,
which we refer to as a Magnus amplitude:%
\footnote{For convenience  in all matrix elements we  include all the particles in the initial state, with all incoming momenta.}
\begin{align}\label{N1phicubed}
iN^{(1)}_3 \coloneq \braket{0|iN^{(1)}|\phi(p_1)\phi(p_2)\phi(p_3)}\,  , 
\end{align}
where 
\begin{align}
\label{first-order-Mag-exp}
i N^{(1)} &= 
-i\int\! d^4x\,\mathcal{H}_I(x)
=-i\int\! d^4x\,\frac{:\phi^3:}{3!}
\end{align}
is the leading-order term in the Magnus expansion, and we defined $|\phi(p_i)\rangle {\coloneq} a^\dagger (\vec{p}_i)|0\rangle$.
The effect of contracting with external states is simply to produce
\begin{align}
\frac1{3!}\braket{0|:\phi^3:|\phi(p_1)\phi(p_2)\phi(p_3)}=e^{-i(p_1+p_2+p_3)\cdot x}\,,
\end{align}
and, hence, the final matrix element is%
\footnote{We define $\hat{\delta}^{(n)}(p) \coloneq (2\pi)^n \delta^{(n)}(p)$.}
\begin{align}\label{N31}
iN_3^{(1)}=-i\int d^4x\,e^{-i(p_1+p_2+p_3)\cdot x}=-i\, \hat{\delta}^{(4)}(p_1+p_2+p_3)\,.
\end{align}
In this case, due to the absence of propagators there is no
difference to an ordinary $S$-matrix element.

\subsection{The four-point amplitude}
\label{4ptphicubed}

Our first non-trivial example is the four-point Magnus amplitude:
\begin{align}
\label{N2phicubed}
iN^{(2)}_4 \coloneq \braket{0|iN^{(2)}|\phi(p_1)\phi(p_2)\phi(p_3)\phi(p_4)}\,  , 
\end{align}
where 
\begin{align}
\label{second-order-Mag-exp}
\begin{split}
i N^{(2)} &=
\frac{(-i)^2}{2} \int\! d^4x_1d^4x_2\, \theta(t_1 - t_2)\,    \left[ \mathcal{H}_I(x_1), \mathcal{H}_I(x_2) \right]\, , 
\end{split}
\end{align}
is the second-order term in the Magnus expansion.
Performing Wick contractions using \eqref{firstW},
the relevant term in the commutator for computing the tree-level matrix element in \eqref{N2phicubed} is 
\begin{align}\label{fpfi}
[\mathcal{H}_I(x_1), \mathcal{H}_I(x_2)] &= \frac{1}{(3!)^2} [:\phi_1^3:, :\phi_2^3:] 
\ni \frac{9}{(3!)^2}  :\phi_1^2\wick{\c1 \phi_1 \c1 \phi_2}\phi_2^2: - (x_1 \leftrightarrow x_2) \nonumber \\
&= 
\frac{1}{(2!)^2} (i\Delta_{12}^{(+)}) :\phi_1^2\phi_2^2: - (x_1 \leftrightarrow x_2) 
= 
\frac{1}{(2!)^2} (i\Delta_{12}^{(+)} - i\Delta_{21}^{(+)}) :\phi_1^2\phi_2^2:\, 
\nonumber \\ & = \frac{1}{(2!)^2} (i\Delta_{12}) :\phi_1^2\phi_2^2:\, , 
\end{align}
where we recall that  $\Delta_{12} \coloneq  \Delta^{(+)}_{12}  + \Delta^{(-)}_{12} = 
\Delta^{(+)}_{12}  - \Delta^{(+)}_{21}$ is the commutation function.
Other possible Wick contractions will not contribute when we contract with external states.
We can also obtain this result using
\begin{subequations}
\begin{align}
[\phi_1^3,\phi_2^3 ] &= \phi_1^2 [\phi_1, \phi_2^3] + \phi_1 [\phi_1, \phi_2^3] \phi_1 + [\phi_1, \phi_2^3] \phi_1^2\, , \\
[\phi_1,\phi_2^3 ] &=[\phi_1, \phi_2]\phi_2^2 + \phi_2 [\phi_1, \phi_2]\phi_2 + \phi_2^2[\phi_1, \phi_2] = 
3 i\Delta_{12}   \phi_2^2 
\, , 
\end{align}
\end{subequations}
which together imply that
\begin{align}
\label{useful-rewriting}
\frac1{(3!)^2}[\phi_1^3,\phi_2^3 ] = \frac{1}{3\cdot (2!)^2} i\Delta_{12}   (\phi_1^2 \phi_2^2 + \phi_1 
\phi_2^2 \phi_1 + \phi_2^2 \phi_1^2)\, . 
\end{align}
Upon applying normal ordering this matches \eqref{fpfi}.
In this second approach, the function $i\Delta_{12}$ appears almost immediately,
and further multiplying it by $\theta_{12}$ gives rise to a retarded propagator $i\Delta^R_{12}$.
Ultimately,
\begin{align}\label{N2}
i N^{(2)} &= 
\frac{(-i)^2}{2}\int\! d^4x_1d^4x_2\, (i\Delta^R_{12})\left(\frac{1}{(2!)^2}:\phi_1^2\phi_2^2:+\cdots\right)\,,
\end{align}
is the relevant term in the Magnus series.
The dots indicate the remaining Wick contractions 
which do not enter the tree-level Magnus amplitude and that  are therefore omitted.
Crucially, we have now absorbed all factors of $\theta_{ij}$ to obtain a result
that is manifestly Lorentz invariant.

We may then consider the effect of contractions on external states. These give
\begin{align}
\label{tme4}
\frac{1}{(2!)^2}\braket{0|:\phi_1^2\phi_2^2:|\phi(p_1)\phi(p_2)\phi(p_3)\phi(p_4)} =
E(x_1,x_2) + E(x_2,x_1) \, ,
\end{align}
where
\begin{equation}
E (x_1,x_2) = 
e^{-i(p_1+p_2)\cdot x_1 - i(p_3+p_4)\cdot x_2} + e^{-i(p_2+p_3)\cdot x_1 - i(p_4+p_1)\cdot x_2} + e^{-i(p_1+p_3)\cdot x_1 - i(p_4+p_2)\cdot x_2}
.
\end{equation}
The full tree-level Magnus amplitude then becomes
\begin{align}
\begin{split}
\label{firstobs2}
iN^{(2)}_{4} &= 
\frac{(-i)^2}{2} \int d^4x_1 d^4x_2 \, (i \Delta^R_{12}) \Big[ E(x_1,x_2) + E(x_2, x_1) \Big] \\
&= \frac{(-i)^2}{2} \int d^4x_1 d^4x_2 \, (i\Delta^{R}_{12} + i\Delta^{A}_{12}) E(x_1,x_2)\, . 
\end{split}
\end{align}
where in the second line we have applied the interchange symmetry $x_1\leftrightarrow x_2$.
Fourier transforming yields
\begin{align}\label{4pttreeres}
iN^{(2)}_{4} &=  \hat{\delta}^{(4)}(p_1+p_2+p_3+p_4) \,  \mathcal{M}\, , \\
\mathcal{M} &= \frac{(-i)^2}{2} \left[ (i\Delta^{R} {+} i\Delta^{A})(p_1{+}p_2) + (i\Delta^{R} {+} i\Delta^{A})(p_2{+}p_3) + (i\Delta^{R} {+} i\Delta^{A})(p_1{+}p_3) \right]\, .\nn
\end{align}
Graphically, we can represent \eqref{4pttreeres} as 
\begin{align}\label{eq:4ptMomentumSpaceDiags}
\mathcal{M}=&\frac{(-i)^2}{2}\Biggl(
\begin{tikzpicture}[font=\small,baseline={([yshift=-.5ex]current bounding box.center)},thick] 
    \draw[massive] (0,0) -- (135:1) node[left]{$p_2$};
    \draw[massive] (0,0) -- (-135:1) node[left]{$p_1$};
    \draw[Rnew] (0,0) -- (1,0);
    \draw[massive] (1,0) -- ++(45:1) node[right]{$p_3$};
    \draw[massive] (1,0) -- ++(-45:1) node[right]{$p_4$};
\end{tikzpicture}
+
\begin{tikzpicture}[font=\small,baseline={([yshift=-.5ex]current bounding box.center)},thick] 
    \draw[massive] (0,0) -- (135:1) node[left]{$p_2$};
    \draw[massive] (0,0) -- (-135:1) node[left]{$p_1$};
    \draw[Anew] (0,0) -- (1,0);
    \draw[massive] (1,0) -- ++(45:1) node[right]{$p_3$};
    \draw[massive] (1,0) -- ++(-45:1) node[right]{$p_4$};
\end{tikzpicture}
\nn\\[2ex] +&
\begin{tikzpicture}[font=\small,baseline={([yshift=-.5ex]current bounding box.center)},thick] 
    \draw[massive] (0,0.5) -- ++(135:1) node[left]{$p_2$};
    \draw[massive] (0,0.5) -- ++(45:1) node[right]{$p_3$};
    \draw[Rnew] (0,0.5) -- (0,-0.5);
    \draw[massive] (0,-0.5) -- ++(-135:1) node[left]{$p_1$};
    \draw[massive] (0,-0.5) -- ++(-45:1) node[right]{$p_4$};
\end{tikzpicture}
+
\begin{tikzpicture}[font=\small,baseline={([yshift=-.5ex]current bounding box.center)},thick] 
    \draw[massive] (0,0.5) -- ++(135:1) node[left]{$p_2$};
    \draw[massive] (0,0.5) -- ++(45:1) node[right]{$p_3$};
    \draw[Anew] (0,0.5) -- (0,-0.5);
    \draw[massive] (0,-0.5) -- ++(-135:1) node[left]{$p_1$};
    \draw[massive] (0,-0.5) -- ++(-45:1) node[right]{$p_4$};
\end{tikzpicture}
+
\begin{tikzpicture}[font=\small,baseline={([yshift=-.5ex]current bounding box.center)},thick] 
    \draw[massive] (0,0.5) -- ++(135:1) node[left]{$p_2$};
    \draw[massive] (0,-0.5) -- ++(-135:1) node[left]{$p_1$};
    \draw[Rnew] (0,-0.5) -- (0,0.5);
    \path (0,-0.5) -- ($(0,0.5)+(45:1)$) node[right] (B){$p_3$};
    \draw[massive] (B) -- ($(B)!0.6!(0,-0.5)$);
    \draw[massive] ($(B)!0.8!(0,-0.5)$) -- (0,-0.5);
    \path (0,0.5) -- ($(0,-0.5)+(-45:1)$) node[right] (A){$p_4$};
    \draw[massive] (A) -- (0,0.5);
\end{tikzpicture}
+
\begin{tikzpicture}[font=\small,baseline={([yshift=-.5ex]current bounding box.center)},thick] 
    \draw[massive] (0,0.5) -- ++(135:1) node[left]{$p_2$};
    \draw[massive] (0,-0.5) -- ++(-135:1) node[left]{$p_1$};
    \draw[Anew] (0,-0.5) -- (0,0.5);
    \path (0,-0.5) -- ($(0,0.5)+(45:1)$) node[right] (B){$p_3$};
    \draw[massive] (B) -- ($(B)!0.6!(0,-0.5)$);
    \draw[massive] ($(B)!0.8!(0,-0.5)$) -- (0,-0.5);
    \path (0,0.5) -- ($(0,-0.5)+(-45:1)$) node[right] (A){$p_4$};
    \draw[massive] (A) -- (0,0.5);
\end{tikzpicture}
\Biggr)
\end{align}
where the direction of the arrow indicates the direction of causality flow.

Two observations are in order here:

{\bf 1.}~First, anticipating a point we will make later, we notice that the factor of $1/2$ in \eqref{eq:4ptMomentumSpaceDiags} corresponds precisely to the Murua coefficient which will be introduced 
in~\eqref{easyMurua2V}.

{\bf 2.}~We also note that  at tree level the four-point Magnus amplitude is obtained by writing the same Feynman diagrams as for the standard amplitude, except for the replacement 
\begin{align}
    i \Delta^F (p)  \longrightarrow \frac{1}{2}\big(i\Delta^{R} + i\Delta^{A} \big)(p)\, .  
\end{align}
One may also note that 
\begin{align}
\frac{1}{2}\big( i\Delta^{R} (p)+ i\Delta^{A}(p) \big) = i \Delta^F (p) - \frac{1}{2}\Delta^{(1)} (p)\, ,  
\end{align}
where \begin{align}
    \Delta^{(1)}(p) \coloneq i\Delta^{(+)}(p) - i\Delta^{(-)}(p)
= (2\pi) \delta(p^2 - m^2)\, ,     
\end{align}
    is  the Hadamard function introduced in \eqref{Hadamard}. The latter function has support only on shell, and hence in general it will not contribute at tree level for generic kinematics. 

\subsection{The five-point  amplitude}
\label{sec:5pttree}

Next we compute the five-point Magnus amplitude:
\begin{align}
    iN_{5}^{(3)} = \langle 0 |i N^{(3)} | \phi(p_1) \cdots \phi(p_5) \rangle\, , 
    \end{align}
    where
    \begin{align}
    \label{N3phicubed}
i N^{(3)} = 
  \frac{ (-i)^3}{6}\int\!d^4x_1 d^4x_2 d^4x_3 \ \theta_{12} \
  \theta_{23} \ \left([\cH_1,[\cH_2,\cH_3]]+[\cH_3,[\cH_2,\cH_1]]\right) \, .   
    \end{align}
    As usual,  $\cH_i$ stands for $\cH(x_i)$ and $\theta_{ij} \coloneq \theta (t_i - t_j)$. A short computation shows that%
\footnote{One may arrive at different-looking but in fact identical expressions for commutators due to   the commutator of two fields being  equal to a c-number. For instance, 
$[\phi_1^3, \phi_2^3] = 3i \Delta_{12} (\phi_1^2 \phi_2^2 + \phi_2^2 \phi_1^2 + \phi_1 \phi_2^2 \phi_1)$, and $[\phi_1^3, \phi_2^3] = 3i \Delta_{12} (\phi_1^2 \phi_2^2 + \phi_2^2 \phi_1^2 + \phi_2 \phi_1^2 \phi_2)$. Both expressions are correct since $\phi_1 \phi_2^2 \phi_1 -\phi_2 \phi_1^2 \phi_2= [\phi_1, \phi_2] \phi_2\phi_1   - \phi_2 \phi_1 [\phi_1, \phi_2]  =0$, since $[\phi_1, \phi_2]$ is a c-number.
}
\begin{align}\label{123phi}
    [\cH_1,[\cH_2,\cH_3]]
    &= \frac{9}{(3!)^3}  (i\Delta_{23}) \\
\Big[ &(i\Delta_{12}) \big(\phi_2 \phi_1^2 \phi_3^2 + \phi_1^2 \phi_2 \phi_3^2 + \phi_3^2 \phi_2 \phi_1^2 + \phi_3^2 \phi_1^2 \phi_2+\phi_1^2 \phi_3^2 \phi_2 + \phi_2 \phi_3^2 \phi_1^2\big) 
\nonumber\\ +& 
(i\Delta_{13} )\big(\phi_2^2 \phi_3 \phi_1^2 + \phi_2^2\phi_1^2 \phi_3 + \phi_3 \phi_1^2 \phi_2^2 + \phi_1^2 \phi_3 \phi_2^2 + \phi_2 \phi_3 \phi_1^2 \phi_2 + \phi_2 \phi_1^2 \phi_3\phi_2\big) \Big]\, .\nonumber
\end{align}
We will set $N^{(3)} \coloneq N_{123} + N_{321}$, where the two terms denote the contributions arising from the two orderings of the nested commutators in \eqref{N3phicubed}. 
We then find%
\footnote{Note that we can rewrite $\theta_{12} \theta_{23} = \theta_{12}\theta_{23} \theta_{13}$.}
\begin{align}\label{interres}
    iN_{123} &=\frac{(-i)^3}{6}
    \int\!d^4x_1 d^4x_2 d^4 x_3 \ 
    (i\Delta^{R}_{23})
    \frac1{(2!)^2}\Big[ (i \Delta^{R}_{12})  :\!\phi_1^2 \phi_2 \phi_3^2 \!:  +\theta_{12}
    (i \Delta^{R}_{13})  :\!\phi_1^2 \phi_2^2 \phi_3 \!:  \Big] \nonumber\\
    &\qquad+\cdots\,,
\end{align}
again ignoring further field contractions that will not contribute to the tree-level amplitude.

We note that in \eqref{interres} there is an additional $\theta_{12}$ which,
at first glance, is not associated with a retarded or advanced propagator.
In order to achieve manifest Lorentz invariance it is necessary for us to absorb it.
This is not difficult though:
since $\Delta^{R}_{23} \Delta^{R}_{13} :\!\phi_1^2 \phi_2^2 \phi_3 \!: $ is symmetric under the exchange $x_1\leftrightarrow x_2$, we can rewrite \eqref{interres} without the $\theta$-function as%
\footnote{Similar manipulations led Dyson to the introduction of time-ordered products \cite{Dyson:1949bp}.}
\begin{align}\label{disapp}
    iN_{123} &=\frac{(-i)^3}{6}   \int\!d^4x_1 d^4x_2 d^4 x_3 \ 
      (i\Delta^{R}_{23})\frac1{(2!)^2}
    \Big[ (i \Delta^{R}_{12}) :\!\phi_1^2 \phi_2 \phi_3^2 \!:  +\frac{1}{2}
    (i \Delta^{R}_{13})  :\!\phi_1^2 \phi_2^2 \phi_3 \!: \Big]\nonumber\\
    &\qquad+\cdots\,.
\end{align}
The relative factor $1/2$ between the two terms in the above result is our first example of different weights associated to  different routings of advanced and retarded propagators.

We can also describe this simplification diagrammatically. If we represent the points $x_{1,2,3}$ as the vertices of a graph, we can interpret the retarded propagators as directed edges between vertices. We then represent the leftover $\theta$-functions as additional (red) directed edges. We use the arrows to represent the flow of
time, so that if an edge goes from $j$ to $i$ then   $t_i>t_j$,  and we introduce the notation  
\begin{equation}
\label{first-ex-pic}
    \begin{tikzpicture}[font=\small, thick]
        \draw[causArrow] (0,0) -- (1,0);
        \draw[fill=white] (0,0) circle(3pt);
        \draw[fill=white] (1,0) circle(3pt);
        \node at (0,0.4){$j$};
        \node at (1,0.4){$i$};
    \end{tikzpicture}
    = i \Delta^R_{ij}\, , \quad \qquad
    \begin{tikzpicture}[font=\small, thick]
        \draw[thetaLine] (0,0) -- (1,0);
        \draw[fill=white] (0,0) circle(3pt);
        \draw[fill=white] (1,0) circle(3pt);
        \node at (0,0.4){$j$};
        \node at (1,0.4){$i$};
    \end{tikzpicture}
    = \theta_{ij}\, .
\end{equation}
Note that since  we are only working with $\phi^3$ theory, we can omit the leftover normal-ordered fields as they can be recovered by looking at the number of fields at each vertex.  
Applying these diagrammatic rules to the second term in
\eqref{interres}, we have 
\begin{equation}
\label{eq:exdiag5pt}
    (i\Delta^{R}_{23})\, (i\Delta^{R}_{13}) \theta_{12} =
    \begin{tikzpicture}[baseline={([yshift=-.5ex]current bounding box.center)},thick,font=\footnotesize]
        \draw[causArrow] (0,0) -- (0.5,0.5);
        \draw[causArrow] (0,0) -- (1,-0.5);
        \draw[thetaLine] (0.5,0.5) -- (1,-0.5);
        \draw[fill=white] (0,0) circle(3pt) node[yshift=0.4cm]{3};
        \draw[fill=white] (0.5,0.5) circle(3pt) node[yshift=0.4cm]{2};
        \draw[fill=white] (1,-0.5) circle(3pt) node[yshift=-0.4cm]{1};
    \end{tikzpicture} \, .
\end{equation}
 In this notation, each set of contractions, viewed as an arrowed graph, results in a directed acyclic graph (DAG), with black edges representing retarded propagators and red edges representing leftover $\theta$-functions.%
 \footnote{The directedeness of our graphs arises because we have advanced and retarded propagators. Their acyclic nature is due to the fact that a loop (cycle) corresponds to a closed time-like loop, which is vanishing.}
 Moving forward, we will omit the labels on the nodes, as the positions are integrated over. The effect of symmetrising $x_1\leftrightarrow x_2$ diagrammatically corresponds to
\begin{equation}
\label{firstevergraph}
\begin{tikzpicture}[baseline={([yshift=-.5ex]current bounding box.center)}, thick]
    \draw[causArrow] (0,0) -- (0.5,0.5);
    \draw[causArrow] (0,0) -- (1,-0.5);
    \draw[thetaLine] (0.5,0.5) -- (1,-0.5);
    \draw[fill=white] (0,0) circle(3pt);
    \draw[fill=white] (0.5,0.5) circle(3pt);
    \draw[fill=white] (1,-0.5) circle(3pt);
\end{tikzpicture} \rightarrow \frac{1}{2!}\left(
\begin{tikzpicture}[baseline={([yshift=-.5ex]current bounding box.center)}, thick]
\begin{scope}
    \draw[causArrow] (0,0) -- (0.5,0.5);
    \draw[causArrow] (0,0) -- (1,-0.5);
    \draw[thetaLine] (0.5,0.5) -- (1,-0.5);
    \draw[fill=white] (0,0) circle(3pt);
    \draw[fill=white] (0.5,0.5) circle(3pt);
    \draw[fill=white] (1,-0.5) circle(3pt);
    \node at (1.5, 0){$+$};
\end{scope}
\begin{scope}[xshift=2cm]
    \draw[causArrow] (0,0) -- (1,0.5);
    \draw[causArrow] (0,0) -- (0.5,-0.5);
    \draw[thetaLine] (0.5,-0.5) -- (1,0.5);
    \draw[fill=white] (0,0) circle(3pt);
    \draw[fill=white] (0.5,-0.5) circle(3pt);
    \draw[fill=white] (1,0.5) circle(3pt);
\end{scope}
\end{tikzpicture} \, \right) = \frac{1}{2!}\,
\begin{tikzpicture}[baseline={([yshift=-.5ex]current bounding box.center)}, thick]
    \draw[causArrow] (0,0) -- (0.5,0.5);
    \draw[causArrow] (0,0) -- (0.5,-0.5);
    \draw[fill=white] (0,0) circle(3pt);
    \draw[fill=white] (0.5,0.5) circle(3pt);
    \draw[fill=white] (0.5,-0.5) circle(3pt);
\end{tikzpicture}\, ,
\end{equation}
and so the factor $\theta_{12}$ has again been absorbed.%
\footnote{Our notation for the final graph in \eqref{firstevergraph} and similar ones is inspired by \cite{CALAQUE2011282, Kim:2024svw}.  } 

We can understand this as follows: the term in \eqref{eq:exdiag5pt} has support in the integration subregion $t_1>t_2>t_3$, as shown by the arrow orientations. But the product of retarded propagators $(i\Delta^{R}_{23})\, (i\Delta^{R}_{13})$ alone allows for a bigger region, namely $t_1>t_3 \,,\, t_2>t_3$. The latter contains one more subregion, $t_2>t_1>t_3$, but this can be obtained via symmetrising $x_1\leftrightarrow x_2$ in \eqref{firstevergraph}. Adding both subregions we recover the full $t_1>t_3 \,,\, t_2>t_3$ region, and therefore the result is given just in terms of retarded propagators.

The same strategy can be applied to $N_{321}$, with the result,
\begin{align}
    iN_{321} &=\frac{(-i)^3}{6}
    \int\!d^4x_1 d^4x_2 d^4 x_3 \ 
    \frac1{(2!)^2}(i\Delta^{R}_{12})
    \Big[ (i \Delta^{R}_{23}) :\!\phi_1^2 \phi_2 \phi_3^2 \!: +
    \frac{1}{2}
    (i \Delta^{R}_{13})  :\!\phi_1 \phi_2^2 \phi_3^2 \!: \Big]\nonumber\\
    &\qquad+\cdots\,.
\end{align}
The $\theta$-function simplification proceeds as above, in this case via
\begin{equation}
\begin{tikzpicture}[baseline={([yshift=-.5ex]current bounding box.center)}, thick]
    \draw[causArrow] (0.5,0.5) -- (1,0);
    \draw[causArrow] (0,-0.5) -- (1,0);
    \draw[thetaLine] (0,-0.5) -- (0.5,0.5);
    \draw[fill=white] (1,0) circle(3pt);
    \draw[fill=white] (0.5,0.5) circle(3pt);
    \draw[fill=white] (0,-0.5) circle(3pt);
\end{tikzpicture} \rightarrow \frac{1}{2!}\left(
\begin{tikzpicture}[baseline={([yshift=-.5ex]current bounding box.center)}, thick]
\begin{scope}
    \draw[causArrow] (0.5,0.5) -- (1,0);
    \draw[causArrow] (0,-0.5) -- (1,0);
    \draw[thetaLine] (0,-0.5) -- (0.5,0.5);
    \draw[fill=white] (1,0) circle(3pt);
    \draw[fill=white] (0.5,0.5) circle(3pt);
    \draw[fill=white] (0,-0.5) circle(3pt);
    \node at (1.5, 0){$+$};
\end{scope}
\begin{scope}[xshift=2cm]
    \draw[causArrow] (0,0.5) -- (1,0);
    \draw[causArrow] (0.5,-0.5) -- (1,0);
    \draw[thetaLine] (0,0.5) -- (0.5,-0.5);
    \draw[fill=white] (1,0) circle(3pt);
    \draw[fill=white] (0.5,-0.5) circle(3pt);
    \draw[fill=white] (0,0.5) circle(3pt);
\end{scope}
\end{tikzpicture} \, \right) = \frac{1}{2!}\,
\begin{tikzpicture}[baseline={([yshift=-.5ex]current bounding box.center)}, thick]
    \draw[causArrow] (0,0.5) -- (0.5,0);
    \draw[causArrow] (0,-0.5) -- (0.5,0);
    \draw[fill=white] (0.5,0) circle(3pt);
    \draw[fill=white] (0,0.5) circle(3pt);
    \draw[fill=white] (0,-0.5) circle(3pt);
\end{tikzpicture}\, .
\end{equation}
Combining terms, we have found that%
\footnote{This expression will reappear in a one-loop context, see \eqref{one-loop-tri-int}.}
\begin{align}
\label{5pt-tree-int}
iN^{(3)}   &=(-i)^3  \int\!d^4x_1 d^4x_2 d^4 x_3  
    \Big[ \frac{1}{3} (i\Delta^{R}_{23})  (i \Delta^{R}_{31})
    +
    \frac{1}{12}\Big((i\Delta^{R}_{23})(i \Delta^{A}_{31}) + (i\Delta^{A}_{23})(i \Delta^{R}_{31}) \Big)
    \Big]\nonumber\\
    &\qquad\times\bigg(\frac1{(2!)^2}:\!\phi_1^2 \phi_2^2 \phi_3 \!: +\cdots\bigg)\, , 
\end{align} 
where we have also used $\Delta^R_{ij}=\Delta^A_{ji}$ 
and relabelled $x_i$ in order to collect terms.
This corresponds to one diagram topology,
with all possible routings of advanced and retarded propagators weighted differently.

Finally, taking the matrix element
$\langle 0 | \cdots | \phi(p_1)\cdots \phi (p_5)\rangle$, we  generate several  diagrams. For instance, the diagram topology 
\begin{equation}
	\begin{tikzpicture}[baseline={([yshift=-.5ex]current bounding box.center)},font=\small,thick]
		\draw (1, 0) -- ++(135:1);
		\draw (1, 0) -- ++(-135:1);
		\draw (2, 1) -- (2, 0);
		\draw (3, 0) -- ++(45:1);
		\draw (3, 0) -- ++(-45:1);
		
		\draw[massive] (1,0) -- (2, 0);
		\draw[massive] (3, 0) -- (2, 0);
		
		\node[align=center] at ($(1,0)+(135:1.2)$){$p_2$};
		\node[align=center] at ($(1,0)+(-135:1.2)$){$p_1$};
		\node[align=center] at ($(3,0)+(45:1.25)$){$p_3$};
		\node[align=center] at ($(3,0)+(-45:1.25)$){$p_4$};
		\node[align=center] at (2, 1.2){$p_5$};
	\end{tikzpicture}\ ,
\end{equation}
which has  momenta $(p_1, p_2)$ and $(p_3, p_4)$ attached to two distinct vertices, with momentum $p_5$ attached to two internal propagators, gives 
\begin{equation}
\begin{split}
\label{5ptex}
 &(-i)^3  \int\!d^4x_1 d^4x_2 d^4 x_3 \ e^{-i (p_1 + p_2)\cdot x_1 - i p_5\cdot x_2  - i (p_3 + p_4) \cdot x_3}
     \\ &
    \cdot \bigg\{  \frac{1}{3}(i\Delta^{R}_{12})  (i \Delta^{R}_{23}) 
    +
    \frac{1}{6}\Big[  (i\Delta^{R}_{12}) (i \Delta^{A}_{23})  +  (i\Delta^{A}_{12}) (i \Delta^{R}_{23})
     \Big] + 
    \frac{1}{3}(i\Delta^{A}_{12})  (i \Delta^{A}_{23})\bigg\} 
    \\
&=(-i)^3  \int\!d^4x_1 d^4x_2 d^4 x_3 \ e^{-i (p_1 + p_2)\cdot x_1 - i p_5\cdot x_2  - i (p_3 + p_4) \cdot x_3}
      \\
&\left[\frac{1}{3}\
\begin{tikzpicture}[thick,font=\footnotesize]
    \draw[massive, causArrow] (0,0) -- (1, 0);
    \draw[massive, causArrow] (1, 0) -- (2, 0);
    \draw[fill=white] (0,0) circle(3pt) node[yshift=0.4cm]{$1$};
    \draw[fill=white] (1,0) circle(3pt) node[yshift=0.4cm]{$2$};
    \draw[fill=white] (2,0) circle(3pt) node[yshift=0.4cm]{$3$};
\end{tikzpicture}
\ +\frac{1}{6}\
\begin{tikzpicture}[baseline={([yshift=-.5ex]current bounding box.center)}, thick,font=\footnotesize]
    \draw[massive, causArrow] (0,0.5) -- (0.5,0);
    \draw[massive, causArrow] (0,-0.5) -- (0.5,0);
    \draw[fill=white] (0.5,0) circle(3pt) node[xshift=0.3cm]{$2$};
    \draw[fill=white] (0,0.5) circle(3pt) node[yshift=0.4cm]{$1$};
    \draw[fill=white] (0,-0.5) circle(3pt) node[yshift=-0.4cm]{$3$};
\end{tikzpicture}
\ +\frac{1}{6}\
\begin{tikzpicture}[baseline={([yshift=-.5ex]current bounding box.center)}, thick,font=\footnotesize]
    \draw[massive, causArrow] (0.4,0) -- (0.9,0.5);
    \draw[massive, causArrow] (0.4,0) -- (0.9,-0.5);
    \draw[fill=white] (0.4,0) circle(3pt) node[xshift=-0.3cm]{$2$};
    \draw[fill=white] (0.9,0.5) circle(3pt) node[yshift=0.4cm]{$1$};
    \draw[fill=white] (0.9,-0.5) circle(3pt) node[yshift=-0.4cm]{$3$};
\end{tikzpicture}
\ + \frac{1}{3}\
\begin{tikzpicture}[thick,font=\footnotesize]
    \draw[massive, causArrow] (0,0) -- (1, 0);
    \draw[massive, causArrow] (1, 0) -- (2, 0);
    \draw[fill=white] (0,0) circle(3pt) node[yshift=0.4cm]{$3$};
    \draw[fill=white] (1,0) circle(3pt) node[yshift=0.4cm]{$2$};
    \draw[fill=white] (2,0) circle(3pt) node[yshift=0.4cm]{$1$};
\end{tikzpicture}
\ \right]\, ,
\end{split}
\end{equation}
with all other diagrams being obtained from relabelings in \eqref{5ptex}.
Performing the integrals into momentum space, the full matrix element of $N^{(3)}$ can be written as
\begin{align}
\begin{split}\label{eq:5ptMomentumSpaceDiags}
	iN^{(3)}_{5}&=\frac{1}{8}\sum_{S_5}(-i)^3\hat{\delta}^{(4)}(p_1+p_2+p_3+p_4+p_5)\quad \\
&\left[\ \frac{1}{3}
\begin{tikzpicture}[baseline={([yshift=-0.5ex]current bounding box.center)},thick]
    \draw (1, 0) -- ++(135:1);
    \draw (1, 0) -- ++(-135:1);
    \draw (2, 1) -- (2, 0);
    \draw (3, 0) -- ++(45:1);
    \draw (3, 0) -- ++(-45:1);
    \draw[massive, Rnew] (1,0) -- (2,0);
	\draw[massive, Rnew] (2,0) -- (3,0);
    \path (2,-1) -- (2,0); 
\end{tikzpicture}
\ +\frac{1}{6}\
\begin{tikzpicture}[baseline={([yshift=-0.5ex]current bounding box.center)},thick]
    \draw (1, 0) -- ++(135:1);
    \draw (1, 0) -- ++(-135:1);
    \draw (2, 1) -- (2, 0);
    \draw (3, 0) -- ++(45:1);
    \draw (3, 0) -- ++(-45:1);
    \draw[massive, Rnew] (1,0) -- (2,0);
	\draw[massive, Anew] (2,0) -- (3,0);
    \path (2,-1) -- (2,0); 
\end{tikzpicture}
\right. \\[3ex]
&\left. +\frac{1}{6} \
\begin{tikzpicture}[baseline={([yshift=-0.5ex]current bounding box.center)},thick]
    \draw (1, 0) -- ++(135:1);
    \draw (1, 0) -- ++(-135:1);
    \draw (2, 1) -- (2, 0);
    \draw (3, 0) -- ++(45:1);
    \draw (3, 0) -- ++(-45:1);
    \draw[massive, Anew] (1,0) -- (2,0);
	\draw[massive, Rnew] (2,0) -- (3,0);
    \path (2,-1) -- (2,0); 
\end{tikzpicture}
\ +\frac{1}{3} \
\begin{tikzpicture}[baseline={([yshift=-0.5ex]current bounding box.center)},thick]
    \draw (1, 0) -- ++(135:1);
    \draw (1, 0) -- ++(-135:1);
    \draw (2, 1) -- (2, 0);
    \draw (3, 0) -- ++(45:1);
    \draw (3, 0) -- ++(-45:1);
    \draw[massive, Anew] (1,0) -- (2,0);
	\draw[massive, Anew] (2,0) -- (3,0);
    \path (2,-1) -- (2,0); 
\end{tikzpicture}
\ \right].
\end{split}
\end{align}
Note that the position of the external legs in the graphs is encoded in the exponential in \eqref{5ptex}. Moreover, we have omitted the labels of external lines in \eqref{eq:5ptMomentumSpaceDiags} for notational simplicity, but they are to be summed over all possible permutations $\sigma \in S_5$ of $(p_1,p_2,p_3,p_4,p_5)$.  Before concluding this section, two observations are in order. 

{\bf 1.}~The factors inside the square brackets correspond precisely to the Murua coefficients found in \eqref{easyMurua3V}. The factor of $1/8$ outside is to compensate for overcounting in the sum over all permutations.

{\bf 2.}~In the four-point example, we saw that the single $\theta$-function in the Magnus expansion immediately combined with a $\Delta$ function to produce retarded and advanced propagators. In the five-point case, the Magnus expansion \eqref{N3phicubed} contains two $\theta$-functions, only one of which initially combines with a $\Delta$ function; nevertheless, we showed that the second $\theta$-function can also be eliminated, as in \eqref{relretadv}. This behaviour is a general feature of the expansion, although it becomes increasingly challenging to identify the rearrangements needed to remove all remaining $\theta$-functions. The forthcoming six-point example will be particularly instructive in this regard (and also to confirm the pattern of the appearance of Murua coefficients, discussed in more detail in Section~\ref{sec:redofromMurua-trees}).

\subsection{The six-point amplitude}
\label{sect:6pttree}

As a final example at tree level, we now focus on the six-point Magnus amplitude
\begin{align}\label{eq:6ptMatrixElementDef}
    N_{6}^{(4)} = \langle 0 | N^{(4)} | \phi(p_1) \cdots \phi(p_6) \rangle\, , 
\end{align}
where $N^{(4)}$ was defined in \eqref{N4}.
For convenience we define
\begin{equation}
i N^{(4)}_{ijkl} := \frac{(-i)^4}{12}\int d^4x_1 d^4x_2 d^4x_3 d^4x_4\ \theta_{12}\theta_{23}\theta_{34} [\cH_i,[\cH_j,[\cH_k,\cH_l]]] \, ,
\end{equation}
so that 
\begin{align}
    N^{(4)} = N^{(4)}_{1234} + N^{(4)}_{4312} +  N^{(4)}_{2341} + N^{(4)}_{1432}\, ,  
\end{align}
and treat the nested commutators separately. 
Starting from $N^{(4)}_{1234}$,
we find that 
\begin{align}
\label{eq:N4comm1234}
\begin{split}
i N^{(4)}_{1234} := &\frac{-i}{12}\int d^4x_1 d^4x_2 d^4x_3 d^4x_4\ 
\left[ 
    \Delta^{R}_{14} \Delta^{R}_{24} \Delta^{R}_{34} \theta_{12} \theta_{23} \frac{:\phi_1^2 \phi_2^2 \phi_3^2:}{(2!)^3}\right. \\
    &+ (\Delta^{R}_{14} \Delta^{R}_{23} \Delta^{R}_{34} \theta_{12} + \Delta^{R}_{13} \Delta^{R}_{24} \Delta^{R}_{34} \theta_{12} \theta_{23}) \frac{:\phi_1^2 \phi_2^2 \phi_3 \phi_4:}{(2!)^2} \\
    &+ \Delta^{R}_{12} \Delta^{R}_{24} \Delta^{R}_{34} \theta_{23} \frac{:\phi_1^2 \phi_2 \phi_3^2 \phi_4:}{(2!)^2} + \Delta^{R}_{13} \Delta^{R}_{23} \Delta^{R}_{34} \theta_{12} \frac{:\phi_1^2 \phi_2^2 \phi_4^2:}{(2!)^3} \\
    &\left.+ \Delta^{R}_{12} \Delta^{R}_{23} \Delta^{R}_{34} \frac{:\phi_1^2 \phi_2 \phi_3 \phi_4^2:}{(2!)^2}
    \right] +\cdots\,,
    \end{split}
\end{align}
again ignoring further contractions from Wick's theorem that will not contribute to the tree-level amplitude.
Our task is now to eliminate the leftover $\theta$-functions.

We start with the first term in \eqref{eq:N4comm1234} and symmetrise:
\begin{align}\label{noi!}
    \Delta^{R}_{14} \Delta^{R}_{24} \Delta^{R}_{34} \theta_{12} \theta_{23} :\phi_1^2 \phi_2^2 \phi_3^2: 
    \to& \frac{1}{3!} \big( \theta_{12} \theta_{23} + \theta_{23} \theta_{31} + \theta_{31} \theta_{12} + \theta_{13} \theta_{32} + \theta_{32} \theta_{21} + \theta_{21} \theta_{13} \big) \nonumber\\
    &\times \Delta^{R}_{14} \Delta^{R}_{24}  \Delta^{R}_{34} :\phi_1^2 \phi_2^2 \phi_3^2: \nonumber\\ & 
    = \frac{1}{3!} \Delta^{R}_{14} \Delta^{R}_{24}  \Delta^{R}_{34} :\phi_1^2 \phi_2^2 \phi_3^2: .
\end{align}
We can understand this as follows: $N_{1234}^{(4)}$ contains an ordering $t_1 {>} t_2 {>} t_3 {>} t_4$,
but the retarded propagators only imply $t_{1,2,3} > t_4$. The latter region can be divided in six sub-regions $t_{\sigma(1)} > t_{\sigma(2)} > t_{\sigma(3)} > t_4$, where $\sigma$ are all permutations of three indices and only the region $\sigma(i) = i$ appears in \eqref{eq:N4comm1234}. However, since $t_{1,2,3}$ are integration variables, we can fully symmetrise them to recover all six regions. Diagrammatically,\footnote{This notation reintroduces a factor of $i$ per propagator compared to \eqref{noi!}.}
\begin{align}
\label{eq:tree6_3out}
\begin{tikzpicture}[baseline={([yshift=-.5ex]current bounding box.center)},thick]
    \coordinate (1) at (0.41,-0.5);
    \coordinate (2) at (0.4*2,0);
    \coordinate (3) at (0.4*3,0.5);
    \draw[thetaLine=0.5] (1) -- (2);
    \draw[thetaLine=0.5] (2) -- (3);
    \draw[causArrow] (0,0) -- (1);
    \draw[causArrow] (0,0) -- (2);
    \draw[causArrow] (0,0) -- (3);
    \draw[fill=white] (0,0) circle(3pt);        \draw[fill=white] (1) circle(3pt);
    \draw[fill=white] (2) circle(3pt);
    \draw[fill=white] (3) circle(3pt);
\end{tikzpicture}
&\rightarrow\frac{1}{3!}\left(
\begin{tikzpicture}[baseline={([yshift=-.5ex]current bounding box.center)},thick]
%
    \foreach \xa/\xb/\xc/\ta/\tb [count=\n] in {1/2/3/0.5/0.5, 2/3/1/0.25/0.5, 3/1/2/0.5/0.5, 1/3/2/0.25/0.5, 3/2/1/0.5/0.5, 2/1/3/0.5/0.5}{
    \begin{scope}[xshift=\n*2cm]
    \coordinate (\xa) at (0.4*\xa,-0.5);
    \coordinate (\xb) at (0.4*\xb,0);
    \coordinate (\xc) at (0.4*\xc,0.5);
    \draw[thetaLine=\ta] (1) -- (2);
    \draw[thetaLine=\tb] (2) -- (3);
    \draw[causArrow] (0,0) -- (\xa);
    \draw[causArrow] (0,0) -- (\xb);
    \draw[causArrow] (0,0) -- (\xc);
    \draw[fill=white] (0,0) circle(3pt);        \draw[fill=white] (\xa) circle(3pt);
    \draw[fill=white] (\xb) circle(3pt);
    \draw[fill=white] (\xc) circle(3pt);
    \end{scope}
    }
%
    \node at (3.6, 0){\small$+$};
    \node at (5.6, 0){\small$+$};
    \node at (7.6, 0){\small$+$};
    \node at (9.6, 0){\small$+$};
    \node at (11.6, 0){\small$+$};
\end{tikzpicture} \, \right)\nn \\
&= \frac{1}{3!}\, \begin{tikzpicture}[baseline={([yshift=-.5ex]current bounding box.center)},thick]
%
    \begin{scope}[xshift=0.5cm]
    \draw[causArrow] (0,0) -- (0.7,-0.5);
    \draw[causArrow] (0,0) -- (0.7,0);
    \draw[causArrow] (0,0) -- (0.7,0.5);
    \draw[fill=white] (0,0) circle(3pt);
    \draw[fill=white] (0.7,-0.5) circle(3pt);
    \draw[fill=white] (0.7,0) circle(3pt);
    \draw[fill=white] (0.7,0.5) circle(3pt);
    \end{scope}
\end{tikzpicture}\,,
\end{align}
the $1/3!$ manifesting naturally as a symmetry factor.

Then, we consider the next three terms in \eqref{eq:N4comm1234}, namely
\begin{align}\label{eq:ex2thetacancell}
    &2 (\Delta^{R}_{14} \Delta^{R}_{23} \Delta^{R}_{34} \theta_{12} + \Delta^{R}_{13} \Delta^{R}_{24} \Delta^{R}_{34} \theta_{12} \theta_{23}) :\phi_1^2 \phi_2^2 \phi_3 \phi_4: + 2 \Delta^{R}_{12} \Delta^{R}_{24} \Delta^{R}_{34} \theta_{23} :\phi_1^2 \phi_2 \phi_3^2 \phi_4:\nonumber \\
    &\to 2 \Delta^{R}_{14} \Delta^{R}_{23} \Delta^{R}_{34} ( \theta_{12} + \theta_{21} \theta_{13} + \theta_{31}) :\phi_1^2 \phi_2^2 \phi_3 \phi_4:\nonumber \\
    &\qquad= 2 \Delta^{R}_{14} \Delta^{R}_{23} \Delta^{R}_{34} :\phi_1^2 \phi_2^2 \phi_3 \phi_4:\, , 
\end{align}
where the arrow means that the equality holds   upon integration.  
All three terms have the same retarded propagators, up to relabelling of the integration variables $t_i$ that we have performed above. More explicitly: we have relabelled $\{t_1,t_2,t_3,t_4\}\to\{t_2,t_1,t_3,t_4\}$ and $\{t_1,t_2,t_3,t_4\}\to\{t_2,t_3,t_1,t_4\}$, in the second and third term respectively, thus obtaining a common factor $\Delta^{R}_{14} \Delta^{R}_{23} \Delta^{R}_{34}$.
Diagrammatically,
\begin{equation}
\begin{tikzpicture}[baseline={([yshift=-.5ex]current bounding box.center)}, thick]

    \draw[causArrow] (0,0) -- (0.8,0);
    \draw[causArrow] (0,0) -- (0.4,-0.5);
    \draw[causArrow] (0.4,-0.5) -- (0.8,-0.5);
    \draw[fill=white] (0,0) circle(3pt);
    \draw[fill=white] (0.8,0) circle(3pt);
    \draw[fill=white] (0.4,-0.5) circle(3pt);
    \draw[fill=white] (0.8,-0.5) circle(3pt);
    \node at (1.4, -0.25){\small$=$};

    \begin{scope}[xshift=2cm]
    \draw[thetaLine] (0.8,-0.5) -- (1.2,0);
    \draw[causArrow] (0.4,-0.5) -- (0.8,-0.5);
    \draw[causArrow] (0,0) -- (1.2,0);
    \draw[causArrow] (0,0) -- (0.4,-0.5);
    \draw[fill=white] (0.4,-0.5) circle(3pt);
    \draw[fill=white] (0.8,-0.5) circle(3pt);
    \draw[fill=white] (0,0) circle(3pt);
    \draw[fill=white] (1.2,0) circle(3pt);
    \node at (1.6, -0.25){\small$+$};
    \end{scope}

    \begin{scope}[xshift=4cm]
    \draw[causArrow] (0,0) -- (0.4,-0.5);
    \draw[thetaLine] (0.4,-0.5) -- (0.8,0);
    \draw[causArrow] (0,0) -- (0.8,0);
    \draw[causArrow] (0.4,-0.5) -- (1.2,-0.5);
    \draw[thetaLine] (0.8,0) -- (1.2,-0.5);
    \draw[fill=white] (0,0) circle(3pt);
    \draw[fill=white] (0.8,0) circle(3pt);
    \draw[fill=white] (0.4,-0.5) circle(3pt);
    \draw[fill=white] (1.2,-0.5) circle(3pt);
    \node at (1.6, -0.25){\small$+$};
    \end{scope}

    \begin{scope}[xshift=6cm]
    \draw[thetaLine] (0.4,0) -- (0.8,-0.5);
    \draw[causArrow] (0,0) -- (0.4,0);
    \draw[causArrow] (0.8,-0.5) -- (1.2,-0.5);
    \draw[causArrow] (0,0) -- (0.8,-0.5);
    \draw[fill=white] (0,0) circle(3pt);
    \draw[fill=white] (0.4,0) circle(3pt);
    \draw[fill=white] (0.8,-0.5) circle(3pt);
    \draw[fill=white] (1.2,-0.5) circle(3pt);
    \end{scope}
\end{tikzpicture}\,,
\end{equation}
the extra theta functions dropping out due to $\theta_{12}\theta_{23}+\theta_{21}\theta_{13}+\theta_{23}\theta_{31}=\theta_{23}$, using the $\theta$-function identities in \eqref{eq:ThetaIds}.

From now on, we consider relabellings and symmetrisations only diagrammatically.
The next term in \eqref{eq:N4comm1234} is
\begin{equation}
    \Delta^{R}_{13} \Delta^{R}_{23} \Delta^{R}_{34} \theta_{12} :\phi_1^2 \phi_2^2 \phi_4^2: \
    \to \frac{1}{2!} \Delta^{R}_{13} \Delta^{R}_{23} \Delta^{R}_{34} :\phi_1^2 \phi_2^2 \phi_4^2: ,
\end{equation}
where we have applied the identity
\begin{equation}
\label{eq:tree6_2out}
\begin{tikzpicture}[baseline={([yshift=-.5ex]current bounding box.center)}, thick]
    \draw[thetaLine]  (0.4,-0.5) -- (0.8,0.5);
    \draw[causArrow] (0,0) -- (0.8,0.5);
    \draw[causArrow] (0,0) -- (0.4,-0.5);
    \draw[causArrow] (-0.5,0) -- (0,0);
    \draw[fill=white] (0,0) circle(3pt);
    \draw[fill=white] (0.8,0.5) circle(3pt);
    \draw[fill=white] (0.4,-0.5) circle(3pt);
    \draw[fill=white] (-0.5,0) circle(3pt);
\end{tikzpicture}
\rightarrow \frac{1}{2!}\left(
\begin{tikzpicture}[baseline={([yshift=-.5ex]current bounding box.center)}, thick]
    \begin{scope}
    \draw[thetaLine]  (0.4,-0.5) -- (0.8,0.5);
    \draw[causArrow] (0,0) -- (0.8,0.5);
    \draw[causArrow] (0,0) -- (0.4,-0.5);
    \draw[causArrow] (-0.5,0) -- (0,0);
    \draw[fill=white] (0,0) circle(3pt);
    \draw[fill=white] (0.8,0.5) circle(3pt);
    \draw[fill=white] (0.4,-0.5) circle(3pt);
    \draw[fill=white] (-0.5,0) circle(3pt);
    \node at (1.2,0){\small$+$};
    \end{scope}

    \begin{scope}[xshift=2.2cm]
    \draw[thetaLine] (0.4, 0.5) -- (0.8,-0.5);
    \draw[causArrow] (0,0) -- (0.4,0.5);
    \draw[causArrow] (0,0) -- (0.8,-0.5);
    \draw[causArrow] (-0.5,0) -- (0,0);
    \draw[fill=white] (0,0) circle(3pt);
    \draw[fill=white] (0.4,0.5) circle(3pt);
    \draw[fill=white] (0.8,-0.5) circle(3pt);
    \draw[fill=white] (-0.5,0) circle(3pt);
    \end{scope}
\end{tikzpicture} \,\right) = \frac{1}{2!} \,
\begin{tikzpicture}[baseline={([yshift=-.5ex]current bounding box.center)}, thick]
    \draw[causArrow] (0,0) -- (0.5,0.5);
    \draw[causArrow] (0,0) -- (0.5,-0.5);
    \draw[causArrow] (-0.5,0) -- (0,0);
    \draw[fill=white] (0,0) circle(3pt);
    \draw[fill=white] (0.5,0.5) circle(3pt);
    \draw[fill=white] (0.5,-0.5) circle(3pt);
    \draw[fill=white] (-0.5,0) circle(3pt);
\end{tikzpicture}
\end{equation}
With all $\theta$-functions in \eqref{eq:N4comm1234} now absorbed, we have
\begin{align}
i N^{(4)}_{1234} := &\frac{-i}{12}\int d^4x_1 d^4x_2 d^4x_3 d^4x_4\ 
\left[ 
    \frac{1}{6} \Delta^{R}_{14} \Delta^{R}_{24}  \Delta^{R}_{34} \frac{:\phi_1^2 \phi_2^2 \phi_3^2:}{(2!)^3}\nonumber\right.\\
    &+\frac{1}{2} \Delta^{R}_{13} \Delta^{R}_{23} \Delta^{R}_{34}  \frac{:\phi_1^2 \phi_2^2 \phi_4^2:}{(2!)^3}
    +\Delta^{R}_{14} \Delta^{R}_{23} \Delta^{R}_{34} \frac{:\phi_1^2 \phi_2^2 \phi_3 \phi_4:}{(2!)^2}\nonumber\\
    &\left.+\Delta^{R}_{12} \Delta^{R}_{23} \Delta^{R}_{34} \frac{:\phi_1^2 \phi_2 \phi_3 \phi_4^2:}{(2!)^2}
    \right]+\cdots\,.
\end{align}
The last term, which we can depict as 
\tikz[font=\small, thick]{
\draw[causArrow] (0,0) -- (0.5,0);
\draw[causArrow] (0.5,0) -- (1,0);
\draw[causArrow] (1,0) -- (1.5,0);
\draw[fill=white] (0,0) circle(3pt);
\draw[fill=white] (0.5,0) circle(3pt);
\draw[fill=white] (1,0) circle(3pt);
\draw[fill=white] (1.5,0) circle(3pt);
},
is proportional to $\Delta^{R}_{12} \Delta^{R}_{23} \Delta^{R}_{34}$
and does not require  further simplifications. 

Then we move onto the nested commutator $N^{(4)}_{4312}$ and we have
\begin{align}
\begin{split}
i N^{(4)}_{4312} := &\frac{-i}{12}\int d^4x_1 d^4x_2 d^4x_3 d^4x_4\ 
\left[ 
\Delta^{R}_{12} \Delta^{R}_{13} \Delta^{R}_{14} \theta_{23} \theta_{34} \frac{:\phi_2^2 \phi_3^2 \phi_4^2:}{(2!)^3}\right. \\
&+ (\Delta^{R}_{12} \Delta^{R}_{14} \Delta^{R}_{23} \theta_{34} 
+ \Delta^{R}_{12} \Delta^{R}_{13} \Delta^{R}_{24} \theta_{23} \theta_{34}) \frac{:\phi_1 \phi_2 \phi_3^2 \phi_4^2:}{(2!)^2} \\
&+ \Delta^{R}_{12} \Delta^{R}_{13} \Delta^{R}_{34} \theta_{23} \frac{:\phi_1 \phi_2^2 \phi_3 \phi_4^2:}{(2!)^2} 
+ \Delta^{R}_{12} \Delta^{R}_{23} \Delta^{R}_{24} \theta_{34} \frac{:\phi_1^2 \phi_3^2 \phi_4^2:}{(2!)^3} \\
&\left.+ \Delta^{R}_{12} \Delta^{R}_{23} \Delta^{R}_{34} \frac{:\phi_1^2 \phi_2 \phi_3 \phi_4^2:}{(2!)^2}
\right]\, .
\end{split}
\end{align}
The first term gives
\begin{equation}
    \Delta^{R}_{12} \Delta^{R}_{13} \Delta^{R}_{14} \theta_{23} \theta_{34} :\phi_2^2 \phi_3^2 \phi_4^2: 
    = \frac{1}{6} \Delta^{R}_{12} \Delta^{R}_{13} \Delta^{R}_{14} :\phi_2^2 \phi_3^2 \phi_4^2:\, ,
\end{equation}
using the identity
\begin{align}
\label{eq:tree6_3in}
\begin{tikzpicture}[baseline={([yshift=-.5ex]current bounding box.center)},thick]
    \coordinate (1) at (1.2-0.4,-0.5);
    \coordinate (2) at (1.2-0.4*2,0);
    \coordinate (3) at (1.2-0.4*3,0.5);
    \draw[thetaLine=0.5] (2) -- (1);
    \draw[thetaLine=0.5] (3) -- (2);
    \draw[causArrow] (1) -- (1.2,0);
    \draw[causArrow] (2) -- (1.2,0);
    \draw[causArrow] (3) -- (1.2,0);
    \draw[fill=white] (1.2,0) circle(3pt);        \draw[fill=white] (1) circle(3pt);
    \draw[fill=white] (2) circle(3pt);
    \draw[fill=white] (3) circle(3pt);
\end{tikzpicture}
&\rightarrow \frac{1}{3!}\left(
\begin{tikzpicture}[baseline={([yshift=-.5ex]current bounding box.center)},thick]
    \foreach \xa/\xb/\xc/\ta/\tb [count=\n] in {1/2/3/0.5/0.5, 2/3/1/0.25/0.5, 3/1/2/0.5/0.5, 1/3/2/0.25/0.5, 3/2/1/0.5/0.5, 2/1/3/0.5/0.5}{
    \begin{scope}[xshift=\n*2cm]
    \coordinate (\xa) at (1.2-0.4*\xa,-0.5);
    \coordinate (\xb) at (1.2-0.4*\xb,0);
    \coordinate (\xc) at (1.2-0.4*\xc,0.5);
    \draw[thetaLine=\ta] (2) -- (1);
    \draw[thetaLine=\tb] (3) -- (2);
    \draw[causArrow] (\xa) -- (1.2,0);
    \draw[causArrow] (\xb) -- (1.2,0);
    \draw[causArrow] (\xc) -- (1.2,0);
    \draw[fill=white] (1.2,0) circle(3pt);        \draw[fill=white] (\xa) circle(3pt);
    \draw[fill=white] (\xb) circle(3pt);
    \draw[fill=white] (\xc) circle(3pt);
    \end{scope}
    }
%
    \node at (3.6, 0){\small$+$};
    \node at (5.6, 0){\small$+$};
    \node at (7.6, 0){\small$+$};
    \node at (9.6, 0){\small$+$};
    \node at (11.6, 0){\small$+$};
\end{tikzpicture}\, \right) \nn \\
&= \frac{1}{3!} \,
\begin{tikzpicture}[baseline={([yshift=-.5ex]current bounding box.center)},thick]
    \begin{scope}
    \draw[causArrow] (0,-0.5) -- (0.7,0);
    \draw[causArrow] (0,0) -- (0.7,0);
    \draw[causArrow] (0,0.5) -- (0.7,0);
    \draw[fill=white] (0.7,0) circle(3pt);
    \draw[fill=white] (0,-0.5) circle(3pt);
    \draw[fill=white] (0,0) circle(3pt);
    \draw[fill=white] (0,0.5) circle(3pt);
    \end{scope}
\end{tikzpicture}\,.
\end{align}
The next three terms give
\begin{align}
\begin{split}
    &2 (\Delta^{R}_{12} \Delta^{R}_{14} \Delta^{R}_{23} \theta_{34} + \Delta^{R}_{12} \Delta^{R}_{13} \Delta^{R}_{24} \theta_{23} \theta_{34}) :\phi_1 \phi_2 \phi_3^2 \phi_4^2: + 2 \Delta^{R}_{12} \Delta^{R}_{13} \Delta^{R}_{34} \theta_{23} :\phi_1 \phi_2^2 \phi_3 \phi_4^2: \\
    &= 2 \Delta^{R}_{12} \Delta^{R}_{14} \Delta^{R}_{23} :\phi_1 \phi_2 \phi_3^2 \phi_4^2:\, ,
    \end{split}
\end{align}
using the identity

\begin{equation}
\begin{tikzpicture}[baseline={([yshift=-.5ex]current bounding box.center)}, thick]

    \draw[causArrow] (0,0) -- (0.8,0);
    \draw[causArrow] (0,-0.5) -- (0.4,-0.5);
    \draw[causArrow] (0.4,-0.5) -- (0.8,0);
    \draw[fill=white] (0,0) circle(3pt);
    \draw[fill=white] (0.8,0) circle(3pt);
    \draw[fill=white] (0.4,-0.5) circle(3pt);
    \draw[fill=white] (0,-0.5) circle(3pt);
    \node at (1.4, -0.25){\small$=$};

    \begin{scope}[xshift=2cm]
    \draw[thetaLine] (0,0) -- (0.4,-0.5);
    \draw[causArrow] (0.4,-0.5) -- (0.8,-0.5);
    \draw[causArrow] (0,0) -- (1.2,0);
    \draw[causArrow] (0.8,-0.5) -- (1.2,0);
    \draw[fill=white] (0.4,-0.5) circle(3pt);
    \draw[fill=white] (0.8,-0.5) circle(3pt);
    \draw[fill=white] (0,0) circle(3pt);
    \draw[fill=white] (1.2,0) circle(3pt);
    \node at (1.6, -0.25){\small$+$};
    \end{scope}

    \begin{scope}[xshift=4cm]
    \draw[thetaLine] (0,-0.5) -- (0.4,0);
    \draw[thetaLine] (0.4,0) -- (0.8,-0.5);
    \draw[causArrow] (0,-0.5) -- (0.8,-0.5);
    \draw[causArrow] (0.4,0) -- (1.2,0);
    \draw[causArrow] (0.8,-0.5) -- (1.2,0);
    \draw[fill=white] (0,-0.5) circle(3pt);
    \draw[fill=white] (0.8,-0.5) circle(3pt);
    \draw[fill=white] (0.4,0) circle(3pt);
    \draw[fill=white] (1.2,0) circle(3pt);
    \node at (1.6, -0.25){\small$+$};
    \end{scope}

    \begin{scope}[xshift=6cm]
    \draw[thetaLine] (0.4,-0.5) -- (0.8,0);
    \draw[causArrow] (0,-0.5) -- (0.4,-0.5);
    \draw[causArrow] (0.8,0) -- (1.2,0);
    \draw[causArrow] (0.4,-0.5) -- (1.2,0);
    \draw[fill=white] (0,-0.5) circle(3pt);
    \draw[fill=white] (0.4,-0.5) circle(3pt);
    \draw[fill=white] (0.8,0) circle(3pt);
    \draw[fill=white] (1.2,0) circle(3pt);
    \end{scope}
\end{tikzpicture}\,.
\end{equation}
Lastly, the next term gives
\begin{equation}
    \Delta^{R}_{12} \Delta^{R}_{23} \Delta^{R}_{24} \theta_{34} :\phi_1^2 \phi_3^2 \phi_4^2: 
    = \frac{1}{2} \Delta^{R}_{12} \Delta^{R}_{23} \Delta^{R}_{24} :\phi_1^2 \phi_3^2 \phi_4^2:\, ,
\end{equation}
using the identity
\begin{equation}
\label{eq:tree6_2in}
\begin{tikzpicture}[baseline={([yshift=-.5ex]current bounding box.center)}, thick]
    \draw[thetaLine] (-0.8,0.5) -- (-0.4,-0.5);
    \draw[causArrow] (-0.8,0.5) -- (0,0);
    \draw[causArrow] (-0.4,-0.5) -- (0,0);
    \draw[causArrow] (0,0) -- (0.5,0);
    \draw[fill=white] (0,0) circle(3pt);
    \draw[fill=white] (-0.8,0.5) circle(3pt);
    \draw[fill=white] (-0.4,-0.5) circle(3pt);
    \draw[fill=white] (0.5,0) circle(3pt);
\end{tikzpicture}
\rightarrow \frac{1}{2!}\left(
\begin{tikzpicture}[baseline={([yshift=-.5ex]current bounding box.center)}, thick]
    \begin{scope}[xshift=2.2cm]
    \draw[thetaLine] (-0.8,0.5) -- (-0.4,-0.5);
    \draw[causArrow] (-0.8,0.5) -- (0,0);
    \draw[causArrow] (-0.4,-0.5) -- (0,0);
    \draw[causArrow] (0,0) -- (0.5,0);
    \draw[fill=white] (0,0) circle(3pt);
    \draw[fill=white] (-0.8,0.5) circle(3pt);
    \draw[fill=white] (-0.4,-0.5) circle(3pt);
    \draw[fill=white] (0.5,0) circle(3pt);
    \node at (0.9,0){\small$+$};
    \end{scope}

    \begin{scope}[xshift=4.4cm]
    \draw[thetaLine] (-0.8,-0.5) -- (-0.4, 0.5);
    \draw[causArrow] (-0.4,0.5) -- (0,0);
    \draw[causArrow] (-0.8,-0.5) -- (0,0);
    \draw[causArrow] (0,0) -- (0.5,0);
    \draw[fill=white] (0,0) circle(3pt);
    \draw[fill=white] (-0.4,0.5) circle(3pt);
    \draw[fill=white] (-0.8,-0.5) circle(3pt);
    \draw[fill=white] (0.5,0) circle(3pt);
    \end{scope}
\end{tikzpicture}\, \right) = \frac{1}{2!}\,
\begin{tikzpicture}[baseline={([yshift=-.5ex]current bounding box.center)}, thick]
    \draw[causArrow] (-0.5,0.5) -- (0,0);
    \draw[causArrow] (-0.5,-0.5) -- (0,0);
    \draw[causArrow] (0,0) -- (0.5,0);
    \draw[fill=white] (0,0) circle(3pt);
    \draw[fill=white] (-0.5,0.5) circle(3pt);
    \draw[fill=white] (-0.5,-0.5) circle(3pt);
    \draw[fill=white] (0.5,0) circle(3pt);
\end{tikzpicture}\,.
\end{equation}
Now all theta functions have disappeared, and we get
\begin{align}
i N^{(4)}_{4312} := &\frac{-i}{12}\int d^4x_1 d^4x_2 d^4x_3 d^4x_4\
\left[ 
    \frac{1}{6} \Delta^{R}_{12} \Delta^{R}_{13} \Delta^{R}_{14} \frac{:\phi_2^2 \phi_3^2 \phi_4^2:}{(2!)^3}
    +\frac{1}{2} \Delta^{R}_{12} \Delta^{R}_{23} \Delta^{R}_{24} \frac{:\phi_1^2 \phi_3^2 \phi_4^2:}{(2!)^3}\right. \nn\\
    &\left.+\Delta^{R}_{12} \Delta^{R}_{14} \Delta^{R}_{23} \frac{:\phi_1 \phi_2 \phi_3^2 \phi_4^2:}{(2!)^2}
    +\Delta^{R}_{12} \Delta^{R}_{23} \Delta^{R}_{34} \frac{:\phi_1^2 \phi_2 \phi_3 \phi_4^2:}{(2!)^2}
    \right] .
\end{align}
Finally we consider the last two nested commutators $N^{(4)}_{2341}$ and $N^{(4)}_{1432}$. In this case, as we are about to show, we need both of them to eliminate all theta functions~$\theta_{ij}$. We have
\begin{align}
\begin{split}
i &N^{(4)}_{2341} + i N^{(4)}_{1432} = \frac{-i}{12}\int d^4x_1 d^4x_2 d^4x_3 d^4x_4\ 
\left[ 
-\Delta^{R}_{14} \Delta^{R}_{24} \Delta^{R}_{34} \theta_{12} \theta_{23} \frac{:\phi_1^2 \phi_2^2 \phi_3^2:}{(2!)^3}\right. \\ &+\Delta^{R}_{13} \Delta^{R}_{23} \Delta^{R}_{34} \theta_{12} \frac{:\phi_1^2 \phi_2^2 \phi_4^2:}{(2!)^3}
+\Delta^{R}_{12} \Delta^{R}_{23} \Delta^{R}_{24} \theta_{34} \frac{:\phi_1^2 \phi_3^2 \phi_4^2:}{(2!)^3}
-\Delta^{R}_{12} \Delta^{R}_{13} \Delta^{R}_{14} \theta_{23} \theta_{34} \frac{:\phi_2^2 \phi_3^2 \phi_4^2:}{(2!)^3} \\
&+ (\Delta^{R}_{12} \Delta^{R}_{14} \Delta^{R}_{34} \theta_{23} 
{+} \Delta^{R}_{13} \Delta^{R}_{14} \Delta^{R}_{24} \theta_{12} \theta_{23} \theta_{34}) \frac{:\phi_1 \phi_2^2 \phi_3^2 \phi_4:}{(2!)^2}+\Delta^{R}_{13} \Delta^{R}_{14} \Delta^{R}_{23} \theta_{12} \theta_{34} \frac{{:}\,\phi_1 \phi_2^2 \phi_3 \phi_4^2\,{:}}{(2!)^2} \\
&\left.+ (\Delta^{R}_{12} \Delta^{R}_{23} \Delta^{R}_{34}
+ \Delta^{R}_{13} \Delta^{R}_{23} \Delta^{R}_{24} \theta_{12} \theta_{34}) \frac{:\phi_1^2 \phi_2 \phi_3 \phi_4^2:}{(2!)^2}
+ \Delta^{R}_{14} \Delta^{R}_{23} \Delta^{R}_{24} \theta_{12} \theta_{34} \frac{:\phi_1^2 \phi_2 \phi_3^2 \phi_4:}{(2!)^2}
\right] .
\end{split}
\end{align}
Some propagator terms above have a permutation symmetry, such as $\Delta^{R}_{14} \Delta^{R}_{24} \Delta^{R}_{34}$ which is symmetric in $\{1,2,3\}$ or $\Delta^{R}_{13} \Delta^{R}_{23} \Delta^{R}_{34}$ which is symmetric in $\{1,2\}$. Simplifying the $\theta$-functions in these terms is straightforward and follows the same reasoning as before, using the identities \eqref{eq:tree6_3out}, \eqref{eq:tree6_2out}, \eqref{eq:tree6_3in}, and \eqref{eq:tree6_2in}.
Besides them, there are five leftover terms that simplify together as follows,
\begin{align}
\begin{split}
& 2 \Delta^{R}_{12} \Delta^{R}_{14} \Delta^{R}_{34} \theta_{23} :\phi_1 \phi_2^2 \phi_3^2 \phi_4: 
+ 2\Delta^{R}_{13} \Delta^{R}_{14} \Delta^{R}_{24} \theta_{12} \theta_{23} \theta_{34} :\phi_1 \phi_2^2 \phi_3^2 \phi_4: \\
&+ 2 \Delta^{R}_{13} \Delta^{R}_{23} \Delta^{R}_{24} \theta_{12} \theta_{34} :\phi_1^2 \phi_2 \phi_3 \phi_4^2: + 2 \Delta^{R}_{14} \Delta^{R}_{23} \Delta^{R}_{24} \theta_{12} \theta_{34} :\phi_1^2 \phi_2 \phi_3^2 \phi_4: \\
&+ 2 \Delta^{R}_{13} \Delta^{R}_{14} \Delta^{R}_{23} \theta_{12} \theta_{34} :\phi_1 \phi_2^2 \phi_3 \phi_4^2: \\
&\rightarrow 2 \Delta^{R}_{12} \Delta^{R}_{14} \Delta^{R}_{34} \big(\theta_{23} + \theta_{13}\theta_{32}\theta_{24} + \theta_{31}\theta_{42} + \theta_{31}\theta_{24} + \theta_{13}\theta_{42}\big) :\phi_1 \phi_2^2 \phi_3^2 \phi_4: \\
&= 2 \Delta^{R}_{12} \Delta^{R}_{14} \Delta^{R}_{34} :\phi_1 \phi_2^2 \phi_3^2 \phi_4: ,
\end{split}
\end{align}
 using the identity
\begin{equation}
\begin{tikzpicture}[baseline={([yshift=-.5ex]current bounding box.center)},thick]

    \draw[causArrow] (0,0) -- (0.8,0);
    \draw[causArrow] (0,0) -- (0.8,-0.75);
    \draw[causArrow] (0,-0.75) -- (0.8,-0.75);
    \draw[fill=white] (0.8,0) circle(3pt);
    \draw[fill=white] (0,0) circle(3pt);
    \draw[fill=white] (0.8,-0.75) circle(3pt);
    \draw[fill=white] (0,-0.75) circle(3pt);
    \node at (1.4, -0.375){\small$=$};

    \begin{scope}[xshift=2cm]
    \draw[thetaLine=0.3] (0.4,0) -- (0.8,-0.75);
    \draw[causArrow] (0,0) -- (0.4,0);
    \draw[causArrow=0.7] (0,0) -- (1.2,-0.75);
    \draw[causArrow] (0.8,-0.75) -- (1.2,-0.75);
    \draw[fill=white] (0,0) circle(3pt);
    \draw[fill=white] (0.4,0) circle(3pt);
    \draw[fill=white] (0.8,-0.75) circle(3pt);
    \draw[fill=white] (1.2,-0.75) circle(3pt);
    \node at (1.6, -0.375){\small$+$};
    \end{scope}

    \begin{scope}[xshift=4cm]
    \draw[thetaLine] (0,0) -- (0.4,-0.75);
    \draw[thetaLine=0.3] (0.4,-0.75) -- (0.8,0);
    \draw[thetaLine] (0.8,0) -- (1.2,-0.75);
    \draw[causArrow] (0,0) -- (0.8,0);
    \draw[causArrow=0.7] (0,0) -- (1.2,-0.75);
    \draw[causArrow] (0.4,-0.75) -- (1.2,-0.75);
    \draw[fill=white] (0,0) circle(3pt);
    \draw[fill=white] (0.8,0) circle(3pt);
    \draw[fill=white] (0.4,-0.75) circle(3pt);
    \draw[fill=white] (1.2,-0.75) circle(3pt);
    \node at (1.6, -0.375){\small$+$};
    \end{scope}

    \begin{scope}[xshift=6cm]
    \draw[thetaLine] (0,-0.75) -- (0.4,0);
    \draw[thetaLine] (0.8,-0.75) -- (1.2,0);
    \draw[causArrow] (0.4,0) -- (1.2,0);
    \draw[causArrow] (0.4,0) -- (0.8,-0.75);
    \draw[causArrow] (0,-0.75) -- (0.8,-0.75);
    \draw[fill=white] (0.4,0) circle(3pt);
    \draw[fill=white] (1.2,0) circle(3pt);
    \draw[fill=white] (0,-0.75) circle(3pt);
    \draw[fill=white] (0.8,-0.75) circle(3pt);
    \node at (1.6, -0.375){\small$+$};
    \end{scope}

    \begin{scope}[xshift=8cm]
    \draw[thetaLine] (0,-0.75) -- (0.4,0);
    \draw[thetaLine] (0.8,0) -- (1.2,-0.75);
    \draw[causArrow] (0.4,0) -- (0.8,0);
    \draw[causArrow] (0.4,0) -- (1.2,-0.75);
    \draw[causArrow] (0,-0.75) -- (1.2,-0.75);
    \draw[fill=white] (0.4,0) circle(3pt);
    \draw[fill=white] (0.8,0) circle(3pt);
    \draw[fill=white] (0,-0.75) circle(3pt);
    \draw[fill=white] (1.2,-0.75) circle(3pt);
    \node at (1.6, -0.375){\small$+$};
    \end{scope}

    \begin{scope}[xshift=10cm]
    \draw[thetaLine] (0,0) -- (0.4,-0.75);
    \draw[thetaLine] (0.8,-0.75) -- (1.2,0);
    \draw[causArrow] (0,0) -- (1.2,0);
    \draw[causArrow] (0,0) -- (0.8,-0.75);
    \draw[causArrow] (0.4,-0.75) -- (0.8,-0.75);
    \draw[fill=white] (0,0) circle(3pt);
    \draw[fill=white] (1.2,0) circle(3pt);
    \draw[fill=white] (0.4,-0.75) circle(3pt);
    \draw[fill=white] (0.8,-0.75) circle(3pt);
    \end{scope}
\end{tikzpicture}\, .
\end{equation}
Note that to identify each of the five subregions we had to first relabel terms to find a common propagator structure, just like we did in \eqref{eq:ex2thetacancell}. Also note that some of the five terms above come from $N^{(4)}_{2341}$ and some from $N^{(4)}_{1432}$, meaning that we need to consider the sum to see the simplification of theta functions.

Putting everything together, we get
\begin{align}
&i N^{(4)}_{2341} + i N^{(4)}_{1432} = \frac{-i}{12}\int d^4x_1 d^4x_2 d^4x_3 d^4x_4\ 
\left[ 
-\frac{1}{6}\Delta^{R}_{14} \Delta^{R}_{24} \Delta^{R}_{34} \frac{:\phi_1^2 \phi_2^2 \phi_3^2:}{(2!)^3}\right.\nonumber \\
&+\frac{1}{2}\Delta^{R}_{13} \Delta^{R}_{23} \Delta^{R}_{34} \frac{:\phi_1^2 \phi_2^2 \phi_4^2:}{(2!)^3}
+\frac{1}{2}\Delta^{R}_{12} \Delta^{R}_{23} \Delta^{R}_{24} \frac{:\phi_1^2 \phi_3^2 \phi_4^2:}{(2!)^3}
-\frac{1}{6}\Delta^{R}_{12} \Delta^{R}_{13} \Delta^{R}_{14} \frac{:\phi_2^2 \phi_3^2 \phi_4^2:}{(2!)^3}\nonumber \\
&\left.+ \Delta^{R}_{12} \Delta^{R}_{14} \Delta^{R}_{34} \frac{:\phi_1 \phi_2^2 \phi_3^2 \phi_4:}{(2!)^2} + \Delta^{R}_{12} \Delta^{R}_{23} \Delta^{R}_{34} \frac{:\phi_1^2 \phi_2 \phi_3 \phi_4^2:}{(2!)^2}
\right] .
\end{align}
Note that the terms $\Delta^{R}_{14} \Delta^{R}_{24} \Delta^{R}_{34}$ and $\Delta^{R}_{12} \Delta^{R}_{13} \Delta^{R}_{14}$ appear with coefficient $-1/6$ here, but they appear with coefficient $+1/6$ in $N^{(4)}_{1234}$ and $N^{(4)}_{4312}$ and hence they drop out.

Overall, the complete expression for the fourth-order $N$-operator is 
\begin{align}\label{N4-Position}
    iN^{(4)}&=\frac{-i}{12} \int d^4x_1 d^4x_2 d^4x_3 d^4x_4\ \left[(\Delta^R_{34} \Delta^R_{41} \Delta^R_{42}+\Delta^R_{14} \Delta^R_{24} \Delta^R_{43})\frac{:\phi_1^2 \phi_2^2 \phi_3^2:}{(2!)^3} \right. \\
    &\left.+(3 \Delta^R_{12} \Delta^R_{23} \Delta^R_{34} + \Delta^R_{12} \Delta^R_{32} \Delta^R_{34} + \Delta^R_{21} \Delta^R_{32} \Delta^R_{34} + \Delta^R_{12} \Delta^R_{32} \Delta^R_{43})\frac{:\phi_1^2 \phi_2 \phi_3 \phi_4^2}{(2!)^2}:\right] + \cdots . \nn
\end{align} 
Sandwiching this between the states as in \eqref{eq:6ptMatrixElementDef}, we arrive at the six-point Magnus amplitude
\begin{align}\label{N64}
    iN^{(4)}_6&= (-i)^4\hat{\delta}^{(4)}(p_1+\cdots+p_6)\, \cM^{(4)}_6\, ,  \\
\begin{split}
    \cM^{(4)}_6&= 
    \frac{1}{16}\sum_{S_6}\left(\
    \frac{1}{6}
    \begin{tikzpicture}[baseline={([yshift=-0.5ex]current bounding box.center)},thick] 
    \draw[Anew] (0,0) -- (1,0);
    \draw[massive] (1,0) -- ++(60:0.5);
    \draw[massive] (1,0) -- ++(-60:0.5);
    \draw[Rnew] (0,0) -- (120:1);
    \draw[massive] (120:1) -- ++(180:0.5);
    \draw[massive] (120:1) -- ++(60:0.5);
    \draw[Rnew] (0,0) -- (-120:1);
    \draw[massive] (-120:1) -- ++(-180:0.5);
    \draw[massive] (-120:1) -- ++(-60:0.5);
    \end{tikzpicture}
    \ +\frac{1}{6} \
    \begin{tikzpicture}[baseline={([yshift=-0.5ex]current bounding box.center)},thick] 
    \draw[Rnew] (0,0) -- (1,0);
    \draw[massive] (1,0) -- ++(60:0.5);
    \draw[massive] (1,0) -- ++(-60:0.5);
    \draw[Anew] (0,0) -- (120:1);
    \draw[massive] (120:1) -- ++(180:0.5);
    \draw[massive] (120:1) -- ++(60:0.5);
    \draw[Anew] (0,0) -- (-120:1);
    \draw[massive] (-120:1) -- ++(-180:0.5);
    \draw[massive] (-120:1) -- ++(-60:0.5);
    \end{tikzpicture}
    \ \right) \\
    &+\frac{1}{4}\sum_{S_6} \left(\
    \frac{1}{4}
    \begin{tikzpicture}[baseline={([yshift=-0.5ex]current bounding box.center)},thick] 
    \draw[massive] (0,0) -- (135:1);
    \draw[massive] (0,0) -- (-135:1);
    \draw[Rnew] (0,0) -- (1,0);
    \draw[massive] (1,0) -- (1,1);
    \draw[Rnew] (1,0) -- (2,0);
    \draw[massive] (2,0) -- (2,1);
    \draw[Rnew] (2,0) -- (3,0);
    \draw[massive] (3,0) -- ++(45:1);
    \draw[massive] (3,0) -- ++(-45:1);
    \path (2,0) -- (2,-1); 
    \end{tikzpicture}
    \ +\frac{1}{12} \
    \begin{tikzpicture}[baseline={([yshift=-0.5ex]current bounding box.center)},thick] 
    \draw[massive] (0,0) -- (135:1);
    \draw[massive] (0,0) -- (-135:1);
    \draw[Rnew] (0,0) -- (1,0);
    \draw[massive] (1,0) -- (1,1);
    \draw[Anew] (1,0) -- (2,0);
    \draw[massive] (2,0) -- (2,1);
    \draw[Rnew] (2,0) -- (3,0);
    \draw[massive] (3,0) -- ++(45:1);
    \draw[massive] (3,0) -- ++(-45:1);
    \path (2,0) -- (2,-1); 
    \end{tikzpicture}
    \right. \\
    &\left.+\frac{1}{12} \
    \begin{tikzpicture}[baseline={([yshift=-0.5ex]current bounding box.center)},thick] 
    \draw[massive] (0,0) -- (135:1);
    \draw[massive] (0,0) -- (-135:1);
    \draw[Anew] (0,0) -- (1,0);
    \draw[massive] (1,0) -- (1,1);
    \draw[Anew] (1,0) -- (2,0);
    \draw[massive] (2,0) -- (2,1);
    \draw[Rnew] (2,0) -- (3,0);
    \draw[massive] (3,0) -- ++(45:1);
    \draw[massive] (3,0) -- ++(-45:1);
    \path (2,0) -- (2,-1); 
    \end{tikzpicture}
    \ +\frac{1}{12}\
    \begin{tikzpicture}[baseline={([yshift=-0.5ex]current bounding box.center)},thick] 
    \draw[massive] (0,0) -- (135:1);
    \draw[massive] (0,0) -- (-135:1);
    \draw[Rnew] (0,0) -- (1,0);
    \draw[massive] (1,0) -- (1,1);
    \draw[Anew] (1,0) -- (2,0);
    \draw[massive] (2,0) -- (2,1);
    \draw[Anew] (2,0) -- (3,0);
    \draw[massive] (3,0) -- ++(45:1);
    \draw[massive] (3,0) -- ++(-45:1);
    \path (2,0) -- (2,-1); 
    \end{tikzpicture}
    \ \right)\,.
\end{split}
\end{align}
As before, we have omitted the labels of external lines for notational simplicity, but they are to be summed over all possible permutations $\sigma \in S_6$ of $(p_1,p_2,p_3,p_4,p_5,p_6)$. Also, the coefficients in front of each diagram (ignoring the coefficients outside the sum, which are due to overcounting) correspond to the Murua coefficients in \eqref{easyMurua4V}. In particular, notice the absence of two star diagrams with all arrows incoming/outgoing, which indeed have vanishing Murua coefficient.

Finally, the subtle  recombination of 
$\theta$-functions into manifestly Lorentz-invariant retarded and advanced propagators that we have illustrated so far in several examples is actually general, as argued  in Section~\ref{sec:newformula}.

\section{One-loop amplitudes  in the 
\texorpdfstring{$\cH_I=\frac{:\phi^3:}{3!}$}{phicubed}
  theory}
  \label{sec:oneloop}

Still working in our theory of a real scalar of mass $m$ with a cubic interaction, we now show a few examples of one-loop amplitudes. These will highlight intriguing connections to the tree amplitudes computed in the previous section, which we will further investigate and explain in Section~\ref{sec:MagnusLoopsFromFL}.  

\subsection{The bubble}
\label{sec:bubble}

As a first one-loop example,  we consider the two-point matrix element
\begin{equation}
    N^{(2)}_{2} = \langle 0| N^{(2)} | \phi(p_1) \phi(p_2) \rangle\,,
\end{equation}
where $N^{(2)}$ is the second-order term in the Magnus expansion given in \eqref{second-order-Mag-exp}, 
that is 
\begin{align}
\label{olnu}
i N^{(2)}_{2} = 
\frac{(-i)^2}{2}\left( \frac{1}{3!} \right)^2 \int\!d^4x_1 d^4x_2 \, \theta (t_1 - t_2)  \braket{0 | [:\phi^3(x_1):, :\phi^3(x_2):] | \phi(p_1) \phi(p_2)}\, .  
\end{align}
We can  work out directly the commutator $[{:\phi^3(x_1):}, {:\phi^3(x_2):}]$. We do it first 
 without normal ordering which we  reinstate at the end. 
In order to do, so we first recall  \eqref{useful-rewriting}, and then note that   in order to 
get the one-loop bubble we need to perform one more Wick contraction. 
This leads to
\begin{align}\label{eq: oneLoopBracketEx}
\begin{split}
    [\phi_1^3, \phi_2^3] &\rightarrow 3i \Delta_{12} \Big( 
    2\cdot 2 :\phi_1 \phi_2: i \Delta^{(+)}_{12} + 
    2 :\phi_2 \phi_1: 
    i \Delta^{(+)}_{12} \\ &  + 2 :\phi_2 \phi_1: 
    i \Delta^{(+)}_{21}  + 2\cdot 2 :\phi_2 \phi_1: i \Delta^{(+)}_{21}\Big) \\
     & = 
     3i \Delta_{12} \cdot 6  :\phi_1 \phi_2 : \Big( i \Delta_{12}^{(+)} + i \Delta_{21}^{(+)}\Big)  \\ &=
     18 \, i\Delta_{12} \Delta^{(1)}_{12}
     :\phi_1\phi_2:\, . 
     \end{split}
\end{align}
We can extend the tree-level diagrammatic notation in \eqref{first-ex-pic} to include now the Hadamard function $\Delta^{(1)}_{ij}$ introduced in \eqref{Hadamard}:
\begin{equation}
\begin{tikzpicture}[font=\small, thick]
        \draw[cut] (0,0) -- (1,0);
        \draw[fill=white] (0,0) circle(3pt);
        \draw[fill=white] (1,0) circle(3pt);
        \node at (0,0.4){$j$};
        \node at (1,0.4){$i$};
    \end{tikzpicture}
    = \Delta^{(1)}_{ij}\,,
\end{equation}
which allows us to represent \eqref{eq: oneLoopBracketEx} as
\begin{equation}
i\Delta^R_{12}\, \Delta^{(1)}_{12}=
\begin{tikzpicture}[baseline={([yshift=-0.5ex]current bounding box.center)},thick,font=\footnotesize]
    \draw[out=45,causArrow] (0,0) node[yshift=0.4cm]{2} to (1,0) node[yshift=0.4cm]{1};
    \draw[out=-45,in=-135,cut] (0,0) to (1,0);
    \draw[fill=white] (0,0) circle(3pt);
    \draw[fill=white] (1,0) circle(3pt);
\end{tikzpicture} \, .
\end{equation}
For $iN^{(2)}$ we can now write 
\begin{align}
\begin{split}
\label{firstobs1}
    i N^{(2)} &= \frac{(-i)^2}{2}
    \left(\frac{1}{3!}\right)^2 \cdot 18 \int\!d^4x_1 d^4x_2 \, \big( \theta_{12} i \Delta_{12}\big) \, \Delta^{(1)}_{12} \, :\phi_1  \phi_2 : + \dots
    \\ & 
    = 
    \frac{(-i)^2}{4} \int\!d^4x_1 d^4x_2 \, \big( i \Delta^{R}_{12}\big) \, \Delta^{(1)}_{12} \, :\phi_1  \phi_2 : + \dots
    \, ,
   \end{split}
\end{align}
where the terms omitted do not contribute to the matrix element $N^{(2)}_2$. Note that as usual we have symmetrised in $x_1\leftrightarrow x_2$ to get rid of $\theta$ functions.
We then use that 
\begin{align}
\label{extstatebub}
\braket{0 | :\phi(x_1)\phi(x_2): | \phi(p_1) \phi(p_2)} = e^{-i(p_1 \cdot x_1 + p_2 \cdot x_2)} + e^{-i(p_1 \cdot x_2 + p_2 \cdot x_1)}
\, ,  
\end{align}
to get 
\begin{align}
\begin{split}
    i N_{2}^{(2)} &= 
   \frac{(-i)^2}{4}\int\!d^4x_1 d^4x_2 \, \big(  i \Delta^{R}_{12}\big) \, \Delta^{(1)}_{12} \,(e^{-i(p_1 \cdot x_1 + p_2 \cdot x_2)} + e^{-i(p_1 \cdot x_2 + p_2 \cdot x_1)}) \\
   &= 
   \frac{(-i)^2}{4}
   \int\!d^4x_1 d^4x_2 \, e^{-i(p_1\Cdot x_1 + p_2 \Cdot x_2)}\Big( i \Delta^{R}_{12} \Delta^{(1)}_{12} + 
   i \Delta^{R}_{21} \Delta^{(1)}_{21}\Big) 
   \\ & 
   = 
   \frac{(-i)^2}{4}
   \int\!d^4x_1 d^4x_2 \, e^{-i(p_1\Cdot x_1 + p_2 \Cdot x_2)}\, \Delta^{(1)}_{12}\Big( i \Delta^{R}_{12}  + 
   i \Delta^{A}_{12} \Big) \, .
   \end{split}
\end{align}
In this derivation we have used that $i\Delta^R (x) = \theta (x_0) i \Delta (x)$, and changed integration variables $x_1\leftrightarrow x_2$ in the term containing    $e^{-i(p_1 \cdot x_2 + p_2 \cdot x_1)}$. 
Thus we conclude that 
\begin{align}
\begin{split}
\label{bubcut-x}
    i N_{2}^{(2)} &= 
   \frac{(-i)^2}{4}
   \int\!d^4x_1 d^4x_2 \, e^{-i(p_1\Cdot x_1 + p_2 \Cdot x_2)}\, \Delta^{(1)}_{12}\Big( i \Delta^{R}_{12}  + 
   i \Delta^{A}_{12} \Big) \, , 
   \end{split}
\end{align}
that is, 
\begin{align}
\label{bubcut}
i N_{2}^{(2)} &=  \hat{\delta}^{(4)}(p_1 + p_2) \cdot (-i)^2 \cdot \frac{1}{2}\int \frac{d^4q}{(2\pi)^4} \Delta^{(1)}(q) \times \frac{1}{2} \left( i\Delta^{R} + i\Delta^{A} \right)(q-p_1 ) \nn \\
&= \hat{\delta}^{(4)}(p_1+p_2)\frac{(-i)^2}{2}\left[\frac{1}{2}\
\begin{tikzpicture}[baseline={([yshift=-0.5ex]current bounding box.center)},font=\small,thick]
    \draw[massive] (-1.25,0) node[left]{$p_1$} -- (-0.75,0);
    \draw[massive,Rnew] (-0.75,0) arc(180:0:0.75);
    \draw[massive,cut=2] (-0.75,0) arc(-180:0:0.75);
    \draw[massive] (0.75,0) -- (1.25,0) node[right]{$p_2$};
\end{tikzpicture}
\ + \frac{1}{2}\
\begin{tikzpicture}[baseline={([yshift=-0.5ex]current bounding box.center)},font=\small,thick]
    \draw[massive] (-1.25,0) node[left]{$p_1$} -- (-0.75,0);
    \draw[massive,Anew] (-0.75,0) arc(180:0:0.75);
    \draw[massive,cut=2] (-0.75,0) arc(-180:0:0.75);
    \draw[massive] (0.75,0) -- (1.25,0) node[right]{$p_2$};
\end{tikzpicture}
\right]
\end{align}
Three important comments are in order here.

{\bf 1.}~First, a key feature of \eqref{bubcut} and \eqref{firstobs1} is that they  are cut integrals, since $\Delta^{(1)}(q) = (2\pi)\,\delta(q^2 - m^2)$. Consequently, we are effectively performing  phase-space integrals over a tree-level quantity. Note that this representation differs from the Feynman-tree-theorem representation of the $T$-matrix \cite{Feynman:1963ax,Feynman:1972mt} (since there one has to sum over multiple cuts as well) and the $Q$-cuts of \cite{Baadsgaard:2015twa}. 

{\bf 2.}~Furthermore, and importantly, we will see that for a one-loop $n$-point Magnus amplitude, what we are actually integrating is an $(n+2)$-point tree-level Magnus amplitude, where by this we mean the sum of all the tree-level Magnus diagrams at $n+2$ points. 
This correspondence is already manifest  by comparing \eqref{firstobs1} to \eqref{N2}. 
In fact, we can go from the first to the second by simply performing the replacement 
\begin{align}
\label{reppp1}
 :\phi_1^2  \phi_2^2 :  \to 
2 \Delta^{(1)}_{12} :\phi_1\phi_2:
\, . 
\end{align}
We will see a similar relation between the five-point tree amplitude and the one-loop triangle later in Section~\ref{sec:phicubed-triangle}, 
and will discuss this further in Sections~\ref{sec:one-loop-Wick} and \ref{sec:MagnusLoopsFromFL}.

{\bf 3.}~Furthermore, it is instructive  to  perform a comparison  with  the Dyson series. 
The difference between Magnus and Dyson matrix elements for the process at hand  is:
\begin{align}
iN^{(2)}_{2} -  iT^{(2)}_{2} = \frac{1}{2} \cdot \frac{1}{2} \cdot \int d^4x_1 d^4x_2 \, e^{-i(p_1 \cdot x_1 + p_2 \cdot x_2)} \left[ (i\Delta_{12}^{(+)})^2 + (i\Delta_{12}^{(-)})^2 \right]\, .
\end{align}
This  precisely corresponds to the expected relation~\cite{Damgaard:2021ipf}
\begin{align}\label{discontinuity-gone}
iN^{(2)} = iT^{(2)} + \frac{1}{2} T^{(1)} T^{(1)}\, , 
\end{align}
between Magnus and $T$-matrix elements, since%
\footnote{The factor of $1/2$ in \eqref{T1T1} arises due to  the presence of  identical particles.}
\begin{align}
\label{T1T1}
     \langle 0 | T^{(1)} T^{(1)}| \phi (p_1) \phi (p_2) \rangle  = \frac{1}{2} \int\! d^4x_1 d^4x_2 \, e^{-i(p_1 \cdot x_1 + p_2 \cdot x_2)} \left[ (i\Delta_{12}^{(+)})^2 + (i\Delta_{12}^{(-)})^2 \right]\, . 
     \end{align}
     
{\bf 4.}~Finally, let us evaluate directly \eqref{bubcut}, which can be done conveniently using 
\eqref{advplusret} and \eqref{FminusFbar}. We find the result 
\begin{align}
i N_{2}^{(2)} &=  \hat{\delta}^{(4)}(p_1 + p_2)   \frac{(-i)^2}{4}\int\!\frac{d^4q}{(2\pi)^4} \left[ i\Delta^{F}(q)  i\Delta^{F}(q-p_1 ) - 
i\Delta^{\bar{F}}(q)  i\Delta^{\bar{F}}(q-p_1 )
\right] \, ,
\end{align}
which is nothing but the difference of the bubble integral evaluated with the  Feynman prescription and the same integral but with  an  anti-Feynman prescription. To be more concrete let us assume that the internal particles are massless and the external momenta $p_{1,2}$ are off-shell.
Using well-known results for the bubble integrals 
we find
\begin{align}
\begin{split}
i N_2^{(2)} &= \hat{\delta}^{(4)}(p_1 + p_2) \frac{c_\Gamma}{4 \eps (1-2 \eps)} \left[-i \, (-s-i \vareps)^{-\eps} - \text{c.c.}\right] \\
& =  \hat{\delta}^{(4)}(p_1 + p_2) \frac{(-i \, c_\Gamma)}{4 \eps (1-2 \eps)} \left[(-s-i \vareps)^{-\eps} + (-s+i \vareps)^{-\eps}\right] \\
& = \hat{\delta}^{(4)}(p_1 + p_2) \frac{(-i\, c_\Gamma)}{2}\left( \frac{1}{\eps} + \log (|s|) +2 \right) + \mathcal{O}(\eps) \, ,
\end{split}
\end{align}
where $s = p_1^2$ and $c_{\Gamma} = \Gamma(1+\eps)\Gamma^2(1-\eps)/(\Gamma(1-2\eps)(4\pi)^{2-\eps})$.
This means that in this one-loop Magnus matrix element the real part/discontinuity of the $T$-matrix element  has been removed. This is a simple example of a general feature, which arises due to  subtracting  the last term  in \eqref{discontinuity-gone}.

\subsection{The three-point triangle  and bubble topologies}
\label{sec:phicubed-triangle}

 We now compute the Magnus matrix element
\begin{align}
    N_{3}^{(3)} = \langle 0 | N^{(3)} | \phi(p_1) \phi(p_2) \phi(p_3) \rangle\, , 
    \end{align}
    where $N^{(3)}$ is given in \eqref{N3phicubed}.
Note that the operator $N^{(3)}$ contains two types of terms contributing to the matrix element $N_{3}^{(3)}$, which give rise to two different graph topologies, namely
\begin{equation}
\label{eq:tribub}
    N^{(3)} \big|_{\rm tri} \propto \, :\!\phi(p_1)\phi(p_2)\phi(p_3)\!: \,  , \qquad N^{(3)} \big|_{\rm bub} \propto \, :\!\phi(p_i)^2\phi(p_j)\!: \, , 
\end{equation}
where $N^{(3)} \big|_{\rm tri}$ will give rise to one-loop triangle graphs and $N^{(3)} \big|_{\rm bub}$ will give rise to one-loop three-point graphs with a bubble on an external leg (see \eqref{eq:TriangleCut} and \eqref{eq:bub12} below). 
For simplicity, we will focus on the former case and only quote the final result for the latter.
    
For convenience, similarly to Section~\ref{sec:5pttree}, we  set 
\begin{align}
N^{(3)}\big|_{\rm tri} = N_{123}^{(3)}\big|_{\rm tri} + N_{321}^{(3)}\big|_{\rm tri}\, , 
\end{align}
where the two terms correspond to the contributions in $N^{(3)}\big|_{\rm tri}$ arising from the two distinct commutators, respectively.
We then find, using the result \eqref{123phi}, performing one more Wick contraction and then normal-ordering the rest, 
\begin{align}
\begin{split}
&iN_{123}^{(3)} \big|_{\rm tri} = \frac{(-i)^3}{6}\left( \frac{1}{3!}\right)^3 \int\!d^4x_1 d^4x_2 d^4 x_3 \ :\!\phi_1 \phi_2 \phi_3\!: \\
&
    3\cdot 
36   
 (i\Delta^R_{23})\Big[    (i\Delta^{R}_{12}) \Delta^{(1)}_{13}  + 
    (i\Delta^{R}_{13}) (\theta_{12}\Delta^{(1)}_{12}) 
  \Big]\, 
      \, , 
\end{split}
\end{align}
where since $\Delta_{13}^R\Delta_{23}^R\Delta^{(1)}_{12}:\!\phi_1 \phi_2 \phi_3\!:$ is symmetric under the exchange $x_1\leftrightarrow x_2$, $N_{123}^{(3)}\big|_{\rm tri}$ can be rewritten  as 
\begin{align}
\begin{split}
&iN_{3, 123}^{(3)}\big|_{\rm tri} = \frac{(-i)^3}{12}\int\!d^4x_1 d^4x_2 d^4 x_3 \ :\!\phi_1 \phi_2 \phi_3\!: \\
&
(i\Delta^{R}_{23}) \Big[    (i\Delta^{R}_{12}) \Delta^{(1)}_{13}  + 
  \frac{1}{2}   (i\Delta^{R}_{13}) \Delta^{(1)}_{12} 
  \Big]\, 
     \, . 
\end{split}
\end{align}
Diagrammatically, the cancellation of the $\theta$-functions happens in the same way as before,
\begin{equation}
\begin{tikzpicture}[baseline={([yshift=-.5ex]current bounding box.center)}, thick]
    \draw[causArrow] (0,0) -- (0.5,0.5);
    \draw[causArrow] (0,0) -- (1,-0.5);
    \draw[thetaLine,out=0,in=30] (0.5,0.5) to (1,-0.5);
    \draw[cut] (0.5,0.5) -- (1,-0.5);
    \draw[fill=white] (0,0) circle(3pt);
    \draw[fill=white] (0.5,0.5) circle(3pt);
    \draw[fill=white] (1,-0.5) circle(3pt);
\end{tikzpicture} \rightarrow \frac{1}{2!}\left(
\begin{tikzpicture}[baseline={([yshift=-.5ex]current bounding box.center)}, thick]
\begin{scope}
    \draw[causArrow] (0,0) -- (0.5,0.5);
    \draw[causArrow] (0,0) -- (1,-0.5);
    \draw[thetaLine,out=0,in=30] (0.5,0.5) to (1,-0.5);
    \draw[cut] (0.5,0.5) -- (1,-0.5);
    \draw[fill=white] (0,0) circle(3pt);
    \draw[fill=white] (0.5,0.5) circle(3pt);
    \draw[fill=white] (1,-0.5) circle(3pt);
    \node at (1.5, 0){$+$};
\end{scope}
\begin{scope}[xshift=2cm]
    \draw[causArrow] (0,0) -- (1,0.5);
    \draw[causArrow] (0,0) -- (0.5,-0.5);
    \draw[thetaLine,out=0,in=30,yscale=-1] (0.5,0.5) to (1,-0.5);
    \draw[cut] (0.5,-0.5) -- (1,0.5);
    \draw[fill=white] (0,0) circle(3pt);
    \draw[fill=white] (0.5,-0.5) circle(3pt);
    \draw[fill=white] (1,0.5) circle(3pt);
\end{scope}
\end{tikzpicture} \, \right) = \frac{1}{2!}\,
\begin{tikzpicture}[baseline={([yshift=-.5ex]current bounding box.center)}, thick]
    \draw[causArrow] (0,0) -- (0.5,0.5);
    \draw[causArrow] (0,0) -- (0.5,-0.5);
    \draw[cut] (0.5,0.5) -- (0.5,-0.5);
    \draw[fill=white] (0,0) circle(3pt);
    \draw[fill=white] (0.5,0.5) circle(3pt);
    \draw[fill=white] (0.5,-0.5) circle(3pt);
\end{tikzpicture}\, ,
\end{equation}
where the additional cut line is symmetric and therefore it does not change the symmetrisation argument. Similarly we find 
\begin{align}
\begin{split}
&iN_{321}^{(3)}\big|_{\rm tri} = \frac{(-i)^3}{12}\int\!d^4x_1 d^4x_2 d^4 x_3 \ :\!\phi_1 \phi_2 \phi_3\!: \\
&
(i\Delta^{R}_{12}) \Big[    (i\Delta^{R}_{23}) \Delta^{(1)}_{13}  + 
  \frac{1}{2}   (i\Delta^{R}_{13}) \Delta^{(1)}_{23} 
  \Big]\, 
     \, , 
\end{split}
\end{align}
and hence 
\begin{align}
\begin{split}
\label{one-loop-tri-int}
&iN^{(3)}\big|_{\rm tri} = \frac{(-i)^3}{12}\int\!d^4x_1 d^4x_2 d^4 x_3 \  :\!\phi_1 \phi_2 \phi_3\!: \\
&
\Big\{ 2  (i\Delta^{R}_{12}) (i\Delta^{R}_{23}) \Delta^{(1)}_{13}  + \frac{1}{2}\Big[ 
  (i\Delta^{R}_{23})   (i\Delta^{R}_{13}) \Delta^{(1)}_{12} + 
  (i\Delta^{R}_{12})   (i\Delta^{R}_{13}) \Delta^{(1)}_{23}
  \Big]\Big\} \, 
     \, . 
\end{split}
\end{align}
We now pause to observe the similarity of this expression with the corresponding expression for the five-point tree-level amplitude in 
\eqref{5pt-tree-int}, up to relabelling of the integration variables. The two are related by the simple replacements
\begin{subequations}\label{eq:ReplacementsTriangle}
\begin{align}
:\!\phi_1^2 \phi_2 \phi_3^2 \!:  &\to 
2\Delta^{(1)}_{13} \,  :\!\phi_1 \phi_2 \phi_3\!: \, , \\   
:\!\phi_1^2 \phi_2^2 \phi_3 \!:  &\to 2\Delta^{(1)}_{12}\,  :\!\phi_1 \phi_2 \phi_3\!: \, , \\  :\!\phi_1 \phi_2^2 \phi_3^2 \!:  &\to 2\Delta^{(1)}_{23}\,   :\!\phi_1 \phi_2 \phi_3\!: \, . 
\end{align}
\end{subequations}
This is similar to the relation we found between the one-loop bubble and the tree-level four-point amplitude in Section~\ref{sec:bubble}, 
and we will discuss it in more detail in Sections~\ref{sec:one-loop-Wick} and \ref{sec:MagnusLoopsFromFL}. 

Then we can put back the external states and compute
\begin{align}
\begin{split}
&iN_{3}^{(3)}\big|_{\rm tri} = \frac{(-i)^3}{12}\int\!d^4x_1 d^4x_2 d^4 x_3 \  \langle 0 | :\!\phi_1 \phi_2 \phi_3\!: | \phi(p_1) \dots \phi(p_3) \rangle \\
&
\Big\{ 2  (i\Delta^{R}_{12}) (i\Delta^{R}_{23}) \Delta^{(1)}_{13}  + \frac{1}{2}\Big[ 
  (i\Delta^{R}_{23})   (i\Delta^{R}_{13}) \Delta^{(1)}_{12} + 
  (i\Delta^{R}_{12})   (i\Delta^{R}_{13}) \Delta^{(1)}_{23}
  \Big]\Big\} \, 
     \, . 
\end{split}
\end{align}
Performing  Wick contractions, and renaming integration variables so that all terms are multiplied by the exponential $e^{-i(p_1\cdot x_1 + p_2\cdot x_2 + p_3\cdot x_3)}$, 
one finally arrives at 
\begin{align}
\label{triphicubed}
\begin{split}
&iN_{3}^{(3)}\big|_{\rm tri} = 
\frac{(-i)^3}{6}
  \int\!d^4x_1 d^4x_2 d^4 x_3 \ e^{-i(p_1\cdot x_1 + p_2\cdot x_2 + p_3\cdot x_3)} \\
\Bigg\{ &\Delta^{(1)}_{31}\left[      (i\Delta^{R}_{12}) (i\Delta^{R}_{23})+  (i\Delta^{A}_{12}) (i\Delta^{A}_{23})  +\frac{1}{2}\Big(  
  (i\Delta^{R}_{12}) (i\Delta^{A}_{23}) + (i\Delta^{A}_{12}) (i\Delta^{R}_{23})\Big) 
  \right]\\
    + & \Delta^{(1)}_{12}\left[      (i\Delta^{R}_{23}) (i\Delta^{R}_{31})+  (i\Delta^{A}_{23}) (i\Delta^{A}_{31})  + 
 \frac{1}{2} \Big((i\Delta^{R}_{23}) (i\Delta^{A}_{31}) + (i\Delta^{A}_{23}) (i\Delta^{R}_{31})\Big) 
  \right]\\
    + &\Delta^{(1)}_{23}\left[      (i\Delta^{R}_{31}) (i\Delta^{R}_{12})+  (i\Delta^{A}_{31}) (i\Delta^{A}_{12})  + 
 \frac{1}{2} \Big(  (i\Delta^{R}_{31}) (i\Delta^{A}_{12}) + (i\Delta^{A}_{31}) (i\Delta^{R}_{12})\Big) 
  \right]\Bigg\} \, 
     \, . 
\end{split}
\end{align}
Two important comments are in order here. 

{\bf 1.}~First we note the  similarly of \eqref{triphicubed} to our result for the bubble in \eqref{bubcut-x}: both are cut integrals, with one of  the  internal lines of the integral function being replaced by a $\Delta^{(1)}_{ij}$ function. 

{\bf 2.}~Furthermore, we note the appearance of different coefficients for the combinations 
($\Delta^R\times\Delta^A +\Delta^A\times\Delta^R$) and 
($\Delta^R\times\Delta^R +\Delta^A\times\Delta^A$). 
These are the same coefficients we 
saw in the case of a five-point tree-level amplitude, 
providing a second example of the  connection between one-loop and tree-level Magnus amplitudes that we observed in Section~\ref{sec:bubble}. 
This will be  explained in detail  in Sections~\ref{sec:one-loop-Wick} and \ref{sec:MagnusLoopsFromFL}.

The momentum-space diagrams that result from the integrals above are 
shown below, 
\begin{align}\label{eq:TriangleCut}
&iN^{(3)}_{3}\big|_{\rm tri}=\sum_{C_3}(-i)^3\hat{\delta}^{(4)}(p_1+p_2+p_3)\nn\\
&\left[\, \frac{1}{6}
\begin{tikzpicture}[baseline={([yshift=-.5ex]current bounding box.center)},thick]
    \draw (-90:1.5) -- (-90:1);
    \draw (30:1.5) -- (30:1);
    \draw (150:1.5) -- (150:1);
    \draw[massive, cut=2] (30:1) -- (150:1);
    \draw[massive, Rnew] (150:1) -- (-90:1);
    \draw[massive, Rnew] (-90:1) -- (30:1);
\end{tikzpicture}
\ +\frac{1}{12}
\begin{tikzpicture}[baseline={([yshift=-.5ex]current bounding box.center)},thick]
    \draw (-90:1.5) -- (-90:1);
    \draw (30:1.5) -- (30:1);
    \draw (150:1.5) -- (150:1);
    \draw[massive, cut=2] (30:1) -- (150:1);
    \draw[massive, Rnew] (150:1) -- (-90:1);
    \draw[massive, Anew] (-90:1) -- (30:1);
\end{tikzpicture}
\ +\frac{1}{12}
\begin{tikzpicture}[baseline={([yshift=-.5ex]current bounding box.center)},thick]
    \draw (-90:1.5) -- (-90:1);
    \draw (30:1.5) -- (30:1);
    \draw (150:1.5) -- (150:1);
    \draw[massive, cut=2] (30:1) -- (150:1);
    \draw[massive, Anew] (150:1) -- (-90:1);
    \draw[massive, Rnew] (-90:1) -- (30:1);
\end{tikzpicture}
\ +\frac{1}{6}
\begin{tikzpicture}[baseline={([yshift=-.5ex]current bounding box.center)},thick]
    \draw (-90:1.5) -- (-90:1);
    \draw (30:1.5) -- (30:1);
    \draw (150:1.5) -- (150:1);
    \draw[massive, cut=2] (30:1) -- (150:1);
    \draw[massive, Anew] (150:1) -- (-90:1);
    \draw[massive, Anew] (-90:1) -- (30:1);
\end{tikzpicture}
\, \right]\, .
\end{align}
Here we also omit the labels of external lines for simplicity, but they are to be summed over all possible cyclic permutations $\sigma \in C_3$ of $(p_1,p_2,p_3)$.

Now we can come back to the second type of term in \eqref{eq:tribub}, corresponding to graphs with a bubble on an external leg. The computation is analogous to the case of the triangle, so we just quote the final result for the matrix element. We have $N_{3}^{(3)}\big|_{\rm bub} = N_{3}^{(3)}\big|_{\rm bub,12} + N_{3}^{(3)}\big|_{\rm bub,23} + N_{3}^{(3)}\big|_{\rm bub,31}$, where 
\begin{align}
\begin{split}
&iN_{3}^{(3)}\big|_{\rm bub,12} = \frac{(-i)^3}{6}   \int\!d^4x_1 d^4x_2 d^4 x_3 \ e^{-i(p_1+p_2) \cdot x_1 -i p_3\cdot x_3} \\
&\cdot \Delta^{(1)}_{23}\bigg[       (i\Delta^{R}_{12}) (i\Delta^{R}_{23})+  (i\Delta^{A}_{12}) (i\Delta^{A}_{23})  + \frac{1}{2}\Big( 
  (i\Delta^{R}_{12}) (i\Delta^{A}_{23}) + (i\Delta^{A}_{12}) (i\Delta^{R}_{23})\Big) 
  \bigg], 
\end{split}
\end{align}
and   the terms $N_{3}^{(3)}\big|_{\rm bub,23}$ and $N_{3}^{(3)}\big|_{\rm bub,31}$ are obtained from $N_{3}^{(3)}\big|_{\rm bub,12}$ by permuting $(p_1, p_2, p_3)$.  Again we note that the result is expressed a cut integral, and furthermore  the same  peculiar factors in front of the 
$(\Delta^R {\times} \Delta^A + \Delta^A {\times} \Delta^R)$ and 
$(\Delta^R{\times} \Delta^R +\Delta^A {\times} \Delta^A)$ combinations appear as in the triangle case (and in the five-point tree-level amplitude). The momentum space expression is
\begin{align}\label{eq:bub12}
    iN_{3}^{(3)}\big|_{\rm bub,12} &= (-i)^3\hat{\delta}^{(4)}(p_1{+}p_2{+}p_3)\nn
    \left[\frac{1}{6}\
\begin{tikzpicture}[baseline={([yshift=-.5ex]current bounding box.center)}, thick,font=\small]
    \draw[massive] (120:0.5) node[above]{$p_2$} -- (0,0) -- (-120:0.5) node[below]{$p_1$};
    \draw[Rnew] (0,0) -- (1,0);
    \draw[Rnew] (1,0) arc(180:0:0.5);
    \draw[cut=2] (1,0) arc(-180:0:0.5);
    \draw[massive] (2,0) -- ++(0.5,0) node[above]{$p_3$};
\end{tikzpicture}
 +\frac{1}{12}\
\begin{tikzpicture}[baseline={([yshift=-.5ex]current bounding box.center)}, thick,font=\small]
    \draw[massive] (120:0.5) node[above]{$p_2$} -- (0,0) -- (-120:0.5) node[below]{$p_1$};
    \draw[Rnew] (0,0) -- (1,0);
    \draw[Anew] (1,0) arc(180:0:0.5);
    \draw[cut=2] (1,0) arc(-180:0:0.5);
    \draw[massive] (2,0) -- ++(0.5,0) node[above]{$p_3$};
\end{tikzpicture} \right.\\ &
\left. +\frac{1}{12}\
\begin{tikzpicture}[baseline={([yshift=-.5ex]current bounding box.center)}, thick,font=\small]
    \draw[massive] (120:0.5) node[above]{$p_2$} -- (0,0) -- (-120:0.5) node[below]{$p_1$};
    \draw[Anew] (0,0) -- (1,0);
    \draw[Rnew] (1,0) arc(180:0:0.5);
    \draw[cut=2] (1,0) arc(-180:0:0.5);
    \draw[massive] (2,0) -- ++(0.5,0) node[above]{$p_3$};
\end{tikzpicture}
 +\frac{1}{6} \
\begin{tikzpicture}[baseline={([yshift=-.5ex]current bounding box.center)}, thick,font=\small]
    \draw[massive] (120:0.5) node[above]{$p_2$} -- (0,0) -- (-120:0.5) node[below]{$p_1$};
    \draw[Anew] (0,0) -- (1,0);
    \draw[Anew] (1,0) arc(180:0:0.5);
    \draw[cut=2] (1,0) arc(-180:0:0.5);
    \draw[massive] (2,0) -- ++(0.5,0) node[above]{$p_3$};
\end{tikzpicture}
\ \right]\, .
\end{align}
It should be noted that these graphs are pathological, as due to the bubble on the external line, the internal propagator is on-shell and thus gives $1/0$. We comment further on the effect of these types of graphs on the Magnus expansion in Section~\ref{sec: BadGraphs}.

\section{Tree-level and one-loop results from Murua}\label{sec:redoMurua}

At this point, we have computed many examples of Magnus amplitudes at both tree- and one-loop level.
We have also seen that
computing these amplitudes is quite similar to computing $S$-matrix elements.
The main additional subtlety is the presence of different types of propagators --
retarded, advanced, Pauli-Jordan and Hadamard --
appearing in the results.
A key question is therefore how one should ``weight'' contributions to the same integrand
appearing with different types of internal propagators.
At tree level, the answer to this problem is provided by Murua's formula \cite{Murua_2006}, and its extension \cite{Kim:2024svw}. In this section we shall see how these tree-level coefficients carry over into loop level.

\subsection{Tree-level Murua coefficients}
\label{sec:redofromMurua-trees}

As we have seen in Section~\ref{sec:treelevel}, a key aspect of the tree-level $N$ operator is
that it involves only retarded and advanced propagators,
not Hadamard or Pauli-Jordan functions.
Summarising the work of that section,
up to $N^{(4)}$ (six external legs) we may write the $N$-operator in the schematic form
\begin{subequations}\label{schematicN}
\begin{align}
    iN^{(1)} &= -i\int d^4x\,\omega\left(\begin{tikzpicture}[font=\small, thick]
        \draw[fill=white] (0,0) circle(3pt);
    \end{tikzpicture}\right)\frac{:\phi^3:}{3!}\,,\\
    iN^{(2)}&=(-i)^2\int d^4x_1d^4x_2\left(\omega\left(\begin{tikzpicture}[font=\small, thick]
        \draw[causArrow] (0,0) -- (1,0);
        \draw[fill=white] (0,0) circle(3pt);
        \draw[fill=white] (1,0) circle(3pt);
    \end{tikzpicture}\right)
    \begin{tikzpicture}[font=\footnotesize, thick]
        \draw[causArrow] (0,0) -- (1,0);
        \draw[fill=white] (0,0) circle(3pt);
        \draw[fill=white] (1,0) circle(3pt);
        \node at (0,0.4){$1$};
        \node at (1,0.4){$2$};
    \end{tikzpicture}\right)\frac{:\phi_1^2\phi_2^2:}{(2!)^2}+\cdots\, ,   \\
   i N^{(3)}&=(-i)^3\int d^4x_1d^4x_2d^4x_3\bigg(
   \omega\left(\begin{tikzpicture}[font=\small, thick]
        \draw[massive, causArrow] (0,0) -- (1, 0);
        \draw[massive, causArrow] (1, 0) -- (2, 0);
        \draw[fill=white] (0,0) circle(3pt);
        \draw[fill=white] (1,0) circle(3pt);
        \draw[fill=white] (2,0) circle(3pt);
    \end{tikzpicture}\right)
    \begin{tikzpicture}[font=\footnotesize, thick]
        \draw[massive, causArrow] (0,0) -- (1, 0);
        \draw[massive, causArrow] (1, 0) -- (2, 0);
        \draw[fill=white] (0,0) circle(3pt) node[yshift=0.4cm]{$1$};
        \draw[fill=white] (1,0) circle(3pt) node[yshift=0.4cm]{$2$};
        \draw[fill=white] (2,0) circle(3pt) node[yshift=0.4cm]{$3$};
    \end{tikzpicture}\\
    &\!\!\!\!\!\!+
    \frac1{2!}\omega\left(\begin{tikzpicture}[baseline={([yshift=-.5ex]current bounding box.center)}, thick]
        \draw[causArrow] (0,0.4) -- (0.5,0);
        \draw[causArrow] (0,-.4) -- (0.5,0);
        \draw[fill=white] (0.5,0) circle(3pt);
        \draw[fill=white] (0,0.4) circle(3pt);
        \draw[fill=white] (0,-.4) circle(3pt);
    \end{tikzpicture}\right)
    \begin{tikzpicture}[baseline={([yshift=-.5ex]current bounding box.center)}, thick,font=\footnotesize]
        \draw[causArrow] (0,0.4) -- (0.5,0);
        \draw[causArrow] (0,-.4) -- (0.5,0);
        \draw[fill=white] (0.5,0) circle(3pt) node[xshift=0.3cm]{$2$};
        \draw[fill=white] (0,0.4) circle(3pt) node[xshift=-0.3cm]{$1$};
        \draw[fill=white] (0,-.4) circle(3pt) node[xshift=-0.3cm]{$3$};
    \end{tikzpicture}+
    \frac1{2!}\omega\left(\begin{tikzpicture}[baseline={([yshift=-.5ex]current bounding box.center)}, thick]
        \draw[causArrow] (0,0) -- (0.5,0.4);
        \draw[causArrow] (0,0) -- (0.5,-0.4);
        \draw[fill=white] (0,0) circle(3pt);
        \draw[fill=white] (0.5,0.4) circle(3pt);
        \draw[fill=white] (0.5,-0.4) circle(3pt);
    \end{tikzpicture}\right)
    \begin{tikzpicture}[baseline={([yshift=-.5ex]current bounding box.center)}, thick,font=\footnotesize]
        \draw[causArrow] (0,0) -- (0.5,0.4);
        \draw[causArrow] (0,0) -- (0.5,-0.4);
        \draw[fill=white] (0,0) circle(3pt) node[xshift=-0.3cm]{$2$};
        \draw[fill=white] (0.5,0.4) circle(3pt) node[xshift=0.3cm]{$1$};
        \draw[fill=white] (0.5,-0.4) circle(3pt) node[xshift=0.3cm]{$3$};
    \end{tikzpicture}
   \bigg)\frac{:\phi_1^2\phi_2\phi_3^2:}{(2!)^2}+\cdots \,,\nonumber\\
   i N^{(4)}&=(-i)^4\int d^4x_1d^4x_2d^4x_3d^4x_4\bigg[\bigg(
   \omega\left(\begin{tikzpicture}[thick]
        \draw[massive, causArrow] (0,0) -- (1, 0);
        \draw[massive, causArrow] (1, 0) -- (2, 0);
        \draw[massive, causArrow] (2, 0) -- (3, 0);
        \draw[fill=white] (0,0) circle(3pt);
        \draw[fill=white] (1,0) circle(3pt);
        \draw[fill=white] (2,0) circle(3pt);
        \draw[fill=white] (3,0) circle(3pt);
    \end{tikzpicture}\right)
    \begin{tikzpicture}[font=\footnotesize, thick]
        \draw[massive, causArrow] (0,0) -- (1, 0);
        \draw[massive, causArrow] (1, 0) -- (2, 0);
        \draw[massive, causArrow] (2, 0) -- (3, 0);
        \draw[fill=white] (0,0) circle(3pt) node[yshift=0.4cm]{$1$};
        \draw[fill=white] (1,0) circle(3pt) node[yshift=0.4cm]{$2$};
        \draw[fill=white] (2,0) circle(3pt) node[yshift=0.4cm]{$3$};
        \draw[fill=white] (3,0) circle(3pt) node[yshift=0.4cm]{$4$};
    \end{tikzpicture}\nonumber\\
    &\!\!\!\!\!\!\!\!\!\!\!\!+
    \omega\left(\begin{tikzpicture}[baseline={([yshift=-.5ex]current bounding box.center)},thick]
        \draw[causArrow] (0,0) -- (0.8,0);
        \draw[causArrow] (0,-0.5) -- (0.4,-0.5);
        \draw[causArrow] (0.4,-0.5) -- (0.8,0);
        \draw[fill=white] (0,0) circle(3pt);
        \draw[fill=white] (0.8,0) circle(3pt);
        \draw[fill=white] (0.4,-0.5) circle(3pt);
        \draw[fill=white] (0,-0.5) circle(3pt);
    \end{tikzpicture}\right)
    \begin{tikzpicture}[baseline={([yshift=-.5ex]current bounding box.center)},thick,font=\footnotesize]
        \draw[causArrow] (0,0) -- (0.8,0);
        \draw[causArrow] (0,-0.5) -- (0.4,-0.5);
        \draw[causArrow] (0.4,-0.5) -- (0.8,0);
        \draw[fill=white] (0,0) circle(3pt) node[xshift=-0.4cm]{$1$};
        \draw[fill=white] (0.8,0) circle(3pt) node[xshift=0.4cm]{$2$};
        \draw[fill=white] (0.4,-0.5) circle(3pt) node[xshift=0.4cm]{$3$};
        \draw[fill=white] (0,-0.5) circle(3pt) node[xshift=-0.4cm]{$4$};
    \end{tikzpicture}+
    \omega\left(\begin{tikzpicture}[baseline={([yshift=-.5ex]current bounding box.center)},thick]
        \draw[causArrow] (0,0) -- (0.8,0);
        \draw[causArrow] (0,0) -- (0.4,-0.5);
        \draw[causArrow] (0.4,-0.5) -- (0.8,-0.5);
        \draw[fill=white] (0,0) circle(3pt);
        \draw[fill=white] (0.8,0) circle(3pt);
        \draw[fill=white] (0.4,-0.5) circle(3pt);
        \draw[fill=white] (0.8,-0.5) circle(3pt);
    \end{tikzpicture}\right)
    \begin{tikzpicture}[baseline={([yshift=-.5ex]current bounding box.center)},thick,font=\footnotesize]
        \draw[causArrow] (0,0) -- (0.8,0);
        \draw[causArrow] (0,0) -- (0.4,-0.5);
        \draw[causArrow] (0.4,-0.5) -- (0.8,-0.5);
        \draw[fill=white] (0,0) circle(3pt) node[xshift=-0.4cm]{$2$};
        \draw[fill=white] (0.8,0) circle(3pt) node[xshift=0.4cm]{$1$};
        \draw[fill=white] (0.4,-0.5) circle(3pt) node[xshift=-0.4cm]{$3$};
        \draw[fill=white] (0.8,-0.5) circle(3pt) node[xshift=0.4cm]{$4$};
    \end{tikzpicture}
    \nn\\ 
    &\!\!\!\!\!\!\!\!\!\!\!\!+\omega\left(\begin{tikzpicture}[baseline={([yshift=-.5ex]current bounding box.center)},thick]
        \draw[causArrow] (0,0) -- (0.8,0);
        \draw[causArrow] (0,0) -- (0.8,-0.75);
        \draw[causArrow] (0,-0.75) -- (0.8,-0.75);
        \draw[fill=white] (0.8,0) circle(3pt);
        \draw[fill=white] (0,0) circle(3pt);
        \draw[fill=white] (0.8,-0.75) circle(3pt);
        \draw[fill=white] (0,-0.75) circle(3pt);
    \end{tikzpicture}\right)
    \begin{tikzpicture}[baseline={([yshift=-.5ex]current bounding box.center)},thick,font=\footnotesize]
        \draw[causArrow] (0,0) -- (0.8,0);
        \draw[causArrow] (0,0) -- (0.8,-0.75);
        \draw[causArrow] (0,-0.75) -- (0.8,-0.75);
        \draw[fill=white] (0.8,0) circle(3pt)  node[xshift=0.4cm]{$4$};
        \draw[fill=white] (0,0) circle(3pt)  node[xshift=-0.4cm]{$3$};
        \draw[fill=white] (0.8,-0.75) circle(3pt) node[xshift=0.4cm]{$2$};
        \draw[fill=white] (0,-0.75) circle(3pt) node[xshift=-0.4cm]{$1$};
    \end{tikzpicture}\bigg)\frac{:\phi_1^2\phi_2\phi_3\phi_4^2:}{(2!)^2}\nonumber\\
   &\!\!\!\!\!\!\!\!\!\!\!\!+
   \bigg(
   \frac1{3!}
   \omega\left(\begin{tikzpicture}[baseline={([yshift=-.5ex]current bounding box.center)},thick]
        \draw[causArrow] (0,-0.5) -- (0.7,0);
        \draw[causArrow] (0,0) -- (0.7,0);
        \draw[causArrow] (0,0.5) -- (0.7,0);
        \draw[fill=white] (0.7,0) circle(3pt);
        \draw[fill=white] (0,-0.5) circle(3pt);
        \draw[fill=white] (0,0) circle(3pt);
        \draw[fill=white] (0,0.5) circle(3pt);
    \end{tikzpicture}\right)
    \begin{tikzpicture}[baseline={([yshift=-.5ex]current bounding box.center)},thick,font=\footnotesize]
        \draw[causArrow] (0,-0.5) -- (0.7,0);
        \draw[causArrow] (0,0) -- (0.7,0);
        \draw[causArrow] (0,0.5) -- (0.7,0);
        \draw[fill=white] (0.7,0) circle(3pt) node[xshift=0.4cm]{$4$};
        \draw[fill=white] (0,-0.5) circle(3pt) node[xshift=-0.4cm]{$3$};
        \draw[fill=white] (0,0) circle(3pt) node[xshift=-0.4cm]{$2$};
        \draw[fill=white] (0,0.5) circle(3pt) node[xshift=-0.4cm]{$1$};
    \end{tikzpicture}+
    \frac1{3!}
    \omega\left(\begin{tikzpicture}[baseline={([yshift=-.5ex]current bounding box.center)},thick]
        \draw[causArrow] (0,0) -- (0.7,-0.5);
        \draw[causArrow] (0,0) -- (0.7,0);
        \draw[causArrow] (0,0) -- (0.7,0.5);
        \draw[fill=white] (0,0) circle(3pt);
        \draw[fill=white] (0.7,-0.5) circle(3pt);
        \draw[fill=white] (0.7,0) circle(3pt);
        \draw[fill=white] (0.7,0.5) circle(3pt);
    \end{tikzpicture}\right)
    \begin{tikzpicture}[baseline={([yshift=-.5ex]current bounding box.center)},thick,font=\footnotesize]
        \draw[causArrow] (0,0) -- (0.7,-0.5);
        \draw[causArrow] (0,0) -- (0.7,0);
        \draw[causArrow] (0,0) -- (0.7,0.5);
        \draw[fill=white] (0,0) circle(3pt) node[xshift=-0.4cm]{$4$};
        \draw[fill=white] (0.7,-0.5) circle(3pt) node[xshift=0.4cm]{$3$};
        \draw[fill=white] (0.7,0) circle(3pt) node[xshift=0.4cm]{$2$};
        \draw[fill=white] (0.7,0.5) circle(3pt) node[xshift=0.4cm]{$1$};
    \end{tikzpicture}\nonumber\\
    &\!\!\!\!\!\!+
    \frac1{2!}
   \omega\left(\begin{tikzpicture}[baseline={([yshift=-.5ex]current bounding box.center)},thick]
        \draw[causArrow] (-0.5,0.5) -- (0,0);
        \draw[causArrow] (-0.5,-0.5) -- (0,0);
        \draw[causArrow] (0,0) -- (0.5,0);
        \draw[fill=white] (0,0) circle(3pt);
        \draw[fill=white] (-0.5,0.5) circle(3pt);
        \draw[fill=white] (-0.5,-0.5) circle(3pt);
        \draw[fill=white] (0.5,0) circle(3pt);
    \end{tikzpicture}\right)
    \begin{tikzpicture}[baseline={([yshift=-.5ex]current bounding box.center)},thick,font=\footnotesize]
        \draw[causArrow] (-0.5,0.5) -- (0,0);
        \draw[causArrow] (-0.5,-0.5) -- (0,0);
        \draw[causArrow] (0,0) -- (0.5,0);
        \draw[fill=white] (0,0) circle(3pt) node[xshift=-0.4cm]{$4$};
        \draw[fill=white] (-0.5,0.5) circle(3pt) node[xshift=-0.4cm]{$1$};
        \draw[fill=white] (-0.5,-0.5) circle(3pt) node[xshift=-0.4cm]{$2$};
        \draw[fill=white] (0.5,0) circle(3pt) node[xshift=0.4cm]{$3$};
    \end{tikzpicture}+
    \frac1{2!}
    \omega\left(\begin{tikzpicture}[baseline={([yshift=-.5ex]current bounding box.center)},thick]
        \draw[causArrow] (0,0) -- (0.5,0.5);
        \draw[causArrow] (0,0) -- (0.5,-0.5);
        \draw[causArrow] (-0.5,0) -- (0,0);
        \draw[fill=white] (0,0) circle(3pt);
        \draw[fill=white] (0.5,0.5) circle(3pt);
        \draw[fill=white] (0.5,-0.5) circle(3pt);
        \draw[fill=white] (-0.5,0) circle(3pt);
    \end{tikzpicture}\right)
    \begin{tikzpicture}[baseline={([yshift=-.5ex]current bounding box.center)},thick,font=\footnotesize]
        \draw[causArrow] (0,0) -- (0.5,0.5);
        \draw[causArrow] (0,0) -- (0.5,-0.5);
        \draw[causArrow] (-0.5,0) -- (0,0);
        \draw[fill=white] (0,0) circle(3pt) node[xshift=0.4cm]{$4$};
        \draw[fill=white] (0.5,0.5) circle(3pt) node[xshift=0.4cm]{$1$};
        \draw[fill=white] (0.5,-0.5) circle(3pt) node[xshift=0.4cm]{$2$};
        \draw[fill=white] (-0.5,0) circle(3pt) node[xshift=-0.4cm]{$3$};
    \end{tikzpicture}\bigg)
    \frac{:\phi_1^2\phi_2^2\phi_3^2:}{(2!)^3}\bigg]+\cdots\,,
\end{align}
\end{subequations}
where for convenience we always pair up powers of the operator $\phi_i^n$ with a factor of  $1/n!$.
As before, we omitted the terms that do not contribute to tree-level matrix elements.
For each tree $\tau$, we have also separated out  explicit symmetry factors $\sigma(\tau)^{-1}$, for example:
\begin{equation}
    \sigma\left(\begin{tikzpicture}[font=\small, thick]
        \draw[causArrow] (0,0) -- (1,0);
        \draw[fill=white] (0,0) circle(3pt);
        \draw[fill=white] (1,0) circle(3pt);
    \end{tikzpicture}\right)=
    \sigma\left(\begin{tikzpicture}[font=\small, thick]
        \draw[massive, causArrow] (0,0) -- (1, 0);
        \draw[massive, causArrow] (1, 0) -- (2, 0);
        \draw[fill=white] (0,0) circle(3pt);
        \draw[fill=white] (1,0) circle(3pt);
        \draw[fill=white] (2,0) circle(3pt);
    \end{tikzpicture}\right)=1\,, \quad
    \sigma\left(\begin{tikzpicture}[baseline={([yshift=-.5ex]current bounding box.center)}, thick]
        \draw[causArrow] (0,0) -- (0.5,0.4);
        \draw[causArrow] (0,0) -- (0.5,-0.4);
        \draw[fill=white] (0,0) circle(3pt);
        \draw[fill=white] (0.5,0.4) circle(3pt);
        \draw[fill=white] (0.5,-0.4) circle(3pt);
    \end{tikzpicture}\right)=
    \sigma\left(\begin{tikzpicture}[baseline={([yshift=-.5ex]current bounding box.center)}, thick]
        \draw[causArrow] (0,0.4) -- (0.5,0);
        \draw[causArrow] (0,-.4) -- (0.5,0);
        \draw[fill=white] (0.5,0) circle(3pt);
        \draw[fill=white] (0,0.4) circle(3pt);
        \draw[fill=white] (0,-.4) circle(3pt);
    \end{tikzpicture}\right)=2!\,.
\end{equation}
At tree level, $\sigma(\tau)$ is defined as the number of permutations of the vertices (with their attached edges) that leave $\tau$  invariant, taking also into account the orientation of edges. 
The remaining symmetry factors can be read off from \eqref{schematicN}.

The results in  \eqref{schematicN} correspond respectively to
\eqref{first-order-Mag-exp}, \eqref{N2}, \eqref{5pt-tree-int} and \eqref{N4-Position}, and there 
we have determined by explicit calculation that the $\omega$ coefficients have the following values:%
\footnote{Note that although all the Murua coefficients below are positive, this does not continue for higher numbers of vertices.}
\begin{subequations}\label{easyMurua}
\begin{align}
    \omega\left(\begin{tikzpicture}[font=\small, thick]
        \draw[fill=white] (0,0) circle(3pt);
    \end{tikzpicture}\right)&=1\,,\\
    \omega\left(\begin{tikzpicture}[font=\small, thick]
        \draw[causArrow] (0,0) -- (1,0);
        \draw[fill=white] (0,0) circle(3pt);
        \draw[fill=white] (1,0) circle(3pt);
    \end{tikzpicture}\right)&=\frac12\,,\label{easyMurua2V}\\
    \omega\left(\begin{tikzpicture}[font=\small, thick]
        \draw[massive, causArrow] (0,0) -- (1, 0);
        \draw[massive, causArrow] (1, 0) -- (2, 0);
        \draw[fill=white] (0,0) circle(3pt);
        \draw[fill=white] (1,0) circle(3pt);
        \draw[fill=white] (2,0) circle(3pt);
    \end{tikzpicture}\right)&=\frac13\,, \quad
    \omega\left(\begin{tikzpicture}[baseline={([yshift=-.5ex]current bounding box.center)}, thick]
        \draw[causArrow] (0,0) -- (0.5,0.4);
        \draw[causArrow] (0,0) -- (0.5,-0.4);
        \draw[fill=white] (0,0) circle(3pt);
        \draw[fill=white] (0.5,0.4) circle(3pt);
        \draw[fill=white] (0.5,-0.4) circle(3pt);
    \end{tikzpicture}\right)=
    \omega\left(\begin{tikzpicture}[baseline={([yshift=-.5ex]current bounding box.center)}, thick]
        \draw[causArrow] (0,0.4) -- (0.5,0);
        \draw[causArrow] (0,-.4) -- (0.5,0);
        \draw[fill=white] (0.5,0) circle(3pt);
        \draw[fill=white] (0,0.4) circle(3pt);
        \draw[fill=white] (0,-.4) circle(3pt);
    \end{tikzpicture}\right)=\frac16\,, \label{easyMurua3V}\\
    \omega\left(\begin{tikzpicture}[font=\small, thick]
        \draw[massive, causArrow] (0,0) -- (1, 0);
        \draw[massive, causArrow] (1, 0) -- (2, 0);
        \draw[massive, causArrow] (2, 0) -- (3, 0);
        \draw[fill=white] (0,0) circle(3pt);
        \draw[fill=white] (1,0) circle(3pt);
        \draw[fill=white] (2,0) circle(3pt);
        \draw[fill=white] (3,0) circle(3pt);
    \end{tikzpicture}\right)&=\frac14\,, \quad
    \omega\left(\begin{tikzpicture}[baseline={([yshift=-.5ex]current bounding box.center)},thick]
    \draw[causArrow] (0,0) -- (0.8,0);
    \draw[causArrow] (0,0) -- (0.8,-0.75);
    \draw[causArrow] (0,-0.75) -- (0.8,-0.75);
    \draw[fill=white] (0.8,0) circle(3pt);
    \draw[fill=white] (0,0) circle(3pt);
    \draw[fill=white] (0.8,-0.75) circle(3pt);
    \draw[fill=white] (0,-0.75) circle(3pt);
    \end{tikzpicture}\right)=
    \omega\left(\begin{tikzpicture}[baseline={([yshift=-.5ex]current bounding box.center)},thick]
        \draw[causArrow] (0,0) -- (0.8,0);
        \draw[causArrow] (0,0) -- (0.4,-0.5);
        \draw[causArrow] (0.4,-0.5) -- (0.8,-0.5);
        \draw[fill=white] (0,0) circle(3pt);
        \draw[fill=white] (0.8,0) circle(3pt);
        \draw[fill=white] (0.4,-0.5) circle(3pt);
        \draw[fill=white] (0.8,-0.5) circle(3pt);
    \end{tikzpicture}\right)=
    \omega\left(\begin{tikzpicture}[baseline={([yshift=-.5ex]current bounding box.center)},thick]
        \draw[causArrow] (0,0) -- (0.8,0);
        \draw[causArrow] (0,-0.5) -- (0.4,-0.5);
        \draw[causArrow] (0.4,-0.5) -- (0.8,0);
        \draw[fill=white] (0,0) circle(3pt);
        \draw[fill=white] (0.8,0) circle(3pt);
        \draw[fill=white] (0.4,-0.5) circle(3pt);
        \draw[fill=white] (0,-0.5) circle(3pt);
    \end{tikzpicture}\right)=\frac1{12}\, \label{easyMurua4V}\\
    \omega\left(\begin{tikzpicture}[baseline={([yshift=-.5ex]current bounding box.center)},thick]
        \draw[causArrow] (0,0) -- (0.5,0.5);
        \draw[causArrow] (0,0) -- (0.5,-0.5);
        \draw[causArrow] (-0.5,0) -- (0,0);
        \draw[fill=white] (0,0) circle(3pt);
        \draw[fill=white] (0.5,0.5) circle(3pt);
        \draw[fill=white] (0.5,-0.5) circle(3pt);
        \draw[fill=white] (-0.5,0) circle(3pt);
    \end{tikzpicture}\right)&=
    \omega\left(\begin{tikzpicture}[baseline={([yshift=-.5ex]current bounding box.center)},thick]
        \draw[causArrow] (-0.5,0.5) -- (0,0);
        \draw[causArrow] (-0.5,-0.5) -- (0,0);
        \draw[causArrow] (0,0) -- (0.5,0);
        \draw[fill=white] (0,0) circle(3pt);
        \draw[fill=white] (-0.5,0.5) circle(3pt);
        \draw[fill=white] (-0.5,-0.5) circle(3pt);
        \draw[fill=white] (0.5,0) circle(3pt);
    \end{tikzpicture}\right)=\frac16\,, \quad
    \omega\left(\begin{tikzpicture}[baseline={([yshift=-.5ex]current bounding box.center)},thick]
        \draw[causArrow] (0,0) -- (0.7,-0.5);
        \draw[causArrow] (0,0) -- (0.7,0);
        \draw[causArrow] (0,0) -- (0.7,0.5);
        \draw[fill=white] (0,0) circle(3pt);
        \draw[fill=white] (0.7,-0.5) circle(3pt);
        \draw[fill=white] (0.7,0) circle(3pt);
        \draw[fill=white] (0.7,0.5) circle(3pt);
    \end{tikzpicture}\right)=
    \omega\left(\begin{tikzpicture}[baseline={([yshift=-.5ex]current bounding box.center)},thick]
        \draw[causArrow] (0,-0.5) -- (0.7,0);
        \draw[causArrow] (0,0) -- (0.7,0);
        \draw[causArrow] (0,0.5) -- (0.7,0);
        \draw[fill=white] (0.7,0) circle(3pt);
        \draw[fill=white] (0,-0.5) circle(3pt);
        \draw[fill=white] (0,0) circle(3pt);
        \draw[fill=white] (0,0.5) circle(3pt);
    \end{tikzpicture}\right)=0\,.\nonumber
\end{align}
\end{subequations}
Crucially, and as anticipated in Section~\ref{sec:treelevel}, these numerical weighting factors $\omega$ are examples of  Murua coefficients.%
\footnote{The Murua coefficients stated here differ from those in \cite{Murua_2006, CALAQUE2011282,Kim:2024svw} by a factor of $(-1)^{n-1}$ where $n$ is the number of vertices.}
We have not proven in full generality that the weighting factors arising from a quantum field theory calculation are Murua coefficients, but we have checked it up to $n=7$ internal vertices, using the new formula for the Magnus expansion given in \eqref{eq:ReformulationCS}. We will discuss this general structure of Murua coefficients at tree level and beyond in Section~\ref{sec: generalMuruaFormulae}.

Murua provided in \cite{Murua_2006} a direct recursive relationship for the $\omega$ coefficients~\eqref{easyMurua}, which we refer to as Murua's formula, 
from which one may extract their numerical values (we have provided their values here up to four vertices). Murua's original formula applies only to rooted trees, and this formula was extended to non-rooted trees in \cite{Kim:2024svw}, to which we also refer the  
interested reader  for a  review of the explicit construction of Murua's formula.%
\footnote{The authors of \cite{Kim:2024svw} also provided a \texttt{Mathematica} code, which may be used to produce explicit values of Murua coefficients beyond those given here. }

The practical usefulness of Murua's formula is that, with these coefficients known,
one does not need to explicitly compute $N$-matrix elements as we did in Section~\ref{sec:treelevel};
instead, one may simply assemble all possible Feynman diagrams as in \eqref{schematicN},
and then ``weight'' with Murua coefficients to build up a final result~\cite{Kim:2024svw,Murua_2006}.
Building up the $N$-operator in this way is simpler than manually collapsing the $\theta_{ij}$ coefficients
as we did in Section~\ref{sec:treelevel}.
However, as Murua's formula only applies directly at tree level,
this will not be the case for generic loop-level examples and so we fall back on the Magnus series.

We also emphasise that Murua's formula applies to $N$-matrix elements,
in addition to the $N$-matrix itself.
Summarising our results for the former, we find
 \begin{subequations}\label{eq: NMatrixTree}
\begin{align}
iN^{(1)}_3&= (-i)^{1}
\omega\left(\,
\begin{tikzpicture}[baseline={([yshift=-0.5ex]current bounding box.center)},thick]
\draw (0,0) -- (0.5,0);
\draw (0,0) -- (120:0.5);
\draw (0,0) -- (-120:0.5);
\end{tikzpicture}
\,\right)
\begin{tikzpicture}[baseline={([yshift=-0.5ex]current bounding box.center)},thick]
\draw (0,0) -- (0.5,0);
\draw (0,0) -- (120:0.5);
\draw (0,0) -- (-120:0.5);
\end{tikzpicture}\, , \\
iN^{(2)}_4&= 
\sum_{\text{perms}}
(-i)^2\left[
\omega\left(\,
\begin{tikzpicture}[font=\small,baseline={([yshift=-.5ex]current bounding box.center)},thick,scale=0.5] 
    \draw[massive] (0,0) -- (135:1);
    \draw[massive] (0,0) -- (-135:1);
    \draw[causArrow] (1,0) -- (0,0);
    \draw[massive] (1,0) -- ++(45:1);
    \draw[massive] (1,0) -- ++(-45:1);
\end{tikzpicture}
\,\right)
\begin{tikzpicture}[font=\small,baseline={([yshift=-.5ex]current bounding box.center)},thick,scale=0.5] 
    \draw[massive] (0,0) -- (135:1);
    \draw[massive] (0,0) -- (-135:1);
    \draw[causArrow] (1,0) -- (0,0);
    \draw[massive] (1,0) -- ++(45:1);
    \draw[massive] (1,0) -- ++(-45:1);
\end{tikzpicture}
+\omega\left(\,
\begin{tikzpicture}[font=\small,baseline={([yshift=-.5ex]current bounding box.center)},thick,scale=0.5] 
    \draw[massive] (0,0) -- (135:1);
    \draw[massive] (0,0) -- (-135:1);
    \draw[causArrow] (0,0) -- (1,0);
    \draw[massive] (1,0) -- ++(45:1);
    \draw[massive] (1,0) -- ++(-45:1);
\end{tikzpicture}
\,\right)
\begin{tikzpicture}[font=\small,baseline={([yshift=-.5ex]current bounding box.center)},thick,scale=0.5] 
    \draw[massive] (0,0) -- (135:1);
    \draw[massive] (0,0) -- (-135:1);
    \draw[causArrow] (0,0) -- (1,0);
    \draw[massive] (1,0) -- ++(45:1);
    \draw[massive] (1,0) -- ++(-45:1);
\end{tikzpicture}
\, \right]\, , \\
iN^{(3)}_{5} &= 
\sum_{\text{perms}}
(-i)^3\left[
\omega\left(\,
\begin{tikzpicture}[baseline={([yshift=-0.5ex]current bounding box.center)},thick,scale=0.5] 
    \draw (1, 0) -- ++(135:1);
    \draw (1, 0) -- ++(-135:1);
    \draw (2, 1) -- (2, 0);
    \draw (3, 0) -- ++(45:1);
    \draw (3, 0) -- ++(-45:1);
    \draw[massive, causArrow] (1,0) -- (2,0);
	\draw[massive, causArrow] (2,0) -- (3,0);
    \path (2,-1) -- (2,0); 
\end{tikzpicture}
\,\right)
\begin{tikzpicture}[baseline={([yshift=-0.5ex]current bounding box.center)},thick,scale=0.5] 
    \draw (1, 0) -- ++(135:1);
    \draw (1, 0) -- ++(-135:1);
    \draw (2, 1) -- (2, 0);
    \draw (3, 0) -- ++(45:1);
    \draw (3, 0) -- ++(-45:1);
    \draw[massive, causArrow] (1,0) -- (2,0);
	\draw[massive, causArrow] (2,0) -- (3,0);
    \path (2,-1) -- (2,0); 
\end{tikzpicture}
+
\omega\left(\,
\begin{tikzpicture}[baseline={([yshift=-0.5ex]current bounding box.center)},thick,scale=0.5] 
    \draw (1, 0) -- ++(135:1);
    \draw (1, 0) -- ++(-135:1);
    \draw (2, 1) -- (2, 0);
    \draw (3, 0) -- ++(45:1);
    \draw (3, 0) -- ++(-45:1);
    \draw[massive, causArrow] (1,0) -- (2,0);
	\draw[massive, causArrowR] (2,0) -- (3,0);
    \path (2,-1) -- (2,0); 
\end{tikzpicture}
\,\right)
\begin{tikzpicture}[baseline={([yshift=-0.5ex]current bounding box.center)},thick,scale=0.5] 
    \draw (1, 0) -- ++(135:1);
    \draw (1, 0) -- ++(-135:1);
    \draw (2, 1) -- (2, 0);
    \draw (3, 0) -- ++(45:1);
    \draw (3, 0) -- ++(-45:1);
    \draw[massive, causArrow] (1,0) -- (2,0);
	\draw[massive, causArrowR] (2,0) -- (3,0);
    \path (2,-1) -- (2,0); 
\end{tikzpicture}
\right. \nn\\
&\left.+\omega\left(\,
\begin{tikzpicture}[baseline={([yshift=-0.5ex]current bounding box.center)},thick,scale=0.5] 
    \draw (1, 0) -- ++(135:1);
    \draw (1, 0) -- ++(-135:1);
    \draw (2, 1) -- (2, 0);
    \draw (3, 0) -- ++(45:1);
    \draw (3, 0) -- ++(-45:1);
    \draw[massive, causArrowR] (1,0) -- (2,0);
	\draw[massive, causArrow] (2,0) -- (3,0);
    \path (2,-1) -- (2,0); 
\end{tikzpicture}
\,\right)
\begin{tikzpicture}[baseline={([yshift=-0.5ex]current bounding box.center)},thick,scale=0.5] 
    \draw (1, 0) -- ++(135:1);
    \draw (1, 0) -- ++(-135:1);
    \draw (2, 1) -- (2, 0);
    \draw (3, 0) -- ++(45:1);
    \draw (3, 0) -- ++(-45:1);
    \draw[massive, causArrowR] (1,0) -- (2,0);
	\draw[massive, causArrow] (2,0) -- (3,0);
    \path (2,-1) -- (2,0); 
\end{tikzpicture}
+\omega\left(\,
\begin{tikzpicture}[baseline={([yshift=-0.5ex]current bounding box.center)},thick,scale=0.5] 
    \draw (1, 0) -- ++(135:1);
    \draw (1, 0) -- ++(-135:1);
    \draw (2, 1) -- (2, 0);
    \draw (3, 0) -- ++(45:1);
    \draw (3, 0) -- ++(-45:1);
    \draw[massive, causArrowR] (1,0) -- (2,0);
	\draw[massive, causArrowR] (2,0) -- (3,0);
    \path (2,-1) -- (2,0); 
\end{tikzpicture}
\,\right)
\begin{tikzpicture}[baseline={([yshift=-0.5ex]current bounding box.center)},thick,scale=0.5] 
    \draw (1, 0) -- ++(135:1);
    \draw (1, 0) -- ++(-135:1);
    \draw (2, 1) -- (2, 0);
    \draw (3, 0) -- ++(45:1);
    \draw (3, 0) -- ++(-45:1);
    \draw[massive, causArrowR] (1,0) -- (2,0);
	\draw[massive, causArrowR] (2,0) -- (3,0);
    \path (2,-1) -- (2,0); 
\end{tikzpicture}
\,\right]\, , \\
iN^{(4)}_6 &= \sum_{\text{perms}}(-i)^4\left[
\omega\left(\,
\begin{tikzpicture}[baseline={([yshift=-0.5ex]current bounding box.center)},thick,scale=0.5] 
    \draw[causArrowR] (0,0) -- (1,0);
    \draw[massive] (1,0) -- ++(60:0.5);
    \draw[massive] (1,0) -- ++(-60:0.5);
    \draw[causArrow] (0,0) -- (120:1);
    \draw[massive] (120:1) -- ++(180:0.5);
    \draw[massive] (120:1) -- ++(60:0.5);
    \draw[causArrow] (0,0) -- (-120:1);
    \draw[massive] (-120:1) -- ++(-180:0.5);
    \draw[massive] (-120:1) -- ++(-60:0.5);
\end{tikzpicture}
\,\right)
\begin{tikzpicture}[baseline={([yshift=-0.5ex]current bounding box.center)},thick,scale=0.5] 
    \draw[causArrowR] (0,0) -- (1,0);
    \draw[massive] (1,0) -- ++(60:0.5);
    \draw[massive] (1,0) -- ++(-60:0.5);
    \draw[causArrow] (0,0) -- (120:1);
    \draw[massive] (120:1) -- ++(180:0.5);
    \draw[massive] (120:1) -- ++(60:0.5);
    \draw[causArrow] (0,0) -- (-120:1);
    \draw[massive] (-120:1) -- ++(-180:0.5);
    \draw[massive] (-120:1) -- ++(-60:0.5);
\end{tikzpicture}
+\omega\left(\,
\begin{tikzpicture}[baseline={([yshift=-0.5ex]current bounding box.center)},thick,scale=0.5] 
    \draw[causArrow] (0,0) -- (1,0);
    \draw[massive] (1,0) -- ++(60:0.5);
    \draw[massive] (1,0) -- ++(-60:0.5);
    \draw[causArrowR] (0,0) -- (120:1);
    \draw[massive] (120:1) -- ++(180:0.5);
    \draw[massive] (120:1) -- ++(60:0.5);
    \draw[causArrowR] (0,0) -- (-120:1);
    \draw[massive] (-120:1) -- ++(-180:0.5);
    \draw[massive] (-120:1) -- ++(-60:0.5);
\end{tikzpicture}
\,\right)
\begin{tikzpicture}[baseline={([yshift=-0.5ex]current bounding box.center)},thick,scale=0.5] 
    \draw[causArrow] (0,0) -- (1,0);
    \draw[massive] (1,0) -- ++(60:0.5);
    \draw[massive] (1,0) -- ++(-60:0.5);
    \draw[causArrowR] (0,0) -- (120:1);
    \draw[massive] (120:1) -- ++(180:0.5);
    \draw[massive] (120:1) -- ++(60:0.5);
    \draw[causArrowR] (0,0) -- (-120:1);
    \draw[massive] (-120:1) -- ++(-180:0.5);
    \draw[massive] (-120:1) -- ++(-60:0.5);
\end{tikzpicture}
\right. \nn\\
&\hspace{40pt}+\left.
\omega\left(\,
\begin{tikzpicture}[baseline={([yshift=-0.5ex]current bounding box.center)},thick,scale=0.5] 
    \draw[causArrow] (0,0) -- (1,0);
    \draw[massive] (1,0) -- ++(60:0.5);
    \draw[massive] (1,0) -- ++(-60:0.5);
    \draw[causArrow] (0,0) -- (120:1);
    \draw[massive] (120:1) -- ++(180:0.5);
    \draw[massive] (120:1) -- ++(60:0.5);
    \draw[causArrow] (0,0) -- (-120:1);
    \draw[massive] (-120:1) -- ++(-180:0.5);
    \draw[massive] (-120:1) -- ++(-60:0.5);
\end{tikzpicture}
\,\right)
\begin{tikzpicture}[baseline={([yshift=-0.5ex]current bounding box.center)},thick,scale=0.5] 
    \draw[causArrow] (0,0) -- (1,0);
    \draw[massive] (1,0) -- ++(60:0.5);
    \draw[massive] (1,0) -- ++(-60:0.5);
    \draw[causArrow] (0,0) -- (120:1);
    \draw[massive] (120:1) -- ++(180:0.5);
    \draw[massive] (120:1) -- ++(60:0.5);
    \draw[causArrow] (0,0) -- (-120:1);
    \draw[massive] (-120:1) -- ++(-180:0.5);
    \draw[massive] (-120:1) -- ++(-60:0.5);
\end{tikzpicture}
+\omega\left(\,
\begin{tikzpicture}[baseline={([yshift=-0.5ex]current bounding box.center)},thick,scale=0.5] 
    \draw[causArrowR] (0,0) -- (1,0);
    \draw[massive] (1,0) -- ++(60:0.5);
    \draw[massive] (1,0) -- ++(-60:0.5);
    \draw[causArrowR] (0,0) -- (120:1);
    \draw[massive] (120:1) -- ++(180:0.5);
    \draw[massive] (120:1) -- ++(60:0.5);
    \draw[causArrowR] (0,0) -- (-120:1);
    \draw[massive] (-120:1) -- ++(-180:0.5);
    \draw[massive] (-120:1) -- ++(-60:0.5);
\end{tikzpicture}
\,\right)
\begin{tikzpicture}[baseline={([yshift=-0.5ex]current bounding box.center)},thick,scale=0.5] 
    \draw[causArrowR] (0,0) -- (1,0);
    \draw[massive] (1,0) -- ++(60:0.5);
    \draw[massive] (1,0) -- ++(-60:0.5);
    \draw[causArrowR] (0,0) -- (120:1);
    \draw[massive] (120:1) -- ++(180:0.5);
    \draw[massive] (120:1) -- ++(60:0.5);
    \draw[causArrowR] (0,0) -- (-120:1);
    \draw[massive] (-120:1) -- ++(-180:0.5);
    \draw[massive] (-120:1) -- ++(-60:0.5);
\end{tikzpicture}
\,\right] \nn\\
&\hspace{-12pt}+\sum_{\text{perms}}(-i)^4\left[
\omega\left(\,
\begin{tikzpicture}[baseline={([yshift=-0.5ex]current bounding box.center)},thick,scale=0.5] 
    \draw[massive] (0,0) -- (135:1);
    \draw[massive] (0,0) -- (-135:1);
    \draw[causArrow] (0,0) -- (1,0);
    \draw[massive] (1,0) -- (1,1);
    \draw[causArrow] (1,0) -- (2,0);
    \draw[massive] (2,0) -- (2,1);
    \draw[causArrow] (2,0) -- (3,0);
    \draw[massive] (3,0) -- ++(45:1);
    \draw[massive] (3,0) -- ++(-45:1);
    \path (2,0) -- (2,-1); 
\end{tikzpicture}
\,\right)
\begin{tikzpicture}[baseline={([yshift=-0.5ex]current bounding box.center)},thick,scale=0.5] 
    \draw[massive] (0,0) -- (135:1);
    \draw[massive] (0,0) -- (-135:1);
    \draw[causArrow] (0,0) -- (1,0);
    \draw[massive] (1,0) -- (1,1);
    \draw[causArrow] (1,0) -- (2,0);
    \draw[massive] (2,0) -- (2,1);
    \draw[causArrow] (2,0) -- (3,0);
    \draw[massive] (3,0) -- ++(45:1);
    \draw[massive] (3,0) -- ++(-45:1);
    \path (2,0) -- (2,-1); 
\end{tikzpicture}
+\omega\left(\,
\begin{tikzpicture}[baseline={([yshift=-0.5ex]current bounding box.center)},thick,scale=0.5] 
    \draw[massive] (0,0) -- (135:1);
    \draw[massive] (0,0) -- (-135:1);
    \draw[causArrow] (0,0) -- (1,0);
    \draw[massive] (1,0) -- (1,1);
    \draw[causArrow] (1,0) -- (2,0);
    \draw[massive] (2,0) -- (2,1);
    \draw[causArrowR] (2,0) -- (3,0);
    \draw[massive] (3,0) -- ++(45:1);
    \draw[massive] (3,0) -- ++(-45:1);
    \path (2,0) -- (2,-1); 
\end{tikzpicture}
\,\right)
\begin{tikzpicture}[baseline={([yshift=-0.5ex]current bounding box.center)},thick,scale=0.5] 
    \draw[massive] (0,0) -- (135:1);
    \draw[massive] (0,0) -- (-135:1);
    \draw[causArrow] (0,0) -- (1,0);
    \draw[massive] (1,0) -- (1,1);
    \draw[causArrow] (1,0) -- (2,0);
    \draw[massive] (2,0) -- (2,1);
    \draw[causArrowR] (2,0) -- (3,0);
    \draw[massive] (3,0) -- ++(45:1);
    \draw[massive] (3,0) -- ++(-45:1);
    \path (2,0) -- (2,-1); 
\end{tikzpicture}
\right. \nn\\
&\hspace{28pt}+\omega\left(\,
\begin{tikzpicture}[baseline={([yshift=-0.5ex]current bounding box.center)},thick,scale=0.5] 
    \draw[massive] (0,0) -- (135:1);
    \draw[massive] (0,0) -- (-135:1);
    \draw[causArrow] (0,0) -- (1,0);
    \draw[massive] (1,0) -- (1,1);
    \draw[causArrowR] (1,0) -- (2,0);
    \draw[massive] (2,0) -- (2,1);
    \draw[causArrow] (2,0) -- (3,0);
    \draw[massive] (3,0) -- ++(45:1);
    \draw[massive] (3,0) -- ++(-45:1);
    \path (2,0) -- (2,-1); 
\end{tikzpicture}
\,\right)
\begin{tikzpicture}[baseline={([yshift=-0.5ex]current bounding box.center)},thick,scale=0.5] 
    \draw[massive] (0,0) -- (135:1);
    \draw[massive] (0,0) -- (-135:1);
    \draw[causArrow] (0,0) -- (1,0);
    \draw[massive] (1,0) -- (1,1);
    \draw[causArrowR] (1,0) -- (2,0);
    \draw[massive] (2,0) -- (2,1);
    \draw[causArrow] (2,0) -- (3,0);
    \draw[massive] (3,0) -- ++(45:1);
    \draw[massive] (3,0) -- ++(-45:1);
    \path (2,0) -- (2,-1); 
\end{tikzpicture}
+\omega\left(\,
\begin{tikzpicture}[baseline={([yshift=-0.5ex]current bounding box.center)},thick,scale=0.5] 
    \draw[massive] (0,0) -- (135:1);
    \draw[massive] (0,0) -- (-135:1);
    \draw[causArrow] (0,0) -- (1,0);
    \draw[massive] (1,0) -- (1,1);
    \draw[causArrowR] (1,0) -- (2,0);
    \draw[massive] (2,0) -- (2,1);
    \draw[causArrowR] (2,0) -- (3,0);
    \draw[massive] (3,0) -- ++(45:1);
    \draw[massive] (3,0) -- ++(-45:1);
    \path (2,0) -- (2,-1); 
\end{tikzpicture}
\,\right)
\begin{tikzpicture}[baseline={([yshift=-0.5ex]current bounding box.center)},thick,scale=0.5] 
    \draw[massive] (0,0) -- (135:1);
    \draw[massive] (0,0) -- (-135:1);
    \draw[causArrow] (0,0) -- (1,0);
    \draw[massive] (1,0) -- (1,1);
    \draw[causArrowR] (1,0) -- (2,0);
    \draw[massive] (2,0) -- (2,1);
    \draw[causArrowR] (2,0) -- (3,0);
    \draw[massive] (3,0) -- ++(45:1);
    \draw[massive] (3,0) -- ++(-45:1);
    \path (2,0) -- (2,-1); 
\end{tikzpicture} \nn\\
&\hspace{28pt}+\omega\left(\,
\begin{tikzpicture}[baseline={([yshift=-0.5ex]current bounding box.center)},thick,scale=0.5] 
    \draw[massive] (0,0) -- (135:1);
    \draw[massive] (0,0) -- (-135:1);
    \draw[causArrowR] (0,0) -- (1,0);
    \draw[massive] (1,0) -- (1,1);
    \draw[causArrow] (1,0) -- (2,0);
    \draw[massive] (2,0) -- (2,1);
    \draw[causArrow] (2,0) -- (3,0);
    \draw[massive] (3,0) -- ++(45:1);
    \draw[massive] (3,0) -- ++(-45:1);
    \path (2,0) -- (2,-1); 
\end{tikzpicture}
\,\right)
\begin{tikzpicture}[baseline={([yshift=-0.5ex]current bounding box.center)},thick,scale=0.5] 
    \draw[massive] (0,0) -- (135:1);
    \draw[massive] (0,0) -- (-135:1);
    \draw[causArrowR] (0,0) -- (1,0);
    \draw[massive] (1,0) -- (1,1);
    \draw[causArrow] (1,0) -- (2,0);
    \draw[massive] (2,0) -- (2,1);
    \draw[causArrow] (2,0) -- (3,0);
    \draw[massive] (3,0) -- ++(45:1);
    \draw[massive] (3,0) -- ++(-45:1);
    \path (2,0) -- (2,-1); 
\end{tikzpicture}
+\omega\left(\,
\begin{tikzpicture}[baseline={([yshift=-0.5ex]current bounding box.center)},thick,scale=0.5] 
    \draw[massive] (0,0) -- (135:1);
    \draw[massive] (0,0) -- (-135:1);
    \draw[causArrowR] (0,0) -- (1,0);
    \draw[massive] (1,0) -- (1,1);
    \draw[causArrow] (1,0) -- (2,0);
    \draw[massive] (2,0) -- (2,1);
    \draw[causArrowR] (2,0) -- (3,0);
    \draw[massive] (3,0) -- ++(45:1);
    \draw[massive] (3,0) -- ++(-45:1);
    \path (2,0) -- (2,-1); 
\end{tikzpicture}
\,\right)
\begin{tikzpicture}[baseline={([yshift=-0.5ex]current bounding box.center)},thick,scale=0.5] 
    \draw[massive] (0,0) -- (135:1);
    \draw[massive] (0,0) -- (-135:1);
    \draw[causArrowR] (0,0) -- (1,0);
    \draw[massive] (1,0) -- (1,1);
    \draw[causArrow] (1,0) -- (2,0);
    \draw[massive] (2,0) -- (2,1);
    \draw[causArrowR] (2,0) -- (3,0);
    \draw[massive] (3,0) -- ++(45:1);
    \draw[massive] (3,0) -- ++(-45:1);
    \path (2,0) -- (2,-1); 
\end{tikzpicture}
\nn\\
&\hspace{28pt}+\omega\left.\left(\,
\begin{tikzpicture}[baseline={([yshift=-0.5ex]current bounding box.center)},thick,scale=0.5] 
    \draw[massive] (0,0) -- (135:1);
    \draw[massive] (0,0) -- (-135:1);
    \draw[causArrowR] (0,0) -- (1,0);
    \draw[massive] (1,0) -- (1,1);
    \draw[causArrowR] (1,0) -- (2,0);
    \draw[massive] (2,0) -- (2,1);
    \draw[causArrow] (2,0) -- (3,0);
    \draw[massive] (3,0) -- ++(45:1);
    \draw[massive] (3,0) -- ++(-45:1);
    \path (2,0) -- (2,-1); 
\end{tikzpicture}
\,\right)
\begin{tikzpicture}[baseline={([yshift=-0.5ex]current bounding box.center)},thick,scale=0.5] 
    \draw[massive] (0,0) -- (135:1);
    \draw[massive] (0,0) -- (-135:1);
    \draw[causArrowR] (0,0) -- (1,0);
    \draw[massive] (1,0) -- (1,1);
    \draw[causArrowR] (1,0) -- (2,0);
    \draw[massive] (2,0) -- (2,1);
    \draw[causArrow] (2,0) -- (3,0);
    \draw[massive] (3,0) -- ++(45:1);
    \draw[massive] (3,0) -- ++(-45:1);
    \path (2,0) -- (2,-1); 
\end{tikzpicture}
+\omega\left(\,
\begin{tikzpicture}[baseline={([yshift=-0.5ex]current bounding box.center)},thick,scale=0.5] 
    \draw[massive] (0,0) -- (135:1);
    \draw[massive] (0,0) -- (-135:1);
    \draw[causArrowR] (0,0) -- (1,0);
    \draw[massive] (1,0) -- (1,1);
    \draw[causArrowR] (1,0) -- (2,0);
    \draw[massive] (2,0) -- (2,1);
    \draw[causArrowR] (2,0) -- (3,0);
    \draw[massive] (3,0) -- ++(45:1);
    \draw[massive] (3,0) -- ++(-45:1);
    \path (2,0) -- (2,-1); 
\end{tikzpicture}
\,\right)
\begin{tikzpicture}[baseline={([yshift=-0.5ex]current bounding box.center)},thick,scale=0.5] 
    \draw[massive] (0,0) -- (135:1);
    \draw[massive] (0,0) -- (-135:1);
    \draw[causArrowR] (0,0) -- (1,0);
    \draw[massive] (1,0) -- (1,1);
    \draw[causArrowR] (1,0) -- (2,0);
    \draw[massive] (2,0) -- (2,1);
    \draw[causArrowR] (2,0) -- (3,0);
    \draw[massive] (3,0) -- ++(45:1);
    \draw[massive] (3,0) -- ++(-45:1);
    \path (2,0) -- (2,-1); 
\end{tikzpicture}
\,\right]\, ,
\end{align}
\end{subequations}
where the sum is over the set of inequivalent    permutations of the external momenta. By this we mean that each graph labelled with external momenta should not appear more than once. We see precisely the same Murua coefficients \eqref{easyMurua}
appearing in \eqref{N31}, \eqref{4pttreeres}, \eqref{eq:5ptMomentumSpaceDiags} and \eqref{N64},
corresponding with the three-, four-, five- and six-point matrix elements of $N$, respectively.
Murua coefficients depend only on the internal vertices and propagators of a graph,
so for example:
\begin{equation}
\label{ex-corr}
    \omega\left(\begin{tikzpicture}[baseline={([yshift=-0.5ex]current bounding box.center)},thick,font=\small]
    \draw (1, 0) -- ++(135:1) node[left]{$p_2$};
    \draw (1, 0) -- ++(-135:1) node[left]{$p_1$};
    \draw (2, 1) node[above]{$p_3$} -- (2, 0);
    \draw (3, 0) -- ++(45:1) node[right]{$p_4$};
    \draw (3, 0) -- ++(-45:1) node[right]{$p_5$};
    \draw[massive, Rnew] (1,0) -- (2,0);
	\draw[massive, Rnew] (2,0) -- (3,0);
    \path (2,-1) -- (2,0); 
\end{tikzpicture}\right)= \omega(\begin{tikzpicture}[font=\small, thick]
        \draw[massive, causArrow] (0,0) -- (1, 0);
        \draw[massive, causArrow] (1, 0) -- (2, 0);
        \draw[fill=white] (0,0) circle(3pt);
        \draw[fill=white] (1,0) circle(3pt);
        \draw[fill=white] (2,0) circle(3pt);
    \end{tikzpicture})\,.
\end{equation}
For clarity, from now now we will call $\tau$ and $g$ a tree before and after contractions with external states (corresponding to diagrams appearing in the $N$-operator and $N$-matrix elements, respectively). We can then say that 
\begin{align}
    \omega(g) = \omega(\tau)\, ,
\end{align}
where $\tau$ describes the internal propagators of the graph $g$.
Conversely, symmetry factors for trees in the two spaces are different due to the presence of external lines. 
This is familiar from the Dyson expansion where the symmetry factors for graphs appearing in the operator $iT$ are distinct from those appearing in matrix elements $\langle0|iT|\prod_i \phi(p_i)\rangle$.
For example,
\begin{equation}
      1=\sigma\left(\begin{tikzpicture}[baseline={([yshift=-0.5ex]current bounding box.center)},thick,font=\small]
    \draw (1, 0) -- ++(135:1) node[left]{$p_2$};
    \draw (1, 0) -- ++(-135:1) node[left]{$p_1$};
    \draw (2, 1) node[above]{$p_3$} -- (2, 0);
    \draw (3, 0) -- ++(45:1) node[right]{$p_4$};
    \draw (3, 0) -- ++(-45:1) node[right]{$p_5$};
    \draw[massive, Anew] (1,0) -- (2,0);
	\draw[massive, Rnew] (2,0) -- (3,0);
    \path (2,-1) -- (2,0); 
\end{tikzpicture}\right)\neq
\sigma\left(\begin{tikzpicture}[baseline={([yshift=-.5ex]current bounding box.center)}, thick]
        \draw[causArrow] (0,0) -- (0.5,0.4);
        \draw[causArrow] (0,0) -- (0.5,-0.4);
        \draw[fill=white] (0,0) circle(3pt);
        \draw[fill=white] (0.5,0.4) circle(3pt);
        \draw[fill=white] (0.5,-0.4) circle(3pt);
    \end{tikzpicture}\right)= 2\,.
\end{equation}
Note that, due to the labelled external legs, all symmetry factors $\sigma(g)$ are equal to 1 at tree level. Thus, we may assemble tree-level $N$-matrix elements exactly as one would assemble tree-level scattering amplitudes,
the only difference being that we weight contributions carrying retarded and advanced propagators with the Murua coefficients,
and appropriate symmetry factors (which do appear beyond tree level).

\subsection{Some properties of Murua coefficients}

Let us note a few simple properties of the Murua coefficients. 
First: due to the fact that  $N=N^\dagger$, it is clear that
coefficients must equal those with all internal causality arrows reversed.
Secondly, a very useful relationship discussed in \cite{Kim:2024svw} is the edge contraction rule.
This states that the sum of two graphs which differ only by a single internal edge,
where that edge contains both routings of propagators, equals the ``collapsed''
graph where that edge is deleted:
\begin{align}\label{eq:edgecontraction}
\omega\left(\begin{tikzpicture}[baseline={([yshift=-.5ex]current bounding box.center)},thick]
        \draw[causArrow] (0,0) -- (1,0);
        \draw[noncausArrow] (1,0) -- (1.75,0.5);%
        \draw[noncausArrow] (1,0) -- (1.75,0.0);%
        \draw[noncausArrow] (1,0) -- (1.75,-0.5);%
        \draw[noncausArrow] (0,0) -- (-.75,0.5);%
        \draw[noncausArrow] (0,0) -- (-.75,0.0);%
        \draw[noncausArrow] (0,0) -- (-.75,-0.5);%
        \draw[fill=white] (0,0) circle(3pt);
        \draw[fill=white] (1,0) circle(3pt);
        \draw[fill=white,color=white] (-.75,0) ellipse (10pt and 20pt);
        \draw[fill=white,color=white] (1.75,0) ellipse (10pt and 20pt);
        \draw[pattern=north east lines] (-.75,0) ellipse (10pt and 20pt);
        \draw[pattern=north east lines] (1.75,0) ellipse (10pt and 20pt);
        \filldraw[fill=white,color=white] (-.75,0) circle (7pt);
        \filldraw[fill=white,color=white] (1.75,0) circle (7pt);
        \node (middle) at (-.75,0) {A}; 
        \node (middle) at (1.75,0) {B}; 
\end{tikzpicture}\right)+
\omega\left(\begin{tikzpicture}[baseline={([yshift=-.5ex]current bounding box.center)},thick]
        \draw[causArrow] (1,0) -- (0,0);
        \draw[noncausArrow] (1,0) -- (1.75,0.5);%
        \draw[noncausArrow] (1,0) -- (1.75,0.0);%
        \draw[noncausArrow] (1,0) -- (1.75,-0.5);%
        \draw[noncausArrow] (0,0) -- (-.75,0.5);%
        \draw[noncausArrow] (0,0) -- (-.75,0.0);%
        \draw[noncausArrow] (0,0) -- (-.75,-0.5);%
        \draw[fill=white] (0,0) circle(3pt);
        \draw[fill=white] (1,0) circle(3pt);
        \draw[fill=white,color=white] (-.75,0) ellipse (10pt and 20pt);
        \draw[fill=white,color=white] (1.75,0) ellipse (10pt and 20pt);
        \draw[pattern=north east lines] (-.75,0) ellipse (10pt and 20pt);
        \draw[pattern=north east lines] (1.75,0) ellipse (10pt and 20pt);
        \filldraw[fill=white,color=white] (-.75,0) circle (7pt);
        \filldraw[fill=white,color=white] (1.75,0) circle (7pt);
        \node (middle) at (-.75,0) {A}; 
        \node (middle) at (1.75,0) {B}; 
\end{tikzpicture}\right)=
\omega\left(\begin{tikzpicture}[baseline={([yshift=-.5ex]current bounding box.center)},thick]
        \draw[noncausArrow] (0,0) -- (0.75,0.5);%
        \draw[noncausArrow] (0,0) -- (0.75,0.0);%
        \draw[noncausArrow] (0,0) -- (0.75,-0.5);%
        \draw[noncausArrow] (0,0) -- (-.75,0.5);%
        \draw[noncausArrow] (0,0) -- (-.75,0.0);%
        \draw[noncausArrow] (0,0) -- (-.75,-0.5);%
        \draw[fill=white] (0,0) circle(3pt);
        \draw[fill=white,color=white] (-.75,0) ellipse (10pt and 20pt);
        \draw[fill=white,color=white] (0.75,0) ellipse (10pt and 20pt);
        \draw[pattern=north east lines] (-.75,0) ellipse (10pt and 20pt);
        \draw[pattern=north east lines] (0.75,0) ellipse (10pt and 20pt);
        \filldraw[fill=white,color=white] (-.75,0) circle (7pt);
        \filldraw[fill=white,color=white] (0.75,0) circle (7pt);
        \node (middle) at (-.75,0) {A}; 
        \node (middle) at (0.75,0) {B}; 
\end{tikzpicture}\right)\,.
\end{align}
One can easily check that this applies to all the examples in \eqref{easyMurua}. 
An important consequence of the edge contraction rule, which is also discussed in \cite{Kim:2024svw}, is the sum rule
\begin{equation}
    \sum_{\tau\in A(\rho)} \omega(\tau)=1\, ,
\end{equation}
where $\rho$ is an undirected graph with $n$ vertices and $A(\rho)$ is the set of all $2^{n-1}$ graphs generated by applying all possible retarded and advanced arrows to the edges of $\rho$. For example, 
\begin{equation}
    A\left(\begin{tikzpicture}[font=\small, thick]
        \draw[massive] (0,0) -- (1, 0);
        \draw[massive] (1, 0) -- (2, 0);
        \draw[fill=white] (0,0) circle(3pt);
        \draw[fill=white] (1,0) circle(3pt);
        \draw[fill=white] (2,0) circle(3pt);
    \end{tikzpicture}\right)%
    =\{     \begin{tikzpicture}[font=\small, thick]
        \draw[massive, causArrow] (0,0) -- (1, 0);
        \draw[massive, causArrow] (1, 0) -- (2, 0);
        \draw[fill=white] (0,0) circle(3pt);
        \draw[fill=white] (1,0) circle(3pt);
        \draw[fill=white] (2,0) circle(3pt);
    \end{tikzpicture},\,
    \begin{tikzpicture}[font=\small, thick]
        \draw[massive, causArrow] (0,0) -- (1, 0);
        \draw[massive, causArrow] (2, 0) -- (1, 0);
        \draw[fill=white] (0,0) circle(3pt);
        \draw[fill=white] (1,0) circle(3pt);
        \draw[fill=white] (2,0) circle(3pt);
    \end{tikzpicture}\,,
    \begin{tikzpicture}[font=\small, thick]
        \draw[massive, causArrow] (1,0) -- (0, 0);
        \draw[massive, causArrow] (1,0) -- (2, 0);
        \draw[fill=white] (0,0) circle(3pt);
        \draw[fill=white] (1,0) circle(3pt);
        \draw[fill=white] (2,0) circle(3pt);
    \end{tikzpicture}\,,
    \begin{tikzpicture}[font=\small, thick]
        \draw[massive, causArrow] (1,0) -- (0, 0);
        \draw[massive, causArrow] (2, 0) -- (1, 0);
        \draw[fill=white] (0,0) circle(3pt);
        \draw[fill=white] (1,0) circle(3pt);
        \draw[fill=white] (2,0) circle(3pt);
    \end{tikzpicture}
    \}\,,
\end{equation}
so that the sum rule gives
\begin{equation}
    2\,\omega\left(\begin{tikzpicture}[font=\small, thick]
        \draw[massive, causArrow] (0,0) -- (1, 0);
        \draw[massive, causArrow] (1, 0) -- (2, 0);
        \draw[fill=white] (0,0) circle(3pt);
        \draw[fill=white] (1,0) circle(3pt);
        \draw[fill=white] (2,0) circle(3pt);
    \end{tikzpicture}\right)+
    \omega\left(\begin{tikzpicture}[baseline={([yshift=-.5ex]current bounding box.center)}, thick]
        \draw[causArrow] (0,0) -- (0.5,0.4);
        \draw[causArrow] (0,0) -- (0.5,-0.4);
        \draw[fill=white] (0,0) circle(3pt);
        \draw[fill=white] (0.5,0.4) circle(3pt);
        \draw[fill=white] (0.5,-0.4) circle(3pt);
    \end{tikzpicture}\right)+
    \omega\left(\begin{tikzpicture}[baseline={([yshift=-.5ex]current bounding box.center)}, thick]
        \draw[causArrow] (0,0.4) -- (0.5,0);
        \draw[causArrow] (0,-.4) -- (0.5,0);
        \draw[fill=white] (0.5,0) circle(3pt);
        \draw[fill=white] (0,0.4) circle(3pt);
        \draw[fill=white] (0,-.4) circle(3pt);
    \end{tikzpicture}\right)=1\,,
\end{equation}
which is easily confirmed from the explicit factors in \eqref{easyMurua3V}.
This rule implies that (at tree level) when we ignore causality prescriptions
the $N$-matrix elements agree with fully connected $T$-matrix elements. Other sum rules for Murua coefficients can be found in \cite{Kim:2024svw}.

\subsection{One-loop Murua coefficients}
\label{sec:redofromMurua-loops}
Let us now explore the $N$-operator at one loop, and the coefficients that weight the diagrams presenting its expansion before and after contraction with external states. 
We will soon discover that tree-level Murua coefficients will make an  appearance at one loop, 
weighting diagrams in a manner analogous, in fact closely related,  to tree level.
The main difference is that we may now also include cut lines.
At one-loop order, we can represent the elements of the $N$ operator
in a similar manner to tree level:
\begin{subequations}\label{eq: oneloopN}
\begin{align}
    iN^{(2)}&=(-i)^2 \! \int d^4x_1d^4x_2\,\,\omega\left(
    \begin{tikzpicture}[baseline={([yshift=-1ex]current bounding box.center)},thick,font=\footnotesize]
        \draw[out=45,causArrow] (0,0) node[yshift=0.4cm]{} to (1,0) node[yshift=0.4cm]{};
        \draw[out=-45,in=-135,cut] (0,0) to (1,0);
        \draw[fill=white] (0,0) circle(3pt);
        \draw[fill=white] (1,0) circle(3pt);
    \end{tikzpicture}\right)
    \begin{tikzpicture}[baseline={([yshift=-1ex]current bounding box.center)},thick,font=\footnotesize]
        \draw[out=45,causArrow] (0,0) node[yshift=0.4cm]{1} to (1,0) node[yshift=0.4cm]{2};
        \draw[out=-45,in=-135,cut] (0,0) to (1,0);
        \draw[fill=white] (0,0) circle(3pt);
        \draw[fill=white] (1,0) circle(3pt);
    \end{tikzpicture}:\phi_1\phi_2:+\cdots\, ,\\
    iN^{(3)}&=(-i)^3 \! \int d^4x_1d^4x_2d^4x_3\Bigg[\,
    \Bigg(\frac12\,\omega\left(
    \begin{tikzpicture}[baseline={([yshift=-.5ex]current bounding box.center)}, thick]
        \draw[causArrow] (0,0) -- (0.5,0.5);
        \draw[causArrow] (0,0) -- (0.5,-0.5);
        \draw[cut] (0.5,0.5) -- (0.5,-0.5);
        \draw[fill=white] (0,0) circle(3pt);
        \draw[fill=white] (0.5,0.5) circle(3pt);
        \draw[fill=white] (0.5,-0.5) circle(3pt);
    \end{tikzpicture}\right)\!\!
    \begin{tikzpicture}[baseline={([yshift=-.5ex]current bounding box.center)}, thick, font=\footnotesize]
        \draw[causArrow] (0,0) node[xshift=-0.25cm]{1} -- (0.5,0.5) node[xshift=0.3cm]{2};
        \draw[causArrow] (0,0) -- (0.5,-0.5) node[xshift=0.3cm]{3};
        \draw[cut] (0.5,0.5) -- (0.5,-0.5);
        \draw[fill=white] (0,0) circle(3pt);
        \draw[fill=white] (0.5,0.5) circle(3pt);
        \draw[fill=white] (0.5,-0.5) circle(3pt);
    \end{tikzpicture}\!\!+
    \frac12\,\omega\left(\begin{tikzpicture}[baseline={([yshift=-.5ex]current bounding box.center)}, thick]
        \draw[causArrow] (0.5,0.5) -- (0,0);
        \draw[causArrow] (0.5,-0.5) -- (0,0);
        \draw[cut] (0.5,0.5) -- (0.5,-0.5);
        \draw[fill=white] (0,0) circle(3pt);
        \draw[fill=white] (0.5,0.5) circle(3pt);
        \draw[fill=white] (0.5,-0.5) circle(3pt);
    \end{tikzpicture}\right)\!\!
    \begin{tikzpicture}[baseline={([yshift=-.5ex]current bounding box.center)}, thick,font=\footnotesize]
        \draw[causArrow] (0.5,0.5) node[xshift=0.3cm]{2} -- (0,0)  node[xshift=-0.25cm]{1};
        \draw[causArrow] (0.5,-0.5) node[xshift=0.3cm]{3} -- (0,0);
        \draw[cut] (0.5,0.5) -- (0.5,-0.5);
        \draw[fill=white] (0,0) circle(3pt);
        \draw[fill=white] (0.5,0.5) circle(3pt);
        \draw[fill=white] (0.5,-0.5) circle(3pt);
    \end{tikzpicture}\\
    &\qquad+
    \omega\left(\begin{tikzpicture}[baseline={([yshift=-.5ex]current bounding box.center)}, thick]
        \draw[causArrow] (0.5,0.5) -- (0,0);
        \draw[causArrow] (0,0) -- (0.5,-0.5);
        \draw[cut] (0.5,0.5) -- (0.5,-0.5);
        \draw[fill=white] (0,0) circle(3pt);
        \draw[fill=white] (0.5,0.5) circle(3pt);
        \draw[fill=white] (0.5,-0.5) circle(3pt);
    \end{tikzpicture}\right)\!\!
    \begin{tikzpicture}[baseline={([yshift=-.5ex]current bounding box.center)}, thick, font=\footnotesize]
        \draw[causArrow] (0.5,0.5) node[xshift=0.3cm]{2} -- (0,0) node[xshift=-0.25cm]{1};
        \draw[causArrow] (0,0) -- (0.5,-0.5) node[xshift=0.3cm]{3};
        \draw[cut] (0.5,0.5) -- (0.5,-0.5);
        \draw[fill=white] (0,0) circle(3pt);
        \draw[fill=white] (0.5,0.5) circle(3pt);
        \draw[fill=white] (0.5,-0.5) circle(3pt);
    \end{tikzpicture}\,\Bigg):\phi_1\phi_2\phi_3:+\Bigg(
    \omega\left(\begin{tikzpicture}[baseline={([yshift=-1ex]current bounding box.center)},thick,font=\footnotesize]
        \draw[out=45,causArrow] (0,0) to (1,0);
        \draw[out=-45,in=-135,cut] (0,0) to (1,0);
        \draw[causArrow] (-0.7,0) -- (0,0);
        \draw[fill=white] (0,0) circle(3pt);
        \draw[fill=white] (-0.7,0) circle(3pt);
        \draw[fill=white] (1,0) circle(3pt);
    \end{tikzpicture}\right)
    \begin{tikzpicture}[baseline={([yshift=-1ex]current bounding box.center)},thick,font=\footnotesize]
        \draw[out=45,causArrow] (0,0) node[yshift=0.4cm]{2} to (1,0) node[yshift=0.4cm]{3};
        \draw[out=-45,in=-135,cut] (0,0) to (1,0);
        \draw[causArrow] (-0.7,0) node[yshift=0.4cm]{1} -- (0,0);
        \draw[fill=white] (0,0) circle(3pt);
        \draw[fill=white] (-0.7,0) circle(3pt);
        \draw[fill=white] (1,0) circle(3pt);
    \end{tikzpicture}\nn\\
    &\qquad+
    \omega\left(\begin{tikzpicture}[baseline={([yshift=-1ex]current bounding box.center)},thick,font=\footnotesize]
        \draw[out=135,in=45,causArrow] (1,0) to (0,0);
        \draw[out=-45,in=-135,cut] (0,0) to (1,0);
        \draw[causArrow] (0,0) -- (-0.7,0);
        \draw[fill=white] (0,0) circle(3pt);
        \draw[fill=white] (-0.7,0) circle(3pt);
        \draw[fill=white] (1,0) circle(3pt);
    \end{tikzpicture}\right)
    \begin{tikzpicture}[baseline={([yshift=-1ex]current bounding box.center)},thick,font=\footnotesize]
        \draw[out=135,in=45,causArrow] (1,0) node[yshift=0.4cm]{3} to (0,0) node[yshift=0.4cm]{2};
        \draw[out=-45,in=-135,cut] (0,0) to (1,0);
        \draw[causArrow] (0,0) -- (-0.7,0) node[yshift=0.4cm]{1};
        \draw[fill=white] (0,0) circle(3pt);
        \draw[fill=white] (-0.7,0) circle(3pt);
        \draw[fill=white] (1,0) circle(3pt);
    \end{tikzpicture}+
    \omega\left(\begin{tikzpicture}[baseline={([yshift=-1ex]current bounding box.center)},thick,font=\footnotesize]
        \draw[out=45,causArrow] (0,0) to (1,0);
        \draw[out=-45,in=-135,cut] (0,0) to (1,0);
        \draw[causArrow] (0,0) -- (-0.7,0);
        \draw[fill=white] (0,0) circle(3pt);
        \draw[fill=white] (-0.7,0) circle(3pt);
        \draw[fill=white] (1,0) circle(3pt);
    \end{tikzpicture}\right)
    \begin{tikzpicture}[baseline={([yshift=-1ex]current bounding box.center)},thick,font=\footnotesize]
        \draw[out=45,causArrow] (0,0) node[yshift=0.4cm]{2} to (1,0) node[yshift=0.4cm]{3};
        \draw[out=-45,in=-135,cut] (0,0) to (1,0);
        \draw[causArrow] (0,0) -- (-0.7,0) node[yshift=0.4cm]{1};
        \draw[fill=white] (0,0) circle(3pt);
        \draw[fill=white] (-0.7,0) circle(3pt);
        \draw[fill=white] (1,0) circle(3pt);
    \end{tikzpicture}\nn\\
    &\qquad+
    \omega\left(\begin{tikzpicture}[baseline={([yshift=-1ex]current bounding box.center)},thick,font=\footnotesize]
       \draw[out=135,in=45,causArrow] (1,0) to (0,0);
        \draw[out=-45,in=-135,cut] (0,0) to (1,0);
        \draw[causArrow] (-0.7,0) -- (0,0);
        \draw[fill=white] (0,0) circle(3pt);
        \draw[fill=white] (-0.7,0) circle(3pt);
        \draw[fill=white] (1,0) circle(3pt);
    \end{tikzpicture}\right)
    \begin{tikzpicture}[baseline={([yshift=-1ex]current bounding box.center)},thick,font=\footnotesize]
        \draw[out=135,in=45,causArrow] (1,0) node[yshift=0.4cm]{3} to (0,0) node[yshift=0.4cm]{2};
        \draw[out=-45,in=-135,cut] (0,0) to (1,0);
        \draw[causArrow] (-0.7,0) node[yshift=0.4cm]{1} -- (0,0);
        \draw[fill=white] (0,0) circle(3pt);
        \draw[fill=white] (-0.7,0) circle(3pt);
        \draw[fill=white] (1,0) circle(3pt);
    \end{tikzpicture}
    \Bigg)\frac{:\phi_1^2\phi_3:}{2!}\Bigg]+\cdots\, , \nn
\end{align}
\end{subequations}
where the dots denote further Wick contraction that do not contribute at one loop.  For each possible routing of propagators,
which now generically includes a single cut propagator,
we introduce a corresponding Murua coefficient $\omega$.
Using our explicit Magnus-based calculations in Section~\ref{sec:oneloop},
in particular the numerical coefficients appearing in \eqref{bubcut}, \eqref{eq:TriangleCut} and \eqref{eq:bub12},
we have been able to determine that
\begin{subequations}\label{eq:oneloopMurua}
\begin{alignat}{2}
\omega\left(
    \begin{tikzpicture}[baseline={([yshift=-1ex]current bounding box.center)},thick,font=\footnotesize]
        \draw[out=45,causArrow] (0,0) node[yshift=0.4cm]{} to (1,0) node[yshift=0.4cm]{};
        \draw[out=-45,in=-135,cut] (0,0) to (1,0);
        \draw[fill=white] (0,0) circle(3pt);
        \draw[fill=white] (1,0) circle(3pt);
    \end{tikzpicture}\right) &= \frac14\,, &&\\
\omega\left(
    \begin{tikzpicture}[baseline={([yshift=-.5ex]current bounding box.center)}, thick]
        \draw[causArrow] (0,0) -- (0.5,0.5);
        \draw[causArrow] (0,0) -- (0.5,-0.5);
        \draw[cut] (0.5,0.5) -- (0.5,-0.5);
        \draw[fill=white] (0,0) circle(3pt);
        \draw[fill=white] (0.5,0.5) circle(3pt);
        \draw[fill=white] (0.5,-0.5) circle(3pt);
    \end{tikzpicture}\right) &=
\omega\left(\begin{tikzpicture}[baseline={([yshift=-.5ex]current bounding box.center)}, thick]
        \draw[causArrow] (0.5,0.5) -- (0,0);
        \draw[causArrow] (0.5,-0.5) -- (0,0);
        \draw[cut] (0.5,0.5) -- (0.5,-0.5);
        \draw[fill=white] (0,0) circle(3pt);
        \draw[fill=white] (0.5,0.5) circle(3pt);
        \draw[fill=white] (0.5,-0.5) circle(3pt);
    \end{tikzpicture}\right)=\frac{1}{12}\,,
\qquad\qquad\qquad\!\!\!\,\, \omega\left(\begin{tikzpicture}[baseline={([yshift=-.5ex]current bounding box.center)}, thick]
        \draw[causArrow] (0.5,0.5) -- (0,0);
        \draw[causArrow] (0,0) -- (0.5,-0.5);
        \draw[cut] (0.5,0.5) -- (0.5,-0.5);
        \draw[fill=white] (0,0) circle(3pt);
        \draw[fill=white] (0.5,0.5) circle(3pt);
        \draw[fill=white] (0.5,-0.5) circle(3pt);
    \end{tikzpicture}\right) &&= \frac16\,,\\
\omega\left(\begin{tikzpicture}[baseline={([yshift=-1ex]current bounding box.center)},thick,font=\footnotesize]
        \draw[out=45,causArrow] (0,0) node[yshift=0.4cm]{} to (1,0) node[yshift=0.4cm]{};
        \draw[out=-45,in=-135,cut] (0,0) to (1,0);
        \draw[causArrow] (0,0) -- (-0.7,0);
        \draw[fill=white] (0,0) circle(3pt);
        \draw[fill=white] (-0.7,0) circle(3pt);
        \draw[fill=white] (1,0) circle(3pt);
    \end{tikzpicture}\right) \! &= \!
\omega\left(\begin{tikzpicture}[baseline={([yshift=-1ex]current bounding box.center)},thick,font=\footnotesize]
       \draw[out=135,in=45,causArrow] (1,0) node[yshift=0.4cm]{} to (0,0) node[yshift=0.4cm]{};
        \draw[out=-45,in=-135,cut] (0,0) to (1,0);
        \draw[causArrow] (-0.7,0) -- (0,0);
        \draw[fill=white] (0,0) circle(3pt);
        \draw[fill=white] (-0.7,0) circle(3pt);
        \draw[fill=white] (1,0) circle(3pt);
    \end{tikzpicture}\right) \! = \! \frac{1}{12}\, , \,\,\,\,
\omega\left(\begin{tikzpicture}[baseline={([yshift=-1ex]current bounding box.center)},thick,font=\footnotesize]
        \draw[out=45,causArrow] (0,0) node[yshift=0.4cm]{} to (1,0) node[yshift=0.4cm]{};
        \draw[out=-45,in=-135,cut] (0,0) to (1,0);
        \draw[causArrow] (-0.7,0) -- (0,0);
        \draw[fill=white] (0,0) circle(3pt);
        \draw[fill=white] (-0.7,0) circle(3pt);
        \draw[fill=white] (1,0) circle(3pt);
    \end{tikzpicture}\right) \! &&= \!
\omega\left(\begin{tikzpicture}[baseline={([yshift=-1ex]current bounding box.center)},thick,font=\footnotesize]
        \draw[out=135,in=45,causArrow] (1,0) node[yshift=0.4cm]{} to (0,0) node[yshift=0.4cm]{};
        \draw[out=-45,in=-135,cut] (0,0) to (1,0);
        \draw[causArrow] (0,0) -- (-0.7,0);
        \draw[fill=white] (0,0) circle(3pt);
        \draw[fill=white] (-0.7,0) circle(3pt);
        \draw[fill=white] (1,0) circle(3pt);
    \end{tikzpicture}\right) \! = \! \frac16\,.
\end{alignat}
\end{subequations}
As in the tree-level case,
diagrams where all retarded/advanced propagators are reversed
are equal to their non-reversed counterparts: again, a consequence of $N$ being hermitian.
Importantly, we also immediately observe a relationship with the tree-level Murua coefficients listed in \eqref{easyMurua}:
\begin{align}\label{eq:oneLooprelationship}
\boxed{
\omega\left(\begin{tikzpicture}[baseline={([yshift=-.5ex]current bounding box.center)},thick]
        \draw[causArrow] (-.75,0.1) -- (.75,0.1);
        \draw[out=-45,in=-135,cut] (-.75,0) to (.75,0);
        \draw[fill=white,color=white] (-.75,0) ellipse (10pt and 20pt);
        \draw[fill=white,color=white] (.75,0) ellipse (10pt and 20pt);
        \draw[pattern=north east lines] (-.75,0) ellipse (10pt and 20pt);
        \draw[pattern=north east lines] (.75,0) ellipse (10pt and 20pt);
        \filldraw[fill=white,color=white] (-.75,0) circle (7pt);
        \filldraw[fill=white,color=white] (.75,0) circle (7pt);
        \node (middle) at (-.75,0) {A}; 
        \node (middle) at (.75,0) {B}; 
\end{tikzpicture}\right)=
\frac12\,\omega\left(\begin{tikzpicture}[baseline={([yshift=-.5ex]current bounding box.center)},thick]
        \draw[causArrow] (-.75,0) -- (.75,0);
        \draw[fill=white,color=white] (-.75,0) ellipse (10pt and 20pt);
        \draw[fill=white,color=white] (.75,0) ellipse (10pt and 20pt);
        \draw[pattern=north east lines] (-.75,0) ellipse (10pt and 20pt);
        \draw[pattern=north east lines] (.75,0) ellipse (10pt and 20pt);
        \filldraw[fill=white,color=white] (-.75,0) circle (7pt);
        \filldraw[fill=white,color=white] (.75,0) circle (7pt);
        \node (middle) at (-.75,0) {A}; 
        \node (middle) at (.75,0) {B}; 
\end{tikzpicture}\right)\,
}
\,\, .
\end{align}
This relationship turns out to be true in general and we will explain its precise origin in Section~\ref{sec:treestoloops},
and see how it generalises also to higher loop orders.
In words: the loop coefficient is one half of the corresponding tree-level
coefficient, with the cut line deleted.%
\footnote{The factor of $1/2$ is due to the fact the Hadamard function \eqref{Hadamard} is twice a cut propagator. 
}
As a natural consequence,
the edge contraction rule~\eqref{eq:edgecontraction} for retarded/advanced propagators will also hold for the one-loop Murua coefficients.

As in the tree-level case, Murua coefficients can also be used to write down matrix elements of $N$,
\begin{subequations}\label{eq: NMatrixOneLoop}
\begin{align}
iN^{(2)}_2&= (-i)^2\left[
\omega\left(
\begin{tikzpicture}[baseline={([yshift=-0.5ex]current bounding box.center)},font=\small,thick,scale=0.5] 
    \draw[massive] (-1.25,0) -- (-0.75,0);
    \draw[massive,causArrow] (-0.75,0) arc(180:0:0.75);
    \draw[massive,cut=1] (-0.75,0) arc(-180:0:0.75);
    \draw[massive] (0.75,0) -- (1.25,0);
\end{tikzpicture}
\right)
\begin{tikzpicture}[baseline={([yshift=-0.5ex]current bounding box.center)},font=\small,thick,scale=0.5] 
    \draw[massive] (-1.25,0) -- (-0.75,0);
    \draw[massive,causArrow] (-0.75,0) arc(180:0:0.75);
    \draw[massive,cut=1] (-0.75,0) arc(-180:0:0.75);
    \draw[massive] (0.75,0) -- (1.25,0);
\end{tikzpicture}
+ \omega\left(
\begin{tikzpicture}[baseline={([yshift=-0.5ex]current bounding box.center)},font=\small,thick,scale=0.5] 
    \draw[massive] (-1.25,0) -- (-0.75,0);
    \draw[massive,causArrowR] (-0.75,0) arc(180:0:0.75);
    \draw[massive,cut=1] (-0.75,0) arc(-180:0:0.75);
    \draw[massive] (0.75,0) -- (1.25,0);
\end{tikzpicture}
\right)
\begin{tikzpicture}[baseline={([yshift=-0.5ex]current bounding box.center)},font=\small,thick,scale=0.5] 
    \draw[massive] (-1.25,0) -- (-0.75,0);
    \draw[massive,causArrowR] (-0.75,0) arc(180:0:0.75);
    \draw[massive,cut=1] (-0.75,0) arc(-180:0:0.75);
    \draw[massive] (0.75,0) -- (1.25,0);
\end{tikzpicture}
\right] \\
iN^{(3)}_3 &= \sum_{\text{perms}}(-i)^3\left[
\omega\left(\,
\begin{tikzpicture}[baseline={([yshift=-.5ex]current bounding box.center)},thick,scale=0.5] 
    \draw (-90:1.5) -- (-90:1);
    \draw (30:1.5) -- (30:1);
    \draw (150:1.5) -- (150:1);
    \draw[massive, cut] (30:1) -- (150:1);
    \draw[massive, causArrow] (150:1) -- (-90:1);
    \draw[massive, causArrow] (-90:1) -- (30:1);
\end{tikzpicture}
\,\right)
\begin{tikzpicture}[baseline={([yshift=-.5ex]current bounding box.center)},thick,scale=0.5] 
    \draw (-90:1.5) -- (-90:1);
    \draw (30:1.5) -- (30:1);
    \draw (150:1.5) -- (150:1);
    \draw[massive, cut] (30:1) -- (150:1);
    \draw[massive, causArrow] (150:1) -- (-90:1);
    \draw[massive, causArrow] (-90:1) -- (30:1);
\end{tikzpicture}
+\omega\left(\,
\begin{tikzpicture}[baseline={([yshift=-.5ex]current bounding box.center)},thick,scale=0.5] 
    \draw (-90:1.5) -- (-90:1);
    \draw (30:1.5) -- (30:1);
    \draw (150:1.5) -- (150:1);
    \draw[massive, cut] (30:1) -- (150:1);
    \draw[massive, causArrow] (150:1) -- (-90:1);
    \draw[massive, causArrowR] (-90:1) -- (30:1);
\end{tikzpicture}
\,\right)
\begin{tikzpicture}[baseline={([yshift=-.5ex]current bounding box.center)},thick,scale=0.5] 
    \draw (-90:1.5) -- (-90:1);
    \draw (30:1.5) -- (30:1);
    \draw (150:1.5) -- (150:1);
    \draw[massive, cut] (30:1) -- (150:1);
    \draw[massive, causArrow] (150:1) -- (-90:1);
    \draw[massive, causArrowR] (-90:1) -- (30:1);
\end{tikzpicture}
\right. \nn\\
&\left.
\hspace{40pt}+\omega\left(\,
\begin{tikzpicture}[baseline={([yshift=-.5ex]current bounding box.center)},thick,scale=0.5] 
    \draw (-90:1.5) -- (-90:1);
    \draw (30:1.5) -- (30:1);
    \draw (150:1.5) -- (150:1);
    \draw[massive, cut] (30:1) -- (150:1);
    \draw[massive, causArrowR] (150:1) -- (-90:1);
    \draw[massive, causArrow] (-90:1) -- (30:1);
\end{tikzpicture}
\,\right)
\begin{tikzpicture}[baseline={([yshift=-.5ex]current bounding box.center)},thick,scale=0.5] 
    \draw (-90:1.5) -- (-90:1);
    \draw (30:1.5) -- (30:1);
    \draw (150:1.5) -- (150:1);
    \draw[massive, cut] (30:1) -- (150:1);
    \draw[massive, causArrowR] (150:1) -- (-90:1);
    \draw[massive, causArrow] (-90:1) -- (30:1);
\end{tikzpicture}
+\omega\left(\,
\begin{tikzpicture}[baseline={([yshift=-.5ex]current bounding box.center)},thick,scale=0.5] 
    \draw (-90:1.5) -- (-90:1);
    \draw (30:1.5) -- (30:1);
    \draw (150:1.5) -- (150:1);
    \draw[massive, cut] (30:1) -- (150:1);
    \draw[massive, causArrowR] (150:1) -- (-90:1);
    \draw[massive, causArrowR] (-90:1) -- (30:1);
\end{tikzpicture}
\,\right)
\begin{tikzpicture}[baseline={([yshift=-.5ex]current bounding box.center)},thick,scale=0.5] 
    \draw (-90:1.5) -- (-90:1);
    \draw (30:1.5) -- (30:1);
    \draw (150:1.5) -- (150:1);
    \draw[massive, cut] (30:1) -- (150:1);
    \draw[massive, causArrowR] (150:1) -- (-90:1);
    \draw[massive, causArrowR] (-90:1) -- (30:1);
\end{tikzpicture}
\right] \nn\\
&+\sum_{\text{perms}}(-i)^3\left[
\omega\left(\,
\begin{tikzpicture}[baseline={([yshift=-.5ex]current bounding box.center)}, thick, scale=0.5] 
    \draw[massive] (120:0.5) -- (0,0) -- (-120:0.5);
    \draw[causArrow] (0,0) -- (1,0);
    \draw[causArrow] (1,0) arc(180:0:0.5);
    \draw[cut=1] (1,0) arc(-180:0:0.5);
    \draw[massive] (2,0) -- ++(0.5,0);
\end{tikzpicture}
\,\right)
\begin{tikzpicture}[baseline={([yshift=-.5ex]current bounding box.center)}, thick, scale=0.5] 
    \draw[massive] (120:0.5) -- (0,0) -- (-120:0.5);
    \draw[causArrow] (0,0) -- (1,0);
    \draw[causArrow] (1,0) arc(180:0:0.5);
    \draw[cut=1] (1,0) arc(-180:0:0.5);
    \draw[massive] (2,0) -- ++(0.5,0);
\end{tikzpicture}
+ \omega\left(\,
\begin{tikzpicture}[baseline={([yshift=-.5ex]current bounding box.center)}, thick, scale=0.5] 
    \draw[massive] (120:0.5) -- (0,0) -- (-120:0.5);
    \draw[causArrow] (0,0) -- (1,0);
    \draw[causArrowR] (1,0) arc(180:0:0.5);
    \draw[cut=1] (1,0) arc(-180:0:0.5);
    \draw[massive] (2,0) -- ++(0.5,0);
\end{tikzpicture}
\,\right)
\begin{tikzpicture}[baseline={([yshift=-.5ex]current bounding box.center)}, thick, scale=0.5] 
    \draw[massive] (120:0.5) -- (0,0) -- (-120:0.5);
    \draw[causArrow] (0,0) -- (1,0);
    \draw[causArrowR] (1,0) arc(180:0:0.5);
    \draw[cut=1] (1,0) arc(-180:0:0.5);
    \draw[massive] (2,0) -- ++(0.5,0);
\end{tikzpicture}
\right. \nn\\
&\left.\hspace{40pt}
+\omega\left(\,
\begin{tikzpicture}[baseline={([yshift=-.5ex]current bounding box.center)}, thick, scale=0.5] 
    \draw[massive] (120:0.5) -- (0,0) -- (-120:0.5);
    \draw[causArrowR] (0,0) -- (1,0);
    \draw[causArrow] (1,0) arc(180:0:0.5);
    \draw[cut=1] (1,0) arc(-180:0:0.5);
    \draw[massive] (2,0) -- ++(0.5,0);
\end{tikzpicture}
\,\right)
\begin{tikzpicture}[baseline={([yshift=-.5ex]current bounding box.center)}, thick, scale=0.5] 
    \draw[massive] (120:0.5) -- (0,0) -- (-120:0.5);
    \draw[causArrowR] (0,0) -- (1,0);
    \draw[causArrow] (1,0) arc(180:0:0.5);
    \draw[cut=1] (1,0) arc(-180:0:0.5);
    \draw[massive] (2,0) -- ++(0.5,0);
\end{tikzpicture}
+ \omega\left(\,
\begin{tikzpicture}[baseline={([yshift=-.5ex]current bounding box.center)}, thick, scale=0.5] 
    \draw[massive] (120:0.5) -- (0,0) -- (-120:0.5);
    \draw[causArrowR] (0,0) -- (1,0);
    \draw[causArrowR] (1,0) arc(180:0:0.5);
    \draw[cut=1] (1,0) arc(-180:0:0.5);
    \draw[massive] (2,0) -- ++(0.5,0);
\end{tikzpicture}
\,\right)
\begin{tikzpicture}[baseline={([yshift=-.5ex]current bounding box.center)}, thick, scale=0.5] 
    \draw[massive] (120:0.5) -- (0,0) -- (-120:0.5);
    \draw[causArrowR] (0,0) -- (1,0);
    \draw[causArrowR] (1,0) arc(180:0:0.5);
    \draw[cut=1] (1,0) arc(-180:0:0.5);
    \draw[massive] (2,0) -- ++(0.5,0);
\end{tikzpicture}
\right]\, ,
\end{align}
\end{subequations}
where the sum is over the set of inequivalent    permutations of the external momenta.
Again, as at tree level, the Murua coefficients only depend on the internal propagator structure so that external lines can be neglected, for example: 
\begin{equation}
    \omega\left(\,
\begin{tikzpicture}[baseline={([yshift=-.5ex]current bounding box.center)},thick,font=\small,scale=0.75]
    \draw (-90:1.5) node[below]{$p_3$} -- (-90:1);
    \draw (30:1.5) node[above, right]{$p_2$} -- (30:1);
    \draw (150:1.5) node[above, left]{$p_1$} -- (150:1);
    \draw[massive, cut=2] (30:1) -- (150:1);
    \draw[massive, Anew] (150:1) -- (-90:1);
    \draw[massive, Rnew] (-90:1) -- (30:1);
\end{tikzpicture}
\,\right)
=
\omega\left(\,
\begin{tikzpicture}[baseline={([yshift=-.5ex]current bounding box.center)}, thick]
        \draw[causArrow] (0,0) -- (0.5,0.4);
        \draw[causArrow] (0,0) -- (0.5,-0.4);
        \draw[cut] (0.5,0.4) -- (0.5,-0.4);
        \draw[fill=white] (0,0) circle(3pt);
        \draw[fill=white] (0.5,0.4) circle(3pt);
        \draw[fill=white] (0.5,-0.4) circle(3pt);
    \end{tikzpicture}
\,\right)
=\frac{1}{2} \omega\left(\,
\begin{tikzpicture}[baseline={([yshift=-.5ex]current bounding box.center)}, thick]
        \draw[causArrow] (0,0) -- (0.5,0.4);
        \draw[causArrow] (0,0) -- (0.5,-0.4);
        \draw[fill=white] (0,0) circle(3pt);
        \draw[fill=white] (0.5,0.4) circle(3pt);
        \draw[fill=white] (0.5,-0.4) circle(3pt);
\end{tikzpicture}
\,\right)\, ,
\end{equation}
after using \eqref{eq:oneLooprelationship} in the final equality.
Finally, notice that the symmetry factors of $g$ differ from those of $\tau$,
\begin{equation}
    1=\sigma\left(
    \begin{tikzpicture}[baseline={([yshift=-.5ex]current bounding box.center)},thick,font=\small,scale=0.75] 
    \draw (-90:1.5) node[below]{$p_3$} -- (-90:1);
    \draw (30:1.5) node[above, right]{$p_2$} -- (30:1);
    \draw (150:1.5) node[above, left]{$p_1$} -- (150:1);
    \draw[massive, cut=2] (30:1) -- (150:1);
    \draw[massive, Anew] (150:1) -- (-90:1);
    \draw[massive, Rnew] (-90:1) -- (30:1);
\end{tikzpicture}
    \right)\neq
\sigma\left(\begin{tikzpicture}[baseline={([yshift=-.5ex]current bounding box.center)}, thick]
        \draw[causArrow] (0,0) -- (0.5,0.4);
        \draw[causArrow] (0,0) -- (0.5,-0.4);
        \draw[cut] (0.5,0.4) -- (0.5,-0.4);
        \draw[fill=white] (0,0) circle(3pt);
        \draw[fill=white] (0.5,0.4) circle(3pt);
        \draw[fill=white] (0.5,-0.4) circle(3pt);
    \end{tikzpicture}\right)= 2\,.
\end{equation}

This means that, as for tree level, at one loop we can produce $N$-matrix elements just as for scattering amplitudes, taking care to weight graphs appropriately. So far we have shown this explicitly at tree and one-loop level, and in the following subsection we give an outline of the general diagrammatic structure of the $N$-operator and Magnus amplitudes. A detailed description of the loop level structure will be presented in Section \ref{sec:treestoloops}.

\subsection{General diagrammatic structure of the \texorpdfstring{$N$}{N} operator and its matrix elements}\label{sec: generalMuruaFormulae}
As we have done above for tree level and one loop, we can express higher-loop pieces of the $N$ operator and $N$-matrix elements by summing over graphs with Murua coefficients and symmetry factors. Explicitly, at any loop order we can write:
\begin{equation}\label{eq:Nsum}
     iN^{(n)} = \sum_{|\tau|=n}\frac{\omega(\tau)}{\sigma(\tau)}\mathcal{I}(\tau) \, ,
 \end{equation}
where the sum is taken over all possible non-isomorphic connected graphs $\tau$ with $|\tau|{=}n$ vertices and edges corresponding to $\Delta^R$, $\Delta^A$ or $\Delta^{(1)}$. The specifics of what kind of edges can appear at $L$-loops will be discussed in Section~\ref{sec:newwickcontr}, and are summarised in Appendix~\ref{App:summaryrules}. The above equation appears in \cite{Kim:2024svw,Murua_2006} for tree-level graphs, but here we wish to generalise it to multi-loop graphs. We define the symmetry factor $\sigma(\tau)$ in the same way as the Dyson expansion except that we must also preserve propagator prescriptions (see Appendix~\ref{app: SymmFactors} for more details). Thus, we will consider \eqref{eq:Nsum} as the \textit{definition} of $\omega(\tau)$ for multiloop graphs $\tau$, which we will continue to refer to as Murua coefficients. The expanded form of \eqref{eq:Nsum} can be seen in \eqref{sec:redofromMurua-trees} at tree level and in \eqref{eq: oneloopN} at one loop. All the kinematic and theory-dependent information is carried by $\mathcal{I}(\tau)$, including any propagators and integrations over positions. For instance, 
 \begin{align}
\begin{aligned}
     \cI\left(\!%
     \begin{tikzpicture}[baseline={([yshift=-.5ex]current bounding box.center)}, thick,font=\footnotesize]
     \draw[causArrow] (0,0) -- (0.5,0.4);
     \draw[causArrow] (0,0) -- (0.5,-0.4);
     \draw[fill=white] (0,0) circle(3pt);
     \draw[fill=white] (0.5,0.4) circle(3pt);
     \draw[fill=white] (0.5,-0.4) circle(3pt);
     \end{tikzpicture}%
    \right) &= (-i)^3
    \int\! d^4x_1 d^4x_2 d^4x_3\ i\Delta^R_{13}\, i\Delta^R_{23} \frac{:\phi_1^2 \phi_2^2 \phi_3:}{(2!)^2}\\
    &= (-i)^3\int\! d^4x_1 d^4x_2 d^4x_3
    \begin{tikzpicture}[baseline={([yshift=-.5ex]current bounding box.center)}, thick,font=\footnotesize]
    \draw[causArrow] (0,0) -- (0.5,0.4);
    \draw[causArrow] (0,0) -- (0.5,-0.4);
    \draw[fill=white] (0,0) circle(3pt) node[xshift=-0.3cm]{$3$};
    \draw[fill=white] (0.5,0.4) circle(3pt) node[xshift=0.3cm]{$1$};
    \draw[fill=white] (0.5,-0.4) circle(3pt) node[xshift=0.3cm]{$2$};
    \end{tikzpicture}%
    \frac{:\phi_1^2 \phi_2^2 \phi_3:}{(2!)^2}\, .
\end{aligned}
\end{align}

We can write an analogous general formula for the $N$-matrix elements\footnote{This follows from performing external leg contractions on \eqref{eq:Nsum}, see Appendix~\ref{app: SymmFactors} for more details.}
\begin{equation}\label{eq:MuruaFormulaMatrixElement}
    \langle 0| iN |\prod_{i=1}^k\phi(p_i)\rangle = \hat{\delta}^{(4)}\left(\sum_{i=1}^k p_i\right)\sum_{g} \frac{\omega(g)}{\sigma(g)}\mathcal{J}(g) \, .
\end{equation}
Here the sum is over all non-isomorphic graphs $g$, but where $g$ now \textit{has external legs} $\{p_i\}_{i=1}^k$. The tree-level and one-loop versions of the above formula can be seen in \eqref{eq: NMatrixTree} and \eqref{eq: NMatrixOneLoop} respectively. 

As discussed previously, the symmetry factors of a graph with and without external legs differ since we must always fix external particle legs. 
The final element of \eqref{eq:MuruaFormulaMatrixElement}, $\mathcal{J}(g)$ now represents the familiar momentum-space propagators (and possibly loop integrals) associated to the graph~$g$,
\begin{equation}
    \mathcal{J}\left(\begin{tikzpicture}[baseline={([yshift=-0.5ex]current bounding box.center)},thick,font=\small]
    \draw (1, 0) -- ++(135:1) node[left]{$p_2$};
    \draw (1, 0) -- ++(-135:1) node[left]{$p_1$};
    \draw (2, 1) node[above]{$p_3$} -- (2, 0);
    \draw (3, 0) -- ++(45:1) node[right]{$p_4$};
    \draw (3, 0) -- ++(-45:1) node[right]{$p_5$};
    \draw[massive, Anew] (1,0) -- (2,0);
	\draw[massive, Rnew] (2,0) -- (3,0);
    \path (2,-1) -- (2,0); 
\end{tikzpicture}\right) = (-i)^3 i\Delta^A(p_1+p_2)\, i\Delta^R(-p_3-p_4)\, .
\end{equation}
Note that  at tree level and one loop the symmetry factor $\sigma(g)$  in \eqref{eq:MuruaFormulaMatrixElement} is always equal to 1, as can be seen in \eqref{eq: NMatrixTree} and \eqref{eq: NMatrixOneLoop}. At tree level this is a familiar fact of $T$-matrix elements which also applies to $N$-matrix elements. At one loop, for $T$-matrix elements we can have non-trivial symmetry factors coming from, say, a bubble diagram. However, for $N$-matrix elements one propagator is always cut, which breaks the symmetry of the bubble graph.  At two loops and beyond, $N$-matrix elements can have non-trivial symmetry factors as we will see in Section~\ref{sec: MuruaForHigherLoops}.

In principle, the $N$-matrix elements and $N$ operator are sums over all non-isomorphic graphs with and without external legs respectively (and internal edges corresponding to $\Delta^{R}$, $\Delta^{A}$ and $\Delta^{(1)}$). However, at tree level we have seen that no $\Delta^{(1)}$ propagators appear, and at one loop exactly one $\Delta^{(1)}$ must appear. A major topic of the next section will be proving how many $\Delta^{(1)}$ propagators can appear at each loop order.


\section{From trees to loops}
\label{sec:treestoloops}

In Section~\ref{sec:redoMurua} we have discussed Murua's formula and how it can be used to obtain tree-level Magnus amplitudes without the need to perform Wick contractions. Although Murua's formula does not apply at loop level, we find that we can still use it to fix diagram coefficients at one loop due to~\eqref{eq:oneLooprelationship}. 
In this section we will generalise this to higher loops. We also make precise the structure of allowed graphs at loop level.

To do so, we present a new approach for Wick contractions of the $N$ operator in Section~\ref{sec:newwickcontr}, and use it to derive a relation between its tree-level and loop-level components. In Section~\ref{sec:MagnusLoopsFromFL} we use this relation to show that Magnus amplitudes with $L+\ell$ loops and $C+\ell$ cuts are equal to $\ell$ forward limits of Magnus amplitudes with $L$ loops and $C$ cuts, as shown in \eqref{eq: genForwardLimit}. This implies that $\ell$-loop $\ell$-cut amplitudes can be obtained from tree-level amplitudes, hence inheriting their Murua coefficients.

Note that we also present a summary of the diagrammatic rules for the Magnus expansion in Appendix~\ref{App:summaryrules}.

\subsection{Wick contraction rules}
\label{sec:newwickcontr}

In the previous sections, we have seen many examples of matrix elements of the $N$ operator, both at tree and loop level. In particular, at one loop we have seen the emergence of the $\Delta^{(1)}_{ij}$ function  from an additional Wick contraction compared to tree level. 
Unlike for the $\Delta_{ij}$ functions, this was not immediate from the structure of the nested commutators, as it required combining various terms. 
Motivated by this imbalance, we will now  define a new set of Wick contraction rules that treat both $\Delta_{ij}$ and $\Delta^{(1)}_{ij}$ on the same footing.

To this end, we return to the new formula for the Magnus expansion in \eqref{eq:ReformulationCS}, as this provides a general expression for $N^{(n)}$ valid for any order $n$ in perturbation theory. Consider the right-nested commutator
\begin{equation}
    \mathcal{C}_{1\dots n} = \Big[\cH_1,\big[\cH_2,\cdots[\cH_{n-1},\cH_n]\cdots\big]\Big]\, .
\end{equation}
Suppose we would like to do a Wick contraction between two fields labeled by $k$ and $k'$. Without loss of generality, we can pick $k'>k$. We now look at  the nested commutator $\mathcal{C}_{1\dots n}$ and zoom in until we arrive at the commutator starting with $\cH_k$,
\begin{align}\label{eq:ZoomedInNestedCommutator}
\begin{aligned}
    \Big[\cH_k,\cdots\big[\cH_{k'},\cdots[\cH_{n-1},\cH_n]\cdots\big]\cdots\Big] = [\cH_k, \mathcal{C}_{k{+}1\dots n}] = \cH_k \mathcal{C}_{k{+}1\dots n}-\mathcal{C}_{k{+}1\dots n} \cH_k\, .
\end{aligned}
\end{align}
If we perform one  Wick contraction on the first of the two terms above, we get
\begin{align}
\begin{split}
    \cH_k \mathcal{C}_{k{+}1\dots n} \longrightarrow&\, n_{k,k'}i\Delta^{(+)}_{kk'} \cH_{k}^{(1)} \big[\cH_{k+1},\cdots[\cH^{(1)}_{k'},\cdots [\cH_{n-1},\cH_n]\cdots]\cdots\big]\\ :=&\, n_{k,k'}i\Delta^{(+)}_{kk'} \cH_{k}^{(1)} \mathcal{C}^{(1)}_{k{+}1\dots n}\, ,
\end{split}
\end{align}
where $n_{k,k'}$ is a combinatorial factor depending on the specifics of $\cH_k$ and $\cH_{k'}$ and  the superscript in $\cH^{(1)}_{k}$ and  $\cC^{(1)}$  are reminders that there is one field missing from $\cH_k$ and $\cC$.
In principle, one superscript label is needed for each species of particle appearing in $\cH_k$, since each will give rise to independent Wick contractions, but for simplicity we present expressions limited to one single type of field, e.g.~$\cH_k=\phi_k^n/n!$. The second term in \eqref{eq:ZoomedInNestedCommutator} will result in a similar expression, except for the fact that the order of $\cH_{k'}$ and $\cH_k$ is inverted and therefore so are the indices of $\Delta^{(+)}_{ij}$,
\begin{equation}
    \mathcal{C}_{k{+}1\dots n} \cH_k \longrightarrow  n_{k,k'}i\Delta^{(+)}_{k'k} \mathcal{C}^{(1)}_{k{+}1\dots n} \cH_{k}^{(1)}\, .
\end{equation}
Importantly, the numerical factor $n_{k,k'}$ is the same. Using \eqref{eq:CommutatorFunction}-\eqref{eqn:DeltaDelta1Properties}, we can rewrite the $\Delta^{(+)}_{ij}$ functions  in terms of $\Delta_{ij}$ and $\Delta^{(1)}_{ij}$,
\begin{equation}
\label{eq:deltadelta1}
    i\Delta^{(+)}_{kk'}=\frac{1}{2}\left(i\Delta_{kk'}+\Delta^{(1)}_{kk'}\right)\, ,\qquad i\Delta^{(+)}_{k'k}=\frac{1}{2}\left(-i\Delta_{kk'}+\Delta^{(1)}_{kk'}\right)\, .
\end{equation}
Putting all this together, we find that
\begin{align}
    [\cH_k,\mathcal{C}_{k{+}1\dots n}] &\longrightarrow n_{kk'} \left(i\Delta^{(+)}_{kk'}\cH^{(1)}_k\mathcal{C}^{(1)}_{k{+}1\dots n}-i\Delta^{(+)}_{k'k}\mathcal{C}^{(1)}_{k{+}1\dots n} \cH^{(1)}_k\right)\nn\\
    &\quad= \frac{n_{kk'}}{2}\left(i\Delta_{kk'}\{\cH^{(1)}_k,\mathcal{C}^{(1)}_{k{+}1\dots n}\}+\Delta^{(1)}_{kk'}[\cH^{(1)}_k,\mathcal{C}^{(1)}_{k{+}1\dots n}]\right)\, .
\end{align}
This means that, overall for the big nested commutator $\mathcal{C}_{1\dots n}$, after doing one Wick contraction,  we get
\begin{align}\label{eq:OneWickContraction}
    \mathcal{C}_{1\dots n} \rightarrow \ &\frac{n_{kk'}}{2}i\Delta_{kk'} \Bigg[\cH_1,\bigg[\cdots\bigg\{\cH_{k}^{(1)},\Big[\cdots\big[\cH^{(1)}_{k'},\cdots[\cH_{n-1},\cH_n]\cdots\big]\cdots\Big]\cdots\bigg\}\cdots\bigg]\cdots\Bigg]\nn\\
    +&\frac{n_{kk'}}{2}\Delta^{(1)}_{kk'} \Bigg[\cH_1,\bigg[\cdots\bigg[\cH_{k}^{(1)},\Big[\cdots\big[\cH^{(1)}_{k'},\cdots[\cH_{n-1},\cH_n]\cdots\big]\cdots\Big]\cdots\bigg]\cdots\bigg]\cdots\Bigg]\, .
\end{align}
We can run this argument again for the case where we start from $\{\cH_k,\mathcal{C}_{k{+}1\dots n}\}$ instead of $[\cH_k,\mathcal{C}_{k{+}1\dots n}]$ in \eqref{eq:ZoomedInNestedCommutator}. All the steps are identical, but 
in this case, $\Delta^{(1)}_{kk'}$ leaves the anticommutator structure unchanged, and $\Delta_{kk'}$ turns it back into a commutator. We then  arrive at two simple rules that extend the usual Wick contractions to nested commutators:
\begin{itemize}
    \item[{\bf 1.}]~When contracting $k$ and $k'$ in a nested commutator (with $k'$ appearing further in than $k$), two terms are produced, one proportional to $i\Delta_{kk'}$ and the other to $\Delta^{(1)}
    _{kk'}$.
    \item[{\bf 2.}]~The term proportional to $i\Delta_{kk'}$ swaps $[\cH_k,\cdots] \leftrightarrow\{\cH_k,\cdots\}$, whereas the term proportional to $\Delta^{(1)}_{kk'}$ leaves the bracket structure unchanged.
\end{itemize}
Therefore, we can now explicitly see that the natural contraction  basis for nested commutators consists of $\Delta_{ij}$ and $\Delta^{(1)}_{ij}$. Repeated application of these rules, to perform the required number of contractions, will at first produce a huge number of terms, containing all possible combinations of commutators and anti-commutators. However, there is a simple way to drastically reduce the number, as once all the contractions are done, we have to normal order the remaining fields. But notice that ${:[A,B]:}=0$, thus there cannot be any commutators left at the end of this procedure. This constrains the possible combinations of propagators that can appear.

\subsubsection{Tree level}
A tree-level diagram with $n$ vertices has $n-1$ propagators. This means that each of the $n-1$ Wick contractions must flip each of the $n-1$ commutators into anticommutators (otherwise terms containing leftover commutators would vanish upon normal ordering). This leaves no room for terms containing $\Delta^{(1)}_{ij}$, which produces no flips.
Therefore, the general surviving contraction at tree level will look like\footnote{An additional consequence of \eqref{eq:TreeWickContractionRule} is that the allowed Wick contractions always produce connected diagrams.}
\begin{align}\label{eq:TreeWickContractionRule}
&\frac{N}{2^{n-1}} (i\Delta_{1a_1})\cdots (i\Delta_{n-1 a_{n-1}}) \,{:\Big\{\cH_{1}^{(d_1)},\big\{\cH_{2}^{(d_2)},\cdots\{\cH_{n{-}1}^{(d_{n-1})},\cH_{n}^{(d_n)}\}\cdots\big\}\Big\}:} \nn\\
&= N (i\Delta_{1a_1})\cdots(i\Delta_{n-1 a_{n-1}}) :\left(\cH_{1}^{(d_1)} \cdots \cH_{n}^{(d_n)}\right): \, .
\end{align}
Wick contractions are always between a field with label $i$ and another with label $a_i\in\{i+1,\dots,n\}$ (in particular we are forced to contract $n-1$ and $n$). The positive integer $d_i$ refers to the number of 
fields at $x_i$ that have been contracted. 
The final numerical factor $N$ will depend on the specifics of the contractions and the diagram, but can be easily deduced from these. A general method to determine this factor is explained in Appendix~\ref{app:newwickcoeffs}.

For example, we can consider doing two Wick contractions on $[\cH_1,[\cH_2,\cH_3]]$, with $\cH_i{=}:\!\phi_i^3\!:\!/3!$. For simplicity suppose we contract 13 and 23. Before normal ordering, there will be four different terms, corresponding to assignments of $\Delta$ and $\Delta^{(1)}$ to the contractions above:
\begin{align}
    &(i\Delta_{13})(i\Delta_{23})\{\cH_1^{(1)},\{\cH_2^{(1)},\cH_3^{(2)}\}\} + (i\Delta_{13})\Delta^{(1)}_{23}\{\cH_1^{(1)},[\cH_2^{(1)},\cH_3^{(2)}]\} \nn\\
    & +\Delta_{13}^{(1)}(i\Delta_{23})[\cH_1^{(1)},\{\cH_2^{(1)},\cH_3^{(2)}\}] + \Delta^{(1)}_{13}\Delta^{(1)}_{23}[\cH_1^{(1)},[\cH_2^{(1)},\cH_3^{(2)}]]\,, 
\end{align}
with the combinatorial coefficient of this particular set of contractions, determined via Appendix~\ref{app:newwickcoeffs}, is given by
\begin{equation}
\label{eq:wickfactorsexample}
    N
    = \frac{1}{4}\, .
\end{equation}
Notice that after normal ordering only the first term survives, since all other terms contain a commutator. Thus we are left with
\begin{equation}
    \frac{1}{2^2}\frac{1}{4} (i\Delta_{13})(i\Delta_{23}) \,{:\{\phi_1^2,\{\phi_2^2,\phi_3\}\}:} = \frac{1}{4} (i\Delta_{13})(i\Delta_{23}) :\phi_1^2 \phi_2^2 \phi_3:\, .
\end{equation}
Repeating this for every possible pair  of contractions from the list  $\{12,13,23\}$, we find that at the end
\begin{align}
\label{eq:3CommsWickContracted}
\begin{split}
[\cH_1,[\cH_2,\cH_3]] \longrightarrow
\frac{1}{4}\,(i\Delta_{13})(i\Delta_{23})\,:\phi_{1}^{2}\phi_{2}^{2}\phi_{3}:
\;+\;
\frac{1}{4}\,(i\Delta_{12})(i\Delta_{23})\,:\phi_{1}^{2}\phi_{2}\phi_{3}^{2}:\, . 
\end{split}
\end{align}
This is the final result for the tree-level contractions from the nested commutator%
\footnote{This result, when combined with $\theta$-functions, gives rise to \eqref{disapp}.}
$[\cH_1,[\cH_2,\cH_3]]$.

\subsubsection{One loop}
\label{sec:one-loop-Wick}

We now move to consider generic Wick contractions at one loop. One important consequence of this discussion will be that at one loop there is always one $\Delta^{(1)}$ function and hence the loop is cut. This fact will be used later in  Section~\ref{sec:MagnusLoopsFromFL} to establish a precise connection between tree-level and loop-level Magnus matrix elements.%
\footnote{This connection is reminiscent of the work of \cite{Caron-Huot:2010fvq} on causal observables, and specifically response functions.}

A one-loop diagram with $n$ vertices contains $n$ propagators. As before, we need to turn all commutators into anticommutators, and this produces $n-1$ contractions  
$i\Delta_{1a_1}\,i\Delta_{2a_2}\cdots\,i\Delta_{n{-}1a_{n-1}}$. The remaining contraction can be between any two points $i$ and $b_{i}\in\{i+1,\dots,n\}$, but note that the $\Delta_{ib_{i}}$ part will convert one anti-commutator back into a commutator giving a vanishing result after normal ordering. This means that the final contraction is forced to be $\Delta^{(1)}_{i b_{i}}$. A generic contraction at one loop will therefore look like
\begin{align}\label{eq:1LoopWickContractionRule}
    &\frac{N}{2^n} (i\Delta_{1a_1})\cdots(i\Delta_{n-1 a_{n-1}})\Delta^{(1)}_{ib_{i}} :\Big\{\cH_{1}^{(d_1)},\big\{\cH_{2}^{(d_2)},\cdots\{\cH_{n{-}1}^{(d_{n-1})},\cH_{n}^{(d_n)}\}\cdots\big\}\Big\}: \nn\\
&= \frac{N}{2} (i\Delta_{1a_1})\cdots(i\Delta_{n-1 a_{n-1}})
\Delta^{(1)}_{ib_{i}} :\left(\cH_{1}^{(d_1)} \cdots \cH_{n}^{(d_n)}\right): \, .
\end{align}
Focusing as an example on the structure $i\Delta_{13}\, i\Delta_{23}$ arising from   the one-loop contractions from the nested commutator $[\cH_1,[\cH_2,\cH_3]]$, we obtain 
\begin{align}
\label{eq:3CommsWick1LoopExample}
    \begin{split}
        \frac{1}{4}\,i\Delta_{13}\,i\Delta_{23}\,\Delta^{(1)}_{23}\, {:\phi_{1}^{2}\phi_{2}:}
 + \frac{1}{4}\,i\Delta_{13}\,i\Delta_{23}\,\Delta^{(1)}_{13}\,{:\phi_{1}\phi_{2}^{2}:}
 + \frac{1}{2}i\Delta_{13}\,i\Delta_{23}\,\Delta^{(1)}_{12}\,{:\phi_{1}\phi_{2}\phi_{3}:}\, .
    \end{split}
\end{align}
Now we make an important observation. As mentioned, the additional Wick contraction gives rise to a factor of $\Delta^{(1)}_{ij}$, which is symmetric in its two indices. But since it is symmetric, the ordering of the Wick-contracted fields $\phi_i$ and $\phi_j$ does not matter. This means that the tree-level answer \eqref{eq:3CommsWickContracted}, despite being normal-ordered, contains enough information to recover the full one-loop answer.

We can make this more precise by defining Wick-contraction-like rules for normal-ordered products. The rule is to consider all possible pairs of fields $\phi_i$ and $\phi_j$, with $i\neq j$, \textit{within} a normal-ordered product and replace them with $(1/2)\Delta^{(1)}_{ij}$ times the leftover fields.
The factor of $1/2$ is expected for any additional Wick contraction, as can be seen in \eqref{eq:1LoopWickContractionRule}, and it is explained in more detail in \eqref{eq:ContractionsLoopFactor}.
For example, we have
\begin{equation}
    {:\phi_{1}^{2}\phi_{2}^{2}\phi_{3}:} \to 2\Delta^{(1)}_{12}\,{:\phi_{1}\phi_{2}\phi_{3}:} + \Delta^{(1)}_{13}\,{:\phi_{1}\phi_{2}^{2}:} + \Delta^{(1)}_{23}\,{:\phi_{1}^{2}\phi_{2}:}\, .
\end{equation}
These rules explain the replacements we had discovered in Section~\ref{sec:oneloop} (see for example  \eqref{reppp1} and \eqref{eq:ReplacementsTriangle}).  
It is easy to check that these rules, which we refer to as \textit{\NOwick}, applied on the $i\Delta_{13}\,i\Delta_{23}$ term in the tree-level answer~\eqref{eq:3CommsWickContracted}, yield the piece \eqref{eq:3CommsWick1LoopExample} of the one-loop answer. Similarly, we can extract the complete one-loop answer from the tree-level expression \eqref{eq:3CommsWickContracted}:
\begin{align}
\label{eq:3CommsWick1LoopFull}
\begin{split}
& \frac{1}{4}\,i\Delta_{13}\,i\Delta_{23}\,\Delta^{(1)}_{23}\,\ :\phi_{1}^{2}\phi_{2}:
 + \frac{1}{4}\,i\Delta_{13}\,i\Delta_{23}\,\Delta^{(1)}_{13}\,:\phi_{1}\phi_{2}^{2}: + \frac{1}{2}\,i\Delta_{13}\,i\Delta_{23}\,\Delta^{(1)}_{12}\,:\phi_{1}\phi_{2}\phi_{3}: \\
 &
 + \frac{1}{4}\,i\Delta_{12}\,i\Delta_{23}\,\Delta^{(1)}_{23}\,:\phi_{1}^{2}\phi_{3}: + \frac{1}{2}i\Delta_{12}\,i\Delta_{23}\,\Delta^{(1)}_{13}\,:\phi_{1}\phi_{2}\phi_{3}:
 + \frac{1}{4}\,i\Delta_{12}\,i\Delta_{23}\,\Delta^{(1)}_{12}\,:\phi_{1}\phi_{3}^{2}:\, .
\end{split}
\end{align}
We have thus established a precise connection between tree and one-loop Magnus operator expressions which will be used later in Section~\ref{sec:MagnusLoopsFromFL} to relate tree and loop Magnus amplitudes.

A second important comment is in order: in the previous sections we discussed how to simplify the product of $\theta$ functions multiplying the nested commutators, $\theta_{12}\theta_{23}$ in this case, and write the integrand in terms of products of retarded propagators $\Delta^{R}_{ij}$ and cut functions  $\Delta^{(1)}_{ij}$. For example, we showed this in detail in the six-point tree-level case in Section~\ref{sect:6pttree}. Throughout those steps, the normal-ordered products of fields are just spectators. This means that the above rules to relate the tree-level and one-loop operators from the nested commutators are still valid even after the simplification of $\theta$ functions. As a corollary, since we know that the simplification of $\theta$ functions will always work at tree level\footnote{More specifically, we know the tree-level result can be expressed as a sum of graphs with retarded and advanced propagators only, and in order to achieve this all $\theta$ functions must be reabsorbed.} from the work of  \cite{Kim:2024svw}, and we know that the extra one-loop cut function  will always be the symmetric $\Delta^{(1)}_{ij}$, we conclude that the $\theta$ simplification will always work at one loop, thus giving a final manifestly Lorentz-invariant result written in terms of retarded and advanced propagators (in addition to the single cut $\Delta^{(1)}_{ij}$ function).

\subsubsection{Higher loops}
At higher loops we have to do even more contractions beyond those at  tree level. Once again, we need $i\Delta_{1a_1}\,i\Delta_{2a_2}\cdots\,i\Delta_{n{-}1a_{n-1}}$ to start with, and at $L$ loops we will need to do $L$ additional contractions. Let us do the new contractions two at a time. One option to preserve the anticommutators is to do two $\Delta^{(1)}$ contractions $\Delta^{(1)}_{ij}\Delta^{(1)}_{k\ell}$, or we could do two $i\Delta$ contractions $i\Delta_{ij}\,i\Delta_{ik}$ so that an anticommutator is flipped to a commutator and back.  Note that, in the latter case, the first argument of each $i\Delta$ should be the same, so they both affect the same (anti)commutator,
\begin{equation}
    \{\dots\{\cH_i^{(d_i)},\dots\}\dots\} \overset{i\Delta_{ij}}{\longrightarrow} \{\dots[\cH_i^{(d_i{+}1)},\dots]\dots\} \overset{i\Delta_{ik}}{\longrightarrow} \{\dots\{\cH_i^{(d_i{+}2)},\dots\}\dots\}\, .
\end{equation}
For $L=2m$ there can be  $1, 2, \ldots, m$ such pairs of $i\Delta$ contractions, resulting in $2m,2m-2,\dots,4,2,0$ $\Delta^{(1)}$ cuts. For an odd number of loops, $L=2m+1$, doing $m$ pairs of Wick contractions leaves one more to be done, which has to be a $\Delta^{(1)}$. This means that there are $2m+1,2m-1,\dots,5,3,1$ cuts. 

For example, at two loops there are two different possibilities,
\begin{align}\label{eq:2LoopWickContractionRule1}
    &\frac{N}{2^2} (i\Delta_{1a_1})\cdots(i\Delta_{n-1 a_{n-1}})\Delta^{(1)}_{ib_{i}}\Delta^{(1)}_{jb_{j}} :\left(\cH_{1}^{(d_1)} \cdots \cH_{n}^{(d_n)}\right):\, ,\quad \text{or} \\
    \label{eq:2LoopWickContractionRule2}
    &\frac{N}{2^2} (i\Delta_{1a_1})\cdots(i\Delta_{n-1 a_{n-1}})(i\Delta_{ib_{i}})(i\Delta_{ib'_{i}}) :\left(\cH_{1}^{(d_1)} \cdots \cH_{n}^{(d_n)}\right):\, ,
\end{align}
where $b'_i\in\{i+1,\dots,n\}$ like $b_i$, but may take on a different value. For concreteness we come back to the example of the $[\cH_1,[\cH_2,\cH_3]]$ nested commutator. The two-loop contractions we get in this case  are
\begin{align}
\label{eq:3CommsWick2LoopFull}
\begin{split}
& \frac{1}{4}\,i\Delta_{13}\,i\Delta_{23}\,\Delta^{(1)}_{12}\Delta^{(1)}_{23}\,\phi_{1}
+ \frac{1}{4}\,i\Delta_{12}\,i\Delta_{23}\,\Delta^{(1)}_{13}\Delta^{(1)}_{23}\,\phi_{1} 
+ \frac{1}{4}\,i\Delta_{13}\,i\Delta_{23}\,\Delta^{(1)}_{12}\Delta^{(1)}_{13}\,\phi_{2} \\[6pt]
& + \frac{1}{8}\,i\Delta_{12}\,i\Delta_{23}\,\bigl(\Delta^{(1)}_{13}\bigr)^{2}\,\phi_{2}
+ \frac{1}{8}\,i\Delta_{13}\,i\Delta_{23}\,\bigl(\Delta^{(1)}_{12}\bigr)^{2}\,\phi_{3}
+ \frac{1}{4}\,i\Delta_{12}\,i\Delta_{23}\,\Delta^{(1)}_{12}\Delta^{(1)}_{13}\,\phi_{3} \\[6pt]
&
+ \frac{1}{8}\,i\Delta_{13}\,i\Delta_{23}\,(i\Delta_{12})^{2}\,\phi_{3}
+ \frac{1}{8}\,i\Delta_{12}\,i\Delta_{23}\,(i\Delta_{13})^{2}\,\phi_{2} \, .
\end{split}
\end{align}
In this case, the first two lines contain two powers of the symmetric $\Delta^{(1)}_{ij}$, meaning both the additional contractions are symmetric. On the other hand, the last line contains only the antisymmetric $i\Delta_{ij}$.

Now we can generalise the \NOwick\ to the two-loop case: we can start from the tree-level normal-ordered products, identify two pairs in all possible ways and replace each with $(1/2)\Delta^{(1)}_{ij}$. For example we have
\begin{equation}
    {:\phi_{1}^{2}\phi_{2}^{2}\phi_{3}:} \to
    \frac{1}{2} \bigl(\Delta^{(1)}_{12}\bigr)^2 \phi_{3}
    + \Delta^{(1)}_{12}\Delta^{(1)}_{13} \phi_2
    + \Delta^{(1)}_{12}\Delta^{(1)}_{23} \phi_1\,.
\end{equation}
We can then apply these rules to \eqref{eq:3CommsWickContracted} to get the first two lines of \eqref{eq:3CommsWick2LoopFull}. However, we cannot get the last line in this way, since it contains antisymmetric contractions where the ordering of the two contracted fields matter, and therefore information about them cannot be recovered from the normal-ordered tree-level case.

In Section~\ref{sec:redoMurua} we have shown that we can use Murua's formula as a shortcut to compute coefficients of tree-level Magnus diagrams. Furthermore, we have seen that one-loop coefficients can also be obtained from Murua's formula, and this is precisely because one-loop operators are related to tree-level ones through the above normal-ordered contractions.

The steps outlined above rely only on the fact that $\Delta^{(1)}_{ij}$ is symmetric and therefore the order of the contracted fields does not matter. This is true not only in the example described above, but at all orders in the Magnus expansion. We conclude that the $(\Delta^{(1)}_{ij})^L$ part of the $L$-loop Magnus operator can always be obtained from the tree-level operator, for any number of vertices, using the \NOwick\ defined above. However,
terms containing fewer than $L$ powers of $\Delta^{(1)}_{ij}$ cannot be obtained from tree level -- for instance the last line of \eqref{eq:3CommsWick2LoopFull}.

\subsection{Magnus loops from forward limits}
\label{sec:MagnusLoopsFromFL}

In this section we show how some loop-level Magnus amplitudes can be straightforwardly obtained from Magnus amplitudes of lower-loop order, but with the same number of vertices. Namely, we will show that taking $\ell$ forward limits of the tree-level answer with $n$ vertices will produce the part of the $n$-vertex $\ell$-loop answer which is proportional to $(\Delta^{(1)})^\ell$. For $\ell = 1$ this gives the complete answer, since the entire one-loop matrix element is proportional to $\Delta^{(1)}$, as discussed in Section~\ref{sec:one-loop-Wick}. For $\ell>1$ in general there will be terms with $C<\ell$ cuts which cannot be obtained from trees, but nonetheless we will show that these can still be built using forward limits of other starting objects.

\subsubsection{Complete one-loop Magnus amplitudes from trees}

A generic term $\cT^{(0)}$ in the tree-level matrix element $\cA^{(0)}$ has the form
\begin{equation}
\label{eq:treegeneric}
    \cT^{(0)} = \bra{0} F(x_1, \ldots, x_n) :\phi_{a} \phi_{b} \prod_i \phi_{c_i}: \ket{\phi(p_1) \dots \phi(p_{n}) \phi(p_{n+1}) \phi(p_{n+2})}\, ,
\end{equation}
where $\phi_i = \phi(x_i)$ as usual, $F(x_1, \ldots ,x_n)$ is a product of retarded propagators $\Delta^R$ and $a,b,c_i = 1,\dots,n$. Note that, at tree level in $\phi^3$ theory, we have ${n+2}$ external legs for $n$ vertices. 
We have also separated out two fields $\phi_a$, $\phi_b$ in the normal-ordered product for reasons that will become clear in a moment. 

To make the connection to one loop, we  consider the following forward limit,
\begin{equation}
    \cT^{(0)}_{\rm FL} := \lim_{\substack{p_{n+1}\to p \\ \, \, \,  p_{n+2}\,  \to -p}} \cT^{(0)}\, , 
\end{equation}
in other words, we prepare to close up the last two legs into a loop  by assigning to them opposite momenta $p$ and $-p$.

Now, consider Wick contractions involving $\phi(p)$ and $\phi(-p)$. One such term will~be
\begin{equation}
    \cT^{(0)}_{{\rm FL}, ab} = \bra{0} F(x_1, \ldots, x_n) \wick{\c1\phi_{a} \c1 a^\dagger(\vec{p})} \wick{\c1\phi_{b} \c1 a^\dagger(-\vec{p})} :\prod_i \phi_{c_i}: \ket{\phi(p_1) \dots \phi(p_{n})}\, ,
\end{equation}
where%
\footnote{We define $\hat{d}^D p \coloneq d^Dp / (2\pi)^D$.}
\begin{align}
\begin{split}
    \wick{\c1\phi_{a} \c1 a^\dagger(\vec{p})} \wick{\c1\phi_{b} \c1 a^\dagger(-\vec{p})} &= \int \frac{\hat{d}^3p_a}{ 2 E(\vec{p}_a)}\frac{\hat{d}^3p_b}{2 E(\vec{p}_b)}\, e^{-ip_a\cdot x_a - i p_b\cdot x_b} \wick{\c1 a(\vec{p}_a) \c1 a^\dagger(\vec{p})} \wick{\c1 a(\vec{p}_b) \c1 a^\dagger (-\vec{p})} \\
    &= \int \hat{d}^3 p_a \hat{d}^3 p_b \, e^{-ip_a\cdot x_a - i p_b\cdot x_b} \hat\delta^{(3)}(\vec{p}_a-\vec{p}) \hat\delta^{(3)}(\vec{p}_b+\vec{p}) \\
    &= e^{-i p\cdot (x_a-x_b)}\, ,
\end{split}
\end{align}   
 yielding
\begin{equation}
    \cT^{(0)}_{{\rm FL}, ab} = \bra{0} e^{-i p\cdot (x_a-x_b)} F(x_1, \ldots, x_n) :\prod_i \phi_{c_i}: \ket{\phi(p_1) \dots \phi(p_{n})}\, .
\end{equation}
Next, we know we can turn the $(n+2)$-point (piece of a) tree-level matrix element \eqref{eq:treegeneric} into an $n$-point  (piece of a) one-loop matrix element  $\cT^{(1)}$ by  applying the \NOwick\ discussed in Section~\ref{sec:one-loop-Wick} and removing $\phi_{n+1}$ and $\phi_{n+2}$ from the external state.

To perform the   \NOwick\  we must 
 pick all possible pairs of fields $(\phi_i,\phi_j)$ within  the normal-ordered product $:\phi_{a} \phi_{b} \prod_i \phi_{c_i}:$ and replace them by $(1/2)\Delta^{(1)}_{ij}$. For example, we can pick the pair $(\phi_a,\phi_b)$ and get
\begin{equation}
\label{eq:1loopgeneric}
    \cT^{(1)}_{ab} = \frac{1}{2}\bra{0} \Delta^{(1)}_{a b} F(x_1, \ldots, x_n) :\prod_i \phi_{c_i}: \ket{\phi(p_1) \dots \phi(p_{n})}\, ,
\end{equation}
where
\begin{align}
\begin{split}
    \Delta^{(1)}_{a b} = \Delta^{(1)}(x_a-x_b) = \int\!\hat{d}^4 p \, e^{-ip\cdot(x_a-x_b)} \hat{\delta}(p^2-m^2)\, , 
\end{split}
\end{align}
such that \eqref{eq:1loopgeneric} becomes
\begin{equation}
    \cT^{(1)}_{ab} = \frac{1}{2} \int\! \hat{d}^4 p \, \hat{\delta}(p^2-m^2) \bra{0} e^{-ip\cdot(x_a-x_b)} F(x_1, \ldots, x_n) :\prod_i \phi_{c_i}: \ket{\phi(p_1) \dots \phi(p_{n})}\, .
\end{equation}
Therefore, we conclude that
\begin{equation}
    \cT^{(1)}_{ab} = \frac{1}{2} \int \hat{d}^4 p \, \hat{\delta}(p^2-m^2) \cT^{(0)}_{{\rm FL},ab}\, ,
\end{equation}
in other words the one-loop term considered is equal to the on-shell phase-space integral in $p$ of the forward limit of the tree-level term considered.

Now, to extend this statement to the full $\cT^{(0)}$ and $\cT^{(1)}$, we must  repeat the steps above for any choice of $\phi_a$ and $\phi_b$ in the normal-ordered product $:\phi_{a} \phi_{b} \prod_i \phi_{c_i}:$. However, in the $\cT^{(0)}$ case, swapping $a$ and $b$ relates two inequivalent terms, depending on which field contracts with $\phi(p)$ and which with $\phi(-p)$; 
on the other hand, to compute $\cT^{(1)}$ we must consider each unordered pair $(\phi_a,\phi_b)$ only once, and furthermore replace it with a single $(1/2)\Delta^{(1)}_{ab}$ (with this last factor of $1/2$ due to the fact that the Hadamard function is twice a cut propagator). This leads to a two-to-one mapping, so that 
\begin{equation}
\label{eq:treeto1loop}
    \cT^{(1)} = \frac{1}{4} \int \hat{d}^4 p \, \hat{\delta}(p^2-m^2) \cT^{(0)}_{\rm FL}\, .
\end{equation}
Note that $\cT^{(0)}_{\rm FL}$ will also produce tadpoles, i.e.~terms where $\phi(p)$ and $\phi(-p)$ are contracted at the same vertex $\phi_i^3$. On the other hand, tadpoles never appear in $\cT^{(1)}$ because all operators at a point $x_i$ are normal ordered. Therefore, \eqref{eq:treeto1loop} is valid once tadpoles are removed from the right-hand side.
One  example of this relation is
\begin{equation}
\begin{tikzpicture}[baseline={([yshift=-0.5ex]current bounding box.center)},thick,font=\small]
    \draw[massive] (0,0) -- (135:0.75) node[above left]{$p_3$};
    \draw[massive] (0,0) -- (-135:0.75) node[below left]{$p_2$};
    \draw[Rnew] (0,0) -- (1,0);
    \draw[massive] (1,0) -- (1,0.75) node[above]{$p_4$};
    \draw[Anew] (1,0) -- (2,0);
    \draw[massive] (2,0) -- (2,0.75) node[above]{$p_5$};
    \draw[Anew] (2,0) -- (3,0);
    \draw[massive] (3,0) -- ++(45:0.75)node[above right]{$p_6$};
    \draw[massive] (3,0) -- ++(-45:0.75) node[below right]{$p_1$};
    \path (2,0) -- (2,-1); 
    \draw[->] (4.25,0) -- ++(1.5,0) node[pos=0.5,above]{$\text{FL}_{12}$};
\end{tikzpicture}
\quad
\begin{tikzpicture}[baseline={([yshift=-0.5ex]current bounding box.center)},thick,scale=1.5,font=\small]
    \draw (0,0) -- ++(-0.5,0) node[left]{$p_4$};
    \draw (0,-1) -- ++(-0.5,0) node[left]{$p_3$};
    \draw (1,0) -- ++(0.5,0) node[right]{$p_5$};
    \draw (1,-1) -- ++(0.5,0) node[right]{$p_6$};
    \draw[Anew] (0,0) -- (1,0);
    \draw[Anew] (0,0) -- (0,-1);
    \draw[Anew] (1,0) -- (1,-1);
    \draw[cut] (0,-1) -- (1,-1);
\end{tikzpicture}\ ,
\end{equation}
where $p_1\to p$ and $p_2 \to -p$.

Finally, since \eqref{eq:treeto1loop} is valid for all terms $\cT^{(0)}$ and $\cT^{(1)}$ in the tree-level and one-loop matrix elements $\cA^{(0)}$ and $\cA^{(1)}$, we have established the relation for Magnus amplitudes
\begin{equation}
\boxed{
    \cA^{(1)} = \frac{1}{4} \int\!\hat{d}^4 p \, \hat{\delta}(p^2-m^2) \cA^{(0)}_{\rm FL} }\, ,
\end{equation}
still valid up to tadpoles, where $\cA^{(0)}_{\rm FL}$ is the $p_{n+1} {\to} p$, $p_{n+2} {\to} -p$ limit of the tree-level matrix element $\cA^{(0)}$.

\subsubsection{Beyond one loop}
We can also repeat the same steps for $\ell$ loops. Consider the $\ell$-loop Magnus matrix element $\cA^{(\ell)}$, and define
\begin{equation}
    \cA^{(\ell,C)} \coloneq \cA^{(\ell)} \big|_{(\Delta^{(1)})^C}\, ,
\end{equation}
i.e.~$\cA^{(\ell,C)}$ is the part of the $\ell$-loop matrix element proportional to $(\Delta^{(1)})^C$. In particular, we already know from the \NOwick\  that $\cA^{(\ell,\ell)}$ is the only part we may hope to reproduce from the tree-level matrix element. We then  consider $\ell$ forward limits, such that each pair of fields $(\phi(p_k),\phi(-p_k))$, with $k = 1,\ldots,\ell$, produces a $\Delta^{(1)}$ term upon performing the phase-space integral $\int\! \hat{d}^4  p_k \hat{\delta}(p_k^2-m^2)$. Repeating all the steps above, we then get
\begin{equation}
\label{eq:Lloopsforwlim}
    \boxed{\cA^{(\ell,\ell)} = \frac{1}{\ell! \, 2^{2 \ell}} 
    \int \prod_{k=1}^\ell\left(\hat{d}^4 p_k\,  \hat{\delta}(p_k^2-m^2) \right) \cA^{(0,0)}_{\ell-{\rm FL}} }
    \,,
\end{equation}
where the prefactor can be derived as follows: $\ell$ normal-ordered contractions each produce $(1/2) \Delta^{(1)}_{ij}$; an additional $2^\ell$ factor compensates an  overcounting due to each pair $(\phi(p_k),\phi(-p_k))$ being counted twice; and the $\ell!$ compensates an overcounting due to permuting all $\ell$ such pairs among themselves. Note that here $\cA^{(0)}_{\ell-{\rm FL}}$ is obtained by taking the forward limit of $\cA^{(0)}$ for each $p_k$. We also note that \eqref{eq:Lloopsforwlim} is reminiscent of the relations between trees and loops discussed in \cite{Caron-Huot:2010fvq} in the context of the in-in formalism. 

Instead of starting from tree-level Magnus amplitudes, we could have started from a $C$-cut piece of an $L$-loop amplitude $\cA^{(L,C)}$ and taken a number of forward limits as described above. In this case, we get a similar formula:
\begin{equation}\label{eq: genForwardLimit}
\boxed{
    \cA^{(L+\ell,C+\ell)} = \frac{2^{2C}C!}{2^{2(C+\ell)}(C+\ell)!} \int \prod_{k=1}^{\ell}\left(\hat{d}^4 p_k \, \hat{\delta}(p_k^2-m^2)\right) \cA^{(L,C)}_{\ell-\text{FL}} }\ .
\end{equation}
 This result links, for example, the zero-cut part of a two-loop amplitude, $\cA^{(2,0)}$, with the one-cut part of a three-loop amplitude, $\cA^{(3,1)}$. Both formulae have been tested for $\phi^3$-theory Magnus amplitudes with up to six vertices and four loops. This web of relations is illustrated in Figure~\ref{fig:MagnusCutStructureFL}.

Summarising, as a consequence of these forward limit relations, all $\phi^3$ Magnus amplitudes at any loop order and multiplicity can be built out of the zero-cut piece of $2n$-loop Magnus amplitudes, $\cA^{(2n,0)}$. In particular, the tree-level diagrams can be obtained directly using the Murua coefficients reviewed in Section~\ref{sec:redofromMurua-trees}, whereas the zero-cut $2n$-loop ($n{>}0$) pieces still require computations to absorb the $\theta$-functions as described in detail in Sections~\ref{sec:treelevel}~and~\ref{sec:oneloop}. 

\begin{figure}
\centering
\begin{tikzpicture}
\def\cOne{cadmiumgreen}
\def\cTwo{amber}
\def\cThree{RoyalBlue}
    \draw[fill=\cOne!20!White] (0,0) rectangle (1,1) node[pos=0.5]{4C};
    \draw[fill=\cTwo!40!White] (1,0) rectangle (2,1) node[pos=0.5]{2C};
    \draw[text=black,fill=\cThree] (2,0) rectangle (3,1) node[pos=0.5]{0C};

    \draw[fill=\cOne!40!White] (0,1) rectangle (1,2) node[pos=0.5]{3C};
    \draw[fill=\cTwo!60!White] (1,1) rectangle (2,2) node[pos=0.5]{1C};

    \draw[fill=\cOne!60!White] (0,2) rectangle (1,3) node[pos=0.5]{2C};
    \draw[text=black,fill=\cTwo] (1,2) rectangle (2,3) node[pos=0.5]{0C};

    \draw[fill=\cOne!80!White] (0,3) rectangle (1,4) node[pos=0.5]{1C};

    \draw[text=black,fill=\cOne] (0,4) rectangle (1,5) node[pos=0.5]{0C};

    \draw[very thick,red,fill=none] (2,0) rectangle (3,1);
    \draw[very thick,red,fill=none] (1,2) rectangle (2,3);
    \draw[very thick,red,fill=none] (0,4) rectangle (1,5);
    \path (-1,0) rectangle (0,1) node[pos=0.5]{4L:};
    \path (-1,1) rectangle (0,2) node[pos=0.5]{3L:};
    \path (-1,2) rectangle (0,3) node[pos=0.5]{2L:};
    \path (-1,3) rectangle (0,4) node[pos=0.5]{1L:};
    \path (-1,4) rectangle (0,5) node[pos=0.5]{0L:};
\end{tikzpicture}
\caption{Cut structure of Magnus amplitudes. At $2n$-loops there are $0,2,\dots,2n$ cuts and at $(2n{+}1)$-loops there are $1,3,\dots,2n{+}1$ cuts. Blocks in the same column are related by forward limits, meaning that everything can be obtained from the zero-cut parts of $2n$-loop Magnus amplitudes (highlighted in red).}
\label{fig:MagnusCutStructureFL}
\end{figure}

\subsubsection{Murua for 
higher-loop diagrams}\label{sec: MuruaForHigherLoops}
The Murua coefficients at tree level can be derived from Murua's formula. The forward limit relation above then allows us to derive the Murua coefficients of $L$-loop $L$-cut diagrams (the first column in Figure \ref{fig:MagnusCutStructureFL}).
We find that the exact relation is:
\begin{equation}\label{eq: MuruaRelation}
    \omega(\tau^{(L,L)})= \frac{\omega(\tau^{(0,0)})}{2^L}\,,
\end{equation}
where $\tau^{(L,L)}$ is an $L$-loop $L$-cut diagram (with or without external legs) and $\tau^{(0,0)}$ is the corresponding tree diagram with the $L$ cuts removed. This reduces to \eqref{eq:oneLooprelationship} at one loop. The factor of $2^L$ above comes from the fact that each cut appears as $\Delta^{(1)}/2$ in the Wick contraction rules \eqref{eq:OneWickContraction}. Since the Wick contractions above always build on tree level (which is fully connected) when we remove the cuts from a digram it remains fully connected. As an example of the above
\begin{equation}
    \omega\left(\begin{tikzpicture}[baseline={([yshift=-0.5ex]current bounding box.center)},thick] 
    \draw[massive] (0,0) -- (0.75,0);
    \draw[Rnew] (0.75,0) -- (0.75,1);
    \draw[Anew] (0.75,0) -- (1.75,0);
    \draw[Rnew] (1.75,0) -- (1.75,1);
    \draw[massive] (1.75,0) -- (2.5,0);
    \draw[out=-45,in=-135,cut] (0.75,1) to (1.75,1);
    \draw[out=135,in=45,cut]   (1.75,1) to (0.75,1);
    \path (2,0) -- (2,-0.5); 
    \end{tikzpicture}\right)
    =\frac{\omega\left(\begin{tikzpicture}[baseline={([yshift=-.5ex]current bounding box.center)},thick]
        \draw[causArrow] (0,0) -- (0.8,0);
        \draw[causArrow] (0,0) -- (0.4,-0.5);
        \draw[causArrow] (0.4,-0.5) -- (0.8,-0.5);
        \draw[fill=white] (0,0) circle(3pt);
        \draw[fill=white] (0.8,0) circle(3pt);
        \draw[fill=white] (0.4,-0.5) circle(3pt);
        \draw[fill=white] (0.8,-0.5) circle(3pt);
    \end{tikzpicture}\right)}{2^2}=\frac{\frac{1}{12}}{4}=\frac{1}{48}\,,
\end{equation}
where, as stated previously, the Murua coefficient $\omega$ depends only on the graph with the external legs removed.
The prefactor of a specific $L$-loop $L$-cut graph for an $N$-matrix element is the product of a symmetry factor and the Murua coefficient, see the general formula \eqref{eq:MuruaFormulaMatrixElement}.
At two loops and above, $N$-matrix graphs can have \textit{non-trivial symmetry factors}.
Recall the symmetry factor is calculated in the same was as the usual Dyson expansion (keeping external legs fixed), except that one must also preserve the propagator prescriptions. For example, 
\begin{equation}
\sigma\left(\begin{tikzpicture}[baseline={([yshift=-0.5ex]current bounding box.center)},thick] 
    \draw[massive] (0,0) -- (0.75,0);
    \draw[Rnew] (0.75,0) -- (0.75,1);
    \draw[Anew] (0.75,0) -- (1.75,0);
    \draw[Rnew] (1.75,0) -- (1.75,1);
    \draw[massive] (1.75,0) -- (2.5,0);
    \draw[out=-45,in=-135,cut] (0.75,1) to (1.75,1);
    \draw[out=135,in=45,cut]   (1.75,1) to (0.75,1);
    \path (2,0) -- (2,-0.5); 
    \end{tikzpicture}\right)=2!
    \,,\quad
    \sigma\left(\begin{tikzpicture}[baseline={([yshift=-0.5ex]current bounding box.center)},thick] 
    \draw[massive] (0,0) -- (0.75,0);
    \draw[massive] (0.75,0) -- (1.5,0);
    \draw[Rnew] (0.75,0) -- (0.75,0.70);
    \draw[out=-45,in=-135,cut] (0.75/3,1.4) to (0.75+1.5/3,1.4);
    \draw[out=135,in=45,cut]   (0.75+1.5/3,1.4) to (0.75/3,1.4);
    \draw[Rnew] (0.75,0.70) -- (0.75/3,1.4);
    \draw[Rnew] (0.75,0.70) -- (0.75+1.5/3,1.4);
    \path (1,0) -- (1,-0.5); 
    \end{tikzpicture}\right)= (2!)^2\,.
\end{equation}
Note that a factor of 2 in each example appears from swapping the two cut propagators. A more detailed discussion of symmetry factors is given Appendix~\ref{app: SymmFactors}. In summary, we can apply Murua's formula to derive the coefficients of any $L$-loop $L$-cut diagram.

Beyond the maximal cut pieces of Magnus amplitudes we can also consider relations between terms with fewer cuts (the second column and beyond in Figure~\ref{fig:MagnusCutStructureFL}). Again these terms are related to each other using the forward limit \eqref{eq: genForwardLimit}. This, in turn, leads to an observed relation between their Murua coefficients, which generalises \eqref{eq: MuruaRelation},
\begin{equation}\label{eq: MuruaRelationBest}\boxed{
    \omega(\tau^{(L,C)})= \frac{\omega(\tau^{(L-C,0)})}{2^C}}\,,
\end{equation}
where $\tau^{(L,C)}$ is an $L$-loop, $C$-cut diagram and $\tau^{(L-C,0)}$ is an $(L-C)$-loop zero-cut diagram (the terms in the red boxes in Figure~\ref{fig:MagnusCutStructureFL}). We have checked this relation at up to  six vertices and four loops in $\phi^3$-theory, and we conjecture that it holds to all loop orders. 

Note that Murua's formula only works for tree-level diagrams and thus to derive the coefficients of $L$-loop zero-cut diagrams we must perform the Magnus expansion directly. After this we can apply \eqref{eq: MuruaRelationBest} to obtain the coefficients of any other diagram.

\subsubsection{External leg bubbles and pathological graphs}\label{sec: BadGraphs}
Up to this point we have focused on the diagrammatics of the Magnus expansion without worrying about whether the graphs we generate are forbidden in some way. For example the bubble in \eqref{eq:bub12} is on an external leg, and hence one of the retarded propagators is strictly on-shell, giving a divergence. The problem of external bubbles also appears in the Dyson expansion where they are removed by LSZ reduction. In practice when computing usual $T$-matrix elements we can simply drop these terms by hand. For Magnus amplitudes we can adopt the same approach and drop any diagram with a bubble on an external leg. 

However, we have seen previously in this section that we can obtain pieces of higher-loop higher-cut amplitudes from lower-loop lower-cut pieces, through a forward limit. This relation applies when we include all diagrams, including external leg bubbles. Thus a natural question arises: what happens to bubbles on external legs when we take a forward limit? These can generate graphs with no external bubbles \textit{but which are always pathological}. As an example consider the following one-loop external bubble and its forward limit 
\begin{equation}
\begin{tikzpicture}[baseline={([yshift=-0.5ex]current bounding box.center)},thick]
    \draw (0,0) -- (135:0.75) node[left, above]{$p_3$};;
    \draw (0,0) -- (-135:0.75) node[left, below]{$p_2$};
    \draw (1,0) -- ++(0,0.75)  node[above]{$p_4$};
    \draw (2,0) -- ++(0,0.75)  node[above]{$p_5$};
    \draw (4,0) -- ++(0.75,0) node[right]{$p_1$};
    \draw[Rnew] (0,0) -- (1,0);
    \draw[Rnew] (1,0) -- (2,0);
    \draw[Rnew] (2,0) -- (3,0);
    \draw[Anew] (3,0) arc(180:0:0.5);
    \draw[cut=2] (3,0) arc(-180:0:0.5);
    \draw[->] (5.75,0) -- ++(1.5,0) node[pos=0.5,above]{$\text{FL}_{12}$};
\begin{scope}[shift={(8,-0.8)}]
    \draw (60:2) --++(0,0.75) node[above]{$p_4$};
    \draw (0,0) -- ++(-150:0.75) node[left, below]{$p_3$};
    \draw (2,0) -- ++(-30:0.75) node[right, below]{$p_5$};
    \draw[Rnew] (0,0) -- (60:2);
    \draw[Rnew] (60:2) --++(-60:2);
    \draw[cut=2] (0,0) -- (0.5,0);
    \draw[Rnew] (2,0) -- (1.5,0);
    \draw[Rnew] (0.5,0) arc(180:0:0.5);
    \draw[cut=2] (0.5,0) arc(-180:0:0.5);
    \draw (0.25,0) node[right, below]{$\ell$};
\end{scope}
\end{tikzpicture}
\end{equation}
Putting aside the forward limit, the two-loop triangle diagram above \textit{is} na\"ively generated from the Magnus expansion. Indeed the same diagram topology will appear in the Dyson expansion except with Feynman propagators. However, the triangle suffers from the same divergence as the corresponding tree-level graph since the loop integrand will contain
\begin{equation}
    \hat{\delta}(\ell^2-m^2)\frac{1}{\ell^2-m^2 +i\vareps\, \text{sgn}(\ell^0)}= \hat{\delta}(\ell^2-m^2) \frac{1}{0}\,.
\end{equation}
The triangle could also have been generated from the forward limit of a one-loop diagram without an external bubble (just by opening the other cut).

In summary, there are two types of divergent contribution that we believe should be dropped from Magnus amplitudes: the first are bubbles on external legs, familiar from the Dyson expansion; and  the second are divergent loop integrals which can appear from the forward limit of loops on external legs.

We do not yet have a general argument as to why these pathological loop integrals should always be dropped (\`a la LSZ for the Dyson expansion) other than the simple fact that they are divergent and appear from the forward limit of external leg bubbles. We leave this, and a more detailed analysis of LSZ reduction of $N$-matrix elements, to later work. 

\section{Classical limit}
\label{sec:classlim}
Here we present a conjecture for the classical limit of Magnus matrix elements. Namely, we find evidence that the $(\Delta^{(1)})^L$ part of an $L$-loop matrix element, easily obtained from the trees through \eqref{eq:Lloopsforwlim}, contains all the classical information.

To this end, we begin  by considering a scalar quantum field theory appropriate to discuss classical limits,
\begin{equation}
\label{eq:toyclassQFT}
    S_{\rm QFT} = \int\!d^4x \left( \frac{1}{2} 
    (\partial_\mu \phi)( \partial^\mu \phi) - \frac{m^2}{2} \phi^2 + \frac{1}{2} (\partial_\mu h)( \partial^\mu h) + \frac{\lambda}{2} \phi^2 h \right) ,
\end{equation}
where  a massive scalar $\phi$ interacts with a massless scalar $h$.
This theory is a useful toy model to study massive scalars coupled to electromagnetism or gravity. This is because all diagrams in this model also exist in the latter two theories, where the massless scalar $h$ is simply replaced by the photon $A_\mu$ or the graviton $h_{\mu\nu}$, and all three share the same Murua coefficients.%
\footnote{Note that the converse is not true, for example there are diagrams in gravity that involve the three-graviton vertex and hence have no analogue in  \eqref{eq:toyclassQFT}. This could be fixed by adding an $h^3$ vertex to the toy model, but this  goes beyond the scope of this work.}
The classical limit then focuses on a regime where the external momenta are of $\cO(1)$ while the transferred momentum $q$ is of $\cO (\hbar)$. The momenta of the massless mediators are also of $\cO(\hbar)$ so that their wavenumbers are of $\cO(1)$.

It is well-understood that the classical limit of amplitudes from \eqref{eq:toyclassQFT} agrees with tree-level amplitudes from the following worldline theory \cite{Capatti:2024bid}, 
\begin{equation}
\label{eq:toyclassWQFT}
    S_{\rm WQFT} = \frac{1}{2}\int\!d^4x  (\partial_\mu h)( \partial^\mu h) - \frac{m}{2} \int\!d\tau \left(\dot{x}^{2}+\lambda h(x)\right) .
\end{equation}
Here  $x^\mu(\tau)= b^\mu + v^\mu \tau + z^\mu(\tau)$, where $b^\mu$ and $v^\mu$ are constant vectors acting as a background, whereas $z^\mu(\tau)$ are the worldline excitations, i.e.~the propagating degrees of freedom.%
\footnote{See \cite{Damgaard:2023vnx, Ajith:2024fna,Du:2024rkf} for  further work exploring the connection  between field theory amplitudes and  the worldline formalism for scalars coupled to electromagnetism and gravity.}

Let us focus on the two-to-two scattering of massive scalars $\phi$. In this case, classical observables such as the scattering angle  can be computed from the radial action $I_r$, which can be  obtained from the four-point matrix elements of the $N$-operator~\cite{Damgaard:2021ipf}.
Namely,  
the radial action from \eqref{eq:toyclassQFT} is given by 
\begin{equation}
\label{eq:toyradialQFT}
    I_r = \lim_{\hbar\to 0}\ \bra{\phi(p_1')\phi(p_2')} N_{\rm QFT} \ket{\phi(p_1)\phi(p_2)} , 
\end{equation}
while the radial action from
\eqref{eq:toyclassWQFT} is, {\it mutatis mutandis} \cite{Kim:2024svw,Haddad:2025cmw}, 
\begin{equation}
\label{eq:toyradialWQFT}
    I_r = \lim_{\hbar\to 0}\ {}_{\text{WQFT}}\bra{0} N_{\rm WQFT} \ket{0}_{\text{WQFT}} \, .
\end{equation} 
Note that  $N_{\rm QFT}$ and $N_{\rm WQFT}$ are the 
$N$ operators of the two respective theories.

Now we can consider the classical contributions from both theories in more detail. In \eqref{eq:toyradialWQFT}, the classical part is simply given by tree-level diagrams in the propagating fields $h(x)$ and $z^\mu(\tau)$ (see  e.g.~\cite{Haddad:2025cmw}). For example, we can list the classical diagrams at one loop,
\begin{equation}
\label{eq:classdiagWQFT}
\begin{tikzpicture}[baseline={([yshift=-0.5ex]current bounding box.center)},thick,scale=1.4]
    \draw[dashed] (0,0) -- ++(-0.5,0);
    \draw[dashed] (0,-1) -- ++(-0.5,0);
    \draw[dashed] (1,0) -- ++(0.5,0);
    \draw[dashed] (1,-1) -- ++(0.5,0);
    \draw[causArrow] (0,0) -- (1,0);
    \draw[vector] (0,0) -- (0,-1);
    \draw[vector] (1,0) -- (1,-1);
    \draw[dashed] (0,-1) -- (1,-1);
    \draw[fill=black] (0,0) circle(1pt);
    \draw[fill=black] (1,0) circle(1pt);
    \draw[fill=black] (0,-1) circle(1pt);
    \draw[fill=black] (1,-1) circle(1pt);
\end{tikzpicture}
,\quad
\begin{tikzpicture}[baseline={([yshift=-0.5ex]current bounding box.center)},thick,scale=1.4]
    \draw[dashed] (0,0) -- ++(-0.5,0);
    \draw[dashed] (0,-1) -- ++(-0.5,0);
    \draw[dashed] (1,0) -- ++(0.5,0);
    \draw[dashed] (1,-1) -- ++(0.5,0);
    \draw[causArrowR] (0,0) -- (1,0);
    \draw[vector] (0,0) -- (0,-1);
    \draw[vector] (1,0) -- (1,-1);
    \draw[dashed] (0,-1) -- (1,-1);
    \draw[fill=black] (0,0) circle(1pt);
    \draw[fill=black] (1,0) circle(1pt);
    \draw[fill=black] (0,-1) circle(1pt);
    \draw[fill=black] (1,-1) circle(1pt);
\end{tikzpicture}
,\quad
\begin{tikzpicture}[baseline={([yshift=-0.5ex]current bounding box.center)},thick,scale=1.4]
    \draw[dashed] (0,0) -- ++(-0.5,0);
    \draw[dashed] (0,-1) -- ++(-0.5,0);
    \draw[dashed] (1,0) -- ++(0.5,0);
    \draw[dashed] (1,-1) -- ++(0.5,0);
    \draw[dashed] (0,0) -- (1,0);
    \draw[vector] (0,0) -- (0,-1);
    \draw[vector] (1,0) -- (1,-1);
    \draw[causArrow] (0,-1) -- (1,-1);
    \draw[fill=black] (0,0) circle(1pt);
    \draw[fill=black] (1,0) circle(1pt);
    \draw[fill=black] (0,-1) circle(1pt);
    \draw[fill=black] (1,-1) circle(1pt);
\end{tikzpicture}
,\quad
\begin{tikzpicture}[baseline={([yshift=-0.5ex]current bounding box.center)},thick,scale=1.4]
    \draw[dashed] (0,0) -- ++(-0.5,0);
    \draw[dashed] (0,-1) -- ++(-0.5,0);
    \draw[dashed] (1,0) -- ++(0.5,0);
    \draw[dashed] (1,-1) -- ++(0.5,0);
    \draw[dashed] (0,0) -- (1,0);
    \draw[vector] (0,0) -- (0,-1);
    \draw[vector] (1,0) -- (1,-1);
    \draw[causArrowR] (0,-1) -- (1,-1);
    \draw[fill=black] (0,0) circle(1pt);
    \draw[fill=black] (1,0) circle(1pt);
    \draw[fill=black] (0,-1) circle(1pt);
    \draw[fill=black] (1,-1) circle(1pt);
\end{tikzpicture}
\end{equation}
where the dotted lines represent the background worldlines, the solid lines are $z^\mu$ propagators and the wavy lines are $h$ propagators. The  direction of the arrows  determines whether the propagators are retarded or advanced. Note that we omitted all arrows on the $h$ lines since  in this case they cannot go on shell, and hence the propagator prescription is irrelevant.%
\footnote{A  delta function of the form $\delta (v\cdot \ell)$, arises from the dotted vertices, with  $\ell$ being  the momentum of the exchanged $h$ particle. Going to the rest frame of $v$ one gets $\ell_0=0$ and hence $\ell$ can never go on shell.}

Now we can consider the one-loop Magnus diagrams contributing to \eqref{eq:toyradialQFT}. We know from Section~\ref{sec:one-loop-Wick} 
that one-loop matrix elements  always have one cut, so  we can first consider diagrams with a cut $\phi$ propagator,
\begin{align}
\label{eq:classdiagQFT}
&
\begin{tikzpicture}[baseline={([yshift=-0.5ex]current bounding box.center)},thick,scale=1.3,font=\footnotesize]
    \draw (0,0) -- ++(-0.5,0) node[above]{$p_1$};
    \draw (0,-1) -- ++(-0.5,0) node[below]{$p_2$};
    \draw (1,0) -- ++(0.5,0) node[above]{$p_1'$};
    \draw (1,-1) -- ++(0.5,0) node[below]{$p_2'$};
    \draw[causArrow] (0,0) -- (1,0);
    \draw[vector] (0,0) -- (0,-1);
    \draw[vector] (1,0) -- (1,-1);
    \draw[cut] (0,-1) -- (1,-1);
    \draw[causArrow] (-0.2,-0.8) -- (-0.2,-0.2) node[midway, left]{$\ell$};
\end{tikzpicture}
,\quad
\begin{tikzpicture}[baseline={([yshift=-0.5ex]current bounding box.center)},thick,scale=1.3,font=\footnotesize]
    \draw (0,0) -- ++(-0.5,0) node[above]{$p_1$};
    \draw (0,-1) -- ++(-0.5,0) node[below]{$p_2$};
    \draw (1,0) -- ++(0.5,0) node[above]{$p_1'$};
    \draw (1,-1) -- ++(0.5,0) node[below]{$p_2'$};
    \draw[causArrowR] (0,0) -- (1,0);
    \draw[vector] (0,0) -- (0,-1);
    \draw[vector] (1,0) -- (1,-1);
    \draw[cut] (0,-1) -- (1,-1);
    \draw[causArrow] (-0.2,-0.8) -- (-0.2,-0.2) node[midway, left]{$\ell$};
\end{tikzpicture}
,\quad
\begin{tikzpicture}[baseline={([yshift=-0.5ex]current bounding box.center)},thick,scale=1.3,font=\footnotesize]
    \draw (0,0) -- ++(-0.5,0) node[above]{$p_1$};
    \draw (0,-1) -- ++(-0.5,0) node[below]{$p_2$};
    \draw (1,0) -- ++(0.5,0) node[above]{$p_1'$};
    \draw (1,-1) -- ++(0.5,0) node[below]{$p_2'$};
    \draw[cut] (0,0) -- (1,0);
    \draw[vector] (0,0) -- (0,-1);
    \draw[vector] (1,0) -- (1,-1);
    \draw[causArrow] (0,-1) -- (1,-1);
\end{tikzpicture}
,\quad
\begin{tikzpicture}[baseline={([yshift=-0.5ex]current bounding box.center)},thick,scale=1.3,font=\footnotesize]
    \draw (0,0) -- ++(-0.5,0) node[above]{$p_1$};
    \draw (0,-1) -- ++(-0.5,0) node[below]{$p_2$};
    \draw (1,0) -- ++(0.5,0) node[above]{$p_1'$};
    \draw (1,-1) -- ++(0.5,0) node[below]{$p_2'$};
    \draw[cut] (0,0) -- (1,0);
    \draw[vector] (0,0) -- (0,-1);
    \draw[vector] (1,0) -- (1,-1);
    \draw[causArrowR] (0,-1) -- (1,-1);
\end{tikzpicture}
\nn\\
&
\begin{tikzpicture}[baseline={([yshift=-0.5ex]current bounding box.center)},thick,scale=1.3,font=\footnotesize]
    \draw (0,0) -- ++(-0.5,0) node[above]{$p_1$};
    \draw (0,-1) -- ++(-0.5,0) node[below]{$p_2$};
    \draw (1,0) -- ++(0.5,0) node[above]{$p_1'$};
    \draw (1,-1) -- ++(0.5,0) node[below]{$p_2'$};
    \draw[causArrow] (0,0) -- (1,0);
    \draw[vector] (0,0) -- (1,-1);
    \draw[white, fill=white, line width=3] (0.4,-0.4) -- (0.6,-0.6);
    \draw[vector] (0,-1) -- (1,0);
    \draw[cut] (0,-1) -- (1,-1);
    \draw[causArrow] (0,-0.8) -- ++(45:0.4) node[midway, xshift=-0.2cm, yshift=0.2cm]{$\ell$};
\end{tikzpicture}
,\quad
\begin{tikzpicture}[baseline={([yshift=-0.5ex]current bounding box.center)},thick,scale=1.3,font=\footnotesize]
    \draw (0,0) -- ++(-0.5,0) node[above]{$p_1$};
    \draw (0,-1) -- ++(-0.5,0) node[below]{$p_2$};
    \draw (1,0) -- ++(0.5,0) node[above]{$p_1'$};
    \draw (1,-1) -- ++(0.5,0) node[below]{$p_2'$};
    \draw[causArrowR] (0,0) -- (1,0);
    \draw[vector] (0,0) -- (1,-1);
    \draw[white, fill=white, line width=3] (0.4,-0.4) -- (0.6,-0.6);
    \draw[vector] (0,-1) -- (1,0);
    \draw[cut] (0,-1) -- (1,-1);
    \draw[causArrow] (0,-0.8) -- ++(45:0.4) node[midway, xshift=-0.2cm, yshift=0.2cm]{$\ell$};
\end{tikzpicture}
,\quad
\begin{tikzpicture}[baseline={([yshift=-0.5ex]current bounding box.center)},thick,scale=1.3,font=\footnotesize]
    \draw (0,0) -- ++(-0.5,0) node[above]{$p_1$};
    \draw (0,-1) -- ++(-0.5,0) node[below]{$p_2$};
    \draw (1,0) -- ++(0.5,0) node[above]{$p_1'$};
    \draw (1,-1) -- ++(0.5,0) node[below]{$p_2'$};
    \draw[cut] (0,0) -- (1,0);
    \draw[vector] (0,0) -- (1,-1);
    \draw[white, fill=white, line width=3] (0.4,-0.4) -- (0.6,-0.6);
    \draw[vector] (0,-1) -- (1,0);
    \draw[causArrow] (0,-1) -- (1,-1);
\end{tikzpicture}
,\quad
\begin{tikzpicture}[baseline={([yshift=-0.5ex]current bounding box.center)},thick,scale=1.3,font=\footnotesize]
    \draw (0,0) -- ++(-0.5,0) node[above]{$p_1$};
    \draw (0,-1) -- ++(-0.5,0) node[below]{$p_2$};
    \draw (1,0) -- ++(0.5,0) node[above]{$p_1'$};
    \draw (1,-1) -- ++(0.5,0) node[below]{$p_2'$};
    \draw[cut] (0,0) -- (1,0);
    \draw[vector] (0,0) -- (1,-1);
    \draw[white, fill=white, line width=3] (0.4,-0.4) -- (0.6,-0.6);
    \draw[vector] (0,-1) -- (1,0);
    \draw[causArrowR] (0,-1) -- (1,-1);
\end{tikzpicture}
\end{align}
%
%
In this case the cut line represents a $\Delta^{(1)}$ and the arrow direction distinguishes between retarded and advanced propagators. As before, we have omitted arrows on $h$ propagators since in this case too one can show they cannot go on shell. The diagrams above look very similar to their worldline counterparts, and in fact we find that they agree in the $\hbar \to 0$ limit. More specifically, each box and crossed-box pair of diagrams in the columns of \eqref{eq:classdiagQFT} reproduces the worldline diagram in the corresponding column of \eqref{eq:classdiagWQFT}.

Let us pause here and explain directly why hyper-classical contributions from \eqref{eq:classdiagQFT} vanish due to the Magnus expansion. We will combine the first two diagrams in  the first line  of \eqref{eq:classdiagQFT} with the first two diagrams in its   second line. The momentum transfer is $q =p_1^\prime - p_1$ and due to the on-shell conditions we have $p_1 \Cdot q \sim q^2$. In the classical limit we scale $(q, \ell) \to (\hbar q, \hbar \ell)$ so that the scalar propagators in the first two diagrams of \eqref{eq:classdiagQFT} become (keeping only the leading term and assuming $p_1^0>0$)
\begin{align}
\frac{1}{(p_1+\hbar\ell)^2 -m_1^2 \pm i \varepsilon (p_1^0 + \hbar \ell^0)} \Rightarrow
\frac{1}{2 \hbar p_1 \cdot \ell  \pm i \varepsilon} \, ,
\end{align}
while the scalar propagators in the corresponding crossed diagrams become 
\begin{align}
\frac{1}{(p_1+\hbar q-\hbar\ell)^2 -m_1^2 \pm i \varepsilon (p_1^0 +\hbar q^0- \hbar \ell^0)} \Rightarrow
\frac{1}{-2 \hbar p_1 \cdot \ell  \pm i \varepsilon} \, .
\end{align}
Combining the integrands of the four diagrams and setting  $\hbar=1$ we get
\begin{align}
\label{killhyper}
\frac{\delta(2 p_2 \cdot \ell) }{2 \ell^2 (q-\ell)^2} \left[ \frac{1}{p_1 \cdot \ell  + i \varepsilon} + \frac{1}{-p_1 \cdot \ell + i \varepsilon} + \frac{1}{p_1 \cdot \ell  - i \varepsilon} + \frac{1}{-p_1 \cdot \ell  - i \varepsilon}\right] = 0 \, ,
\end{align}
showing the absence of hyper-classical terms. This should be contrasted with a computation using the Feynman prescription where the sum in square brackets in \eqref{killhyper} would turn into
\begin{equation}
\left[ \frac{1}{p_1 \cdot \ell  + i \varepsilon } + \frac{1}{-p_1 \cdot \ell + i \varepsilon} \right] = - 2 \pi i  \delta(p_1 \cdot \ell) \, ,
\end{equation}
giving rise to a hyper-classical contribution.

A second class of diagrams are the ones with cut $h$ propagators, such as
\begin{equation}
\begin{tikzpicture}[baseline={([yshift=-0.5ex]current bounding box.center)},thick,scale=1.3,font=\footnotesize]
    \draw (0,0) -- ++(-0.5,0) node[above]{$p_1$};
    \draw (0,-1) -- ++(-0.5,0) node[below]{$p_2$};
    \draw (1,0) -- ++(0.5,0) node[above]{$p_1'$};
    \draw (1,-1) -- ++(0.5,0) node[below]{$p_2'$};
    \draw[causArrowR] (0,0) -- (1,0);
    \draw[vector] (0,0) -- (0,-1);
    \path[causArrow] (0,0) -- (0,-1);
    \draw[vector] (1,0) -- (1,-1);
    \path[cut] (1,0) -- (1,-1);
    \draw[causArrow] (0,-1) -- (1,-1);
\end{tikzpicture}
,\quad
\begin{tikzpicture}[baseline={([yshift=-0.5ex]current bounding box.center)},thick,scale=1.3,font=\footnotesize]
    \draw (0,0) -- ++(-0.5,0) node[above]{$p_1$};
    \draw (0,-1) -- ++(-0.5,0) node[below]{$p_2$};
    \draw (1,0) -- ++(0.5,0) node[above]{$p_1'$};
    \draw (1,-1) -- ++(0.5,0) node[below]{$p_2'$};
    \draw[causArrowR] (0,0) -- (1,0);
    \draw[vector] (0,0) -- (0,-1);
    \path[causArrowR] (0,0) -- (0,-1);
    \draw[vector] (1,0) -- (1,-1);
    \path[cut] (1,0) -- (1,-1);
    \draw[causArrow] (0,-1) -- (1,-1);
\end{tikzpicture}
, \quad \dots
\end{equation}
%
%
where we have omitted diagrams with alternative arrow orientations, or with a cut on the other $h$ propagator, for simplicity. We find, expanding the propagators in the classical limit, that all these diagrams are quantum contributions, and hence can be ignored in computing the  radial action $I_r$.

We conclude by presenting a conjecture. We have just seen that, in the classical limit,  there is a correspondence between the Magnus diagrams  \eqref{eq:classdiagWQFT} and \eqref{eq:classdiagQFT} in the worldline and quantum field theories, respectively. We believe that  this  correspondence between Magnus diagrams from   \eqref{eq:toyclassQFT} and \eqref{eq:toyclassWQFT} is general. Namely, a worldline diagram with a given arrowed-propagator structure is classically equivalent to the sum of quantum-field-theory diagrams with the same arrowed-propagator structure, $\phi$ cuts in place of dotted lines and all possible crossings of the massless legs. Since worldline classical contributions are always trees, we only ever need to compute diagrams from \eqref{eq:toyclassQFT} with $L$ cuts on $\phi$ propagators, at $L$ loops.  These can, in turn, be obtained easily from tree-level diagrams through \eqref{eq:Lloopsforwlim}. We will study this in more detail in follow-up work.

\section{Conclusions and outlook}
\label{sec:bye}

In this work we studied the Magnus expansion in quantum field theory. Since the coefficients of diagrams in this  expansion only depend on the causality prescription of propagators and not on the specific interactions considered, our considerations can be immediately extended to other theories, but here we focussed on the simple example of scalar $\phi^3$ theory.

Our first step was to consider examples of Magnus amplitudes at tree level and one loop, in Sections~\ref{sec:treelevel} and \ref{sec:oneloop} respectively, computed through explicit Wick contractions. A key step in this was the simplification of $\theta$-functions, which was streamlined through diagrams combining different integration regions into products of retarded and advanced propagators. In Section~\ref{sec:newformula} we also presented a general argument showing why this simplification is always guaranteed to work. 

In the tree-level case, the result reproduces the expected answer from Murua's formula, which gives the correct coefficients for Magnus diagrams without the need to perform Wick contractions. This was studied in detail in \cite{Kim:2024svw} in the case of worldline theories, and we confirmed its validity in the context of quantum field theory in Section~\ref{sec:redofromMurua-trees}. Moreover, we found that all one-loop diagrams have exactly one cut internal edge, and hence one-loop Magnus amplitudes are easily obtained from tree-level ones via a phase-space integral of the forward limit of two external legs. This implies that the coefficients of one-loop diagrams are also given by Murua's formula, as discussed in Section~\ref{sec:redofromMurua-loops}.

We studied the case of two or more loops in Section~\ref{sec:MagnusLoopsFromFL}. There we found that there are always diagrams with $L$ cut edges at $L$ loops, and these are also obtained via $L$ forward limits of tree-level amplitudes. Therefore, for this class of diagrams, Murua's formula still applies, up to a factor of $2^L$. However, for $L > 1$ there are always diagrams with $n < L$ cuts and these cannot be obtained from tree-level forward limits. 
Hence the coefficients of such diagrams cannot be obtained using Murua's formula and must instead be computed through explicit Wick contractions. To facilitate this, in Section~\ref{sec:newformula} we derived a new formula for the Magnus expansion that significantly reduces the number of required contractions. We have also found additional relations between Murua coefficients of diagrams with fewer cuts, as explained in Section~\ref{sec: MuruaForHigherLoops}.

An important motivation to study the Magnus expansion is its connection to the radial action for gravitational binary systems. Namely, it is believed the four-point Magnus amplitude for massive spinless particles coupled to gravity recovers the radial action for an effective theory describing binary systems of black holes or neutron stars, in the classical $\hbar {\to} 0$ limit.%
\footnote{For particles with spin, the correspondence is more subtle \cite{Kim:2025gis}.}
In worldline theories, the classical limit is equivalent to computing only tree-level graphs, and hence Murua's formula is enough to achieve this. In quantum field theory, it is known that loop contributions give higher-order classical contributions and hence they cannot be ignored. However, we expect that all classical contributions arise from the $L$-cut part of the $L$-loop amplitudes, and hence they can easily be extracted from tree amplitudes via \eqref{eq:Lloopsforwlim}. 
More specifically we believe that, in the classical limit, Magnus loop diagrams with cuts on massive lines agree with worldline diagrams with the same retarded and advanced propagators and the background worldline in place of the cuts. We presented this at one loop in Section~\ref{sec:classlim} and we will discuss it in more detail in follow-up work.

\newpage
\section*{Acknowledgements}
Particular thanks go to Gustav Mogull for interesting discussions and comments on this paper. We would also like to thank Lara Battino, Massimo Bianchi, Nathan Moynihan and   Paolo Muratore-Ginanneschi for stimulating   conversations. GT thanks the Physics Department at the University of Rome ``Tor Vergata''
for their warm hospitality and support.
This work was supported by the Science and Technology Facilities Council (STFC) Consolidated Grant ST/X00063X/1 \textit{``Amplitudes, Strings  \& Duality''}.
The work of  PVM is supported by a STFC quota studentship.
The work of GRB is supported by the U.K. Royal Society through Grant URF{\textbackslash}R1 {\textbackslash}20109.
GT was also supported by a Leverhulme research fellowship RF-2023-279$\backslash 9$.
No new data were generated or analysed during this study.

\newpage

\appendix

\section{Summary of Magnus diagrammar}
\label{App:summaryrules}
Here we summarise the rules for which diagrams appear in the Magnus expansion and their relations. The rules here apply beyond the simple $\phi^3$ theory we have considered in this work.
\begin{itemize}
    \item The only propagators that can appear in diagrams are retarded, advanced and cut/Hadamard:
    \begin{align}
            \begin{tikzpicture}[baseline={([yshift=-2.8ex]current bounding box.center)},thick]
        \draw[Rnew] (0,0) -- (2,0);
        \draw[-Latex] (0.3, 0.3) -- (1.7, 0.3) node[midway, above]{$k$};
    \end{tikzpicture}=i\Delta^{R}(k)&= \frac{i}{k^2-m^2+i\varepsilon\,\sgn(k^0)}\, ,\\
    \begin{tikzpicture}[baseline={([yshift=-2.8ex]current bounding box.center)},thick]
        \draw[Anew] (0,0) -- (2,0);
        \draw[-Latex] (0.3, 0.3) -- (1.7, 0.3) node[midway, above]{$k$};
    \end{tikzpicture}=i\Delta^{A}(k)&= \frac{i}{k^2-m^2-i\varepsilon\,\sgn(k^0)}\, , \\
    \begin{tikzpicture}[baseline={([yshift=-2.8ex]current bounding box.center)},thick]
        \draw[cut=2] (0,0) -- (2,0);
        \draw[-Latex] (0.3, 0.3) -- (1.7, 0.3) node[midway, above]{$k$};
    \end{tikzpicture}=\Delta^{(1)}(k)&= 2\pi\delta(k^2-m^2)\, .
    \end{align}
    When the momentum arrow is omitted, we assume it flows in the same direction as the causality arrow.
    \item All Magnus amplitudes are fully connected.
     
    \item At tree level only retarded and advanced propagators can appear and diagrams therefore correspond to connected, directed trees.
    \item Diagrams with even/odd numbers of loops can only have even/odd numbers of cuts, with the maximum  number of cuts being the number of loops, see Figure~\ref{fig:MagnusCutStructureFL}.
    \item When the cuts in a Magnus diagram are removed, the remaining diagram is still fully connected. That is, cuts can never appear in such a way that they divide a diagram in two. For example, 
    \begin{equation}
        \text{Allowed: }\begin{tikzpicture}[baseline={([yshift=-0.1ex]current bounding box.center)},thick] 
    \draw[massive] (0.25,0) -- (0.75,0);
    \draw[out=-45,in=-135,Rnew] (0.75,0) to (1.75,0);
    \draw[out=135,in=45,cut]   (1.75,0) to (0.75,0);
    \draw[Rnew] (1.75,0) -- (2.25,0);
    \path (2,0) -- (2,-0.5); 
    \draw[out=-45,in=-135,Rnew] (2.25,0) to (3.25,0);
    \draw[out=135,in=45,cut]   (3.25,0) to (2.25,0);
    \draw[massive] (3.25,0) -- (3.75,0);
    \end{tikzpicture}\,,\quad \text{Forbidden: }\begin{tikzpicture}[baseline={([yshift=-0.1ex]current bounding box.center)},thick] 
    \draw[massive] (0.25,0) -- (0.75,0);
    \draw[out=-45,in=-135,cut] (0.75,0) to (1.75,0);
    \draw[out=135,in=45,cut]   (1.75,0) to (0.75,0);
    \draw[Rnew] (1.75,0) -- (2.25,0);
    \path (2,0) -- (2,-0.5); 
    \draw[out=-45,in=-135,Rnew] (2.25,0) to (3.25,0);
    \draw[out=135,in=45,Anew]   (3.25,0) to (2.25,0);
    \draw[massive] (3.25,0) -- (3.75,0);
    \end{tikzpicture}\,.
    \end{equation}
    \item Any closed loop of retarded propagators is zero. These can only appear at two loops and beyond since every diagram at one loop requires exactly one cut propagator. As an example at two loops, 
    \begin{equation}
    \text{Nonzero:}\begin{tikzpicture}[baseline={([yshift=-0.5ex]current bounding box.center)},thick] 
    \draw[massive] (0,0) -- (0.75,0);
    \draw[massive] (0.75,0) -- (1.5,0);
    \draw[Rnew] (0.75,0) -- (0.75,0.70);
    \draw[out=-45,in=-135,Rnew] (0.75/3,1.4) to (0.75+1.5/3,1.4);
    \draw[out=135,in=45,Anew]   (0.75+1.5/3,1.4) to (0.75/3,1.4);
    \draw[Rnew] (0.75,0.70) -- (0.75/3,1.4);
    \draw[Rnew] (0.75,0.70) -- (0.75+1.5/3,1.4);
    \path (1,0) -- (1,-0.5); 
    \end{tikzpicture}\,,\quad \text{Zero:}\begin{tikzpicture}[baseline={([yshift=-0.5ex]current bounding box.center)},thick] 
    \draw[massive] (0,0) -- (0.75,0);
    \draw[massive] (0.75,0) -- (1.5,0);
    \draw[Rnew] (0.75,0) -- (0.75,0.70);
    \draw[out=-45,in=-135,Rnew] (0.75/3,1.4) to (0.75+1.5/3,1.4);
    \draw[out=135,in=45,Rnew]   (0.75+1.5/3,1.4) to (0.75/3,1.4);
    \draw[Rnew] (0.75,0.70) -- (0.75/3,1.4);
    \draw[Rnew] (0.75,0.70) -- (0.75+1.5/3,1.4);
    \path (1,0) -- (1,-0.5); 
    \end{tikzpicture}\, .
\end{equation}
    \item Lower-loop, lower-cut pieces of Magnus amplitudes are related to higher-loop, higher-cut pieces through taking the forward limit \eqref{eq: genForwardLimit}, which we reproduce here
    \begin{equation}
    \cA^{(L+\ell,C+\ell)} = \frac{2^{2C}C!}{2^{2(C+\ell)}(C+\ell)!} \int \prod_{k=1}^{\ell}\left(\hat{d}^4 p_k \, \hat{\delta}(p_k^2-m^2)\right) \cA^{(L,C)}_{\ell-\text{FL}}\, .
\end{equation}
    In the above $\cA^{A,B}$ is a $A$-loop $B$-cut piece of a Magnus amplitude.
    When there are multiple particle species one should perform forward limits for each and sum the results.
    \item Bubbles on external legs and divergent loop integrals appearing from their forward limits should be dropped. See Section~\ref{sec: BadGraphs}.
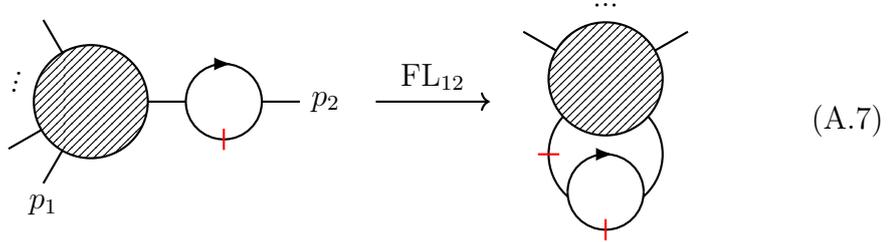
    \begin{equation}
\begin{tikzpicture}[baseline={([yshift=-0.5ex]current bounding box.center)},thick]
    \draw (0.75,0) -- (1.25,0);
    \draw[Rnew] (1.25,0) arc(180:0:0.5);
    \draw[cut=2] (1.25,0) arc(-180:0:0.5);
    \draw (2.25,0) --++(0.5,0) node[right]{$p_2$};
    \draw[pattern=north east lines] (0,0) circle(0.75cm);
    \draw (-120:0.75) -- (-120:1.25) node[below]{$p_1$};
    \draw (-150:0.75) -- (-150:1.25);
    \path[postaction={decorate,decoration={text along path,text align=center,text={...}}}] (-180:1) arc(-180:-210:1);
    \draw (-240:0.75) -- (-240:1.25);
    \draw[->] (3.75,0) -- ++(1.5,0) node[pos=0.5,above]{$\text{FL}_{12}$};
\end{tikzpicture}
\quad
\begin{tikzpicture}[baseline={([yshift=-0.5ex]current bounding box.center)},thick]
    \draw[cut=2] ([shift=(90:0.75)]0,-1) arc(90:270:0.75);
    \draw ([shift=(90:0.75)]0,-1) arc(90:-90:0.75);
    \fill[fill=white,draw=none] (0,0) circle(0.75cm);
    \fill[fill=white,draw=none] (0,-1.5) circle(0.5cm);
    \draw[pattern=north east lines] (0,0) circle(0.75cm);
    \draw (30:0.75) -- (30:1.25);
    \draw (150:0.75) -- (150:1.25);
    \path[postaction={decorate,decoration={text along path,text align=center,text={...}}}] (75:1) arc(75:105:1);
    \draw[Anew] ([shift=(180:0.5)]0,-1.5) arc(180:0:0.5);
    \draw[cut=2] ([shift=(180:0.5)]0,-1.5) arc(-180:0:0.5);
\end{tikzpicture}
\end{equation}
    \item The general diagrammatic expansion of the $N$-operator and $N$-matrix elements in terms of Murua coefficients is given in \eqref{eq:Nsum} and \eqref{eq:MuruaFormulaMatrixElement} respectively, which we reproduce here:
    \begin{equation}
     iN^{(n)} = \sum_{|\tau|=n}\frac{\omega(\tau)}{\sigma(\tau)}\mathcal{I}(\tau) \, ,
 \end{equation}
 \begin{equation}
    \langle 0| iN |\prod_{i=1}^k\phi(p_i)\rangle = \hat{\delta}^{(4)}\left(\sum_{i=1}^k p_i\right)\sum_{g} \frac{\omega(g)}{\sigma(g)}\mathcal{J}(g) \, ,
\end{equation}
    where $g$ and $\tau$ are diagrams with and without external legs respectively, $\omega$ are Murua coefficients and $\sigma$ are symmetry factors. The diagrams which can appear at each loop order are described in the bullet points above. 
    \item The symmetry factor $\sigma$ of each diagram is calculated in the same way as the Dyson expansion, while preserving propagator prescriptions. See Appendix~\ref{app: SymmFactors} for more details.
    \item The Murua coefficients $\omega$ only depend on the diagram with external legs removed. 
    \item At tree level the Murua coefficients can be found by an extension \cite{Kim:2024svw} of Murua's formula \cite{Murua_2006}.
    \item Beyond tree level the Murua coefficients must be calculated by performing Wick contractions in the Magnus expansion explicitly. 
    \item The forward limit relations also relate the Murua coefficients of diagrams with different numbers of cuts through \eqref{eq: MuruaRelationBest} which we restate here:
    \begin{equation}
    \omega(\tau^{(L,C)})= \frac{\omega(\tau^{(L-C,0)})}{2^C}\,.
\end{equation}
    In the above, $\tau^{(A,B)}$ is a diagram with $A$-loops and $B$-cuts. This implies that only the Murua coefficients of zero-cut, even-loop diagrams are enough to derive the Murua coefficients of any other diagram.
\end{itemize}

\section{Proof of the new Magnus formula}\label{sec:ProofNewMagnusFormula}

In the following we present a proof of \eqref{eq:DescentSetThetaIdentity}.
To begin, notice that we can formalise the procedure above of relabelling $x_i$'s as follows (again suppressing the integrations)
\begin{align}
    \mathrm{CS}^D_n = \sum_{\sigma\in S^D_n} \theta_{12}\cdots \theta_{n-1\ n}\, \mathcal{C}_\sigma &\longrightarrow \left(\sum_{\sigma\in S^D_n}\sigma^{-1}(\theta_{12}\cdots \theta_{n-1\ n})\right) \mathcal{C}_{12\dots n} \nn \\
    &\quad= \left(\sum_{\sigma\in S^D_n} \theta_{\sigma^{-1}(1)\sigma^{-1}(2)} \cdots \theta_{\sigma^{-1}(n-1)\sigma^{-1}(n)}\right)\mathcal{C}_{12\dots n}\, , \label{eq:TargetSum}
\end{align}
where $\mathcal{C}_\sigma$ represents the nested commutators with the order of the subscripts indicated by $\sigma$. In this language, the proof is reduced to evaluating the value of the sum.

Let us begin with a concrete example before moving on to the more general case. Let $n$ be arbitrary, and fix $D=\{1\}$, meaning that we are looking at permutations with one descent between the first and second elements. Therefore, the defining condition that all the relevant $\sigma$ must obey is
\begin{equation}
    \sigma(1)>\sigma(2)\, ,\qquad \text{and}\qquad \sigma(2)<\sigma(3)<\cdots <\sigma(n)\, .
\end{equation}
This ensures there is a descent between positions 1 and 2, and that there are no further descents in the rest of the permutation. Let $\tau=\sigma^{-1}$, and let $i=\tau(a_i)$, meaning that the $a_i$th position in $\tau$ maps to $i$. Then the defining condition for $\sigma$ constrains the form of $\tau$ by imposing
\begin{equation}
    a_1 >a_2\, ,\qquad \text{and} \qquad a_2<a_3<\cdots <a_n\, .
\end{equation}
Put differently,  1 must appear in $\tau$ after 2, and the rest of the numbers must be in order. Concretely, the permutations $\tau$ satisfying this property are
\begin{equation}
    {\red2}{\red1}34\dots n,\ {\red2}3{\red1}4\dots n,\ {\red2}34{\red1}\dots n\ ,\dots,\ {\red2}34\dots n{\red1}\, .
\end{equation}
When written as $\theta$-functions, each of these permutations impose a total ordering on $\{1,\dots, n\}$, and adding them corresponds to an OR operation. This results in
\begin{align}
\theta_{21}\theta_{13}\cdots\theta_{n-1\ n} &  \quad + \quad \cdots \quad + \quad \theta_{23}\theta_{34}\cdots\theta_{n1}  \nn\\
    \iff 2>1>3>\dots >n & \quad \text{or} \quad \dots \quad \text{or} \quad 2>3>\dots>n>1\, .
\end{align}
Notice that we always have $2>3>\cdots >n$ but 1 is only constrained to $2>1$, as it appears in all possible positions between the remaining $\{3,\dots,n\}$. Thus we get
\begin{equation}
    \theta_{21}\theta_{23}\theta_{34}\cdots\theta_{n-1\ n} \, .
\end{equation}
We now consider the general case. Suppose that there are $k$ descents at $D=\{i_1,i_2,\dots, i_k\} \subseteq\{1,\dots, n{-}1\}$. This means that, for each $1\leq\ell\leq k-1$,
\begin{equation}
    \sigma(i_\ell)>\sigma(i_\ell+1)\, ,\qquad \sigma(i_\ell+1)<\cdots<\sigma(i_{\ell+1})\, ,
\end{equation}
alongside the conditions at the start and end of the permutation
\begin{equation}
    \sigma(1)<\cdots<\sigma(i_1)\, ,\qquad \sigma(i_k+1)<\cdots <\sigma(n)\, .
\end{equation}
If we again write $\sigma^{-1}=\tau$ and define $\tau(a_j)=j$ as before, we can translate the conditions above into conditions on the positions of $\{1,\dots,n\}$ in $\tau$:
\begin{enumerate}
    \setlength\itemsep{-0.3ex}
    \item[C1] Each $i_\ell$ must appear after $i_\ell+1$ in $\tau$.
    \item[C2] All numbers between $i_\ell+1$ and $i_{\ell+1}$ must appear in order relative to each other, formally including $i_0=0$ and $i_{k+1}=n$.
\end{enumerate}
Therefore, when we apply these permutations to the $\theta$-functions, each term can be written in the form
\begin{equation}
    \underbrace{\left(\theta_{i_1{+}1\ i_1}\cdots \theta_{i_k{+}1\ i_k}\right)}_{\text{C1}} \underbrace{\left(\theta_{12}\cdots\theta_{i_1{-}1\ i_1}\right)
    \cdots \left(\theta_{i_{k{-}1}{+}1\ i_{k{-}1}{+}2}\cdots\theta_{n{-}1\ n}\right)}_{\text{C2}} \ \Theta_\tau\, ,
\end{equation}
where $\Theta_\tau$ represents the leftover $\theta$-functions that are not captured by the factors at the front, which will be specific to each permutation $\tau$. However, when we perform the sum over $\tau$, it amounts to summing over all possible ways to satisfy C1 and C2, so that at the end the extra $\theta$-functions in $\Theta_\tau$ should all cancel and leave only the prefactor. We can write this differently if we introduce the ascent set $A=\{1,\dots,n{-}1\}\setminus D$, so that
\begin{equation}
    \theta_{12}\theta_{23}\cdots\theta_{i_1+1\ i_1}\theta_{i_1+1\ i_1+2}\cdots\theta_{i_2+1\ i_2}\cdots \theta_{n{-}1\ n} = \left(\prod_{i\in D}\theta_{i{+}1\ i}\right)\left(\prod_{j\in A} \theta_{j\ j{+}1}\right)\, .
\end{equation}

\section{Coefficients of new Wick contraction rules}
\label{app:newwickcoeffs}
In Section~\ref{sec:newwickcontr} we have shown that applying the Wick contraction rules to the nested commutator
\begin{equation}
    \mathcal{C}_{1\dots n} = \Big[\cH_1,\big[\cH_2,\cdots[\cH_{n-1},\cH_n]\cdots\big]\Big]
\end{equation}
produces both $i\Delta_{ij}$ and $\Delta^{(1)}_{ij}$ contractions. We dedicate this appendix to explaining how to calculate the combinatorial factor $N$ appearing in front of each contraction, which we choose to decompose into three factors $N=f_V f_L f_P$. We will focus only on $\phi^m$ theory, though it can be easily extended to more general theories with multiple species of interacting scalars. A general Wick contraction looks like
\begin{align}
N(i\Delta_{i_1i'_1})^{c_1}\cdots(\Delta^{(1)}_{i_ki'_k})^{c_k}\ :\left(\cH_{(1)}^{(d_1)} \cdots \cH_{n}^{(d_n)}\right): \, ,
\end{align}
with any number of $i\Delta$ or $\Delta^{(1)}$ functions, and where $d_i$ denotes the number of contractions at spacetime point $x_i$. Suppose in total there are $c_1+\dots+c_k=P$ contractions. If at point $v$ (corresponding to $x_v$) there are $d_v$ contractions, then each vertex will carry a residual factor of 
\begin{equation}
    \frac{m\times(m-1)\times\cdots\times(m-(d_v-1))}{m!} = \frac{m!/(m-d_v)!}{m!} = \frac{1}{(m-d_v)!}\, ,
\end{equation}
arising from the original  $1/m!$ from the Hamiltonian (after performing  $d_v$ Wick contractions). Repeating this for each spacetime point, the vertex factor $f_V$ is
\begin{equation}
    f_V =\prod_{1\leq v\leq n}\frac{1}{(m-d_v)!}\, .
\end{equation}
As we have seen in \eqref{eq:OneWickContraction}, each of the $P$ contractions brings a factor of $1/2$. However, there is also a factor of $2^{n-1}$ coming from the normal ordering of the nested anticommutators, so overall we have\footnote{We have used the well-known fact that the number of loops $L$ is related to the number of propagators $P$ and the number of vertices $V=n$ by $L=P-V+1$.}
\begin{equation}\label{eq:ContractionsLoopFactor}
    f_L=\frac{1}{2^{P-(n-1)}}=\frac{1}{2^L}\, .
\end{equation}
Moreover, note that the Wick contraction method in Section~\ref{sec:newwickcontr} is sequential, meaning we perform one Wick contraction at a time until we reach the required number. In principle, when doing Wick contractions in this fashion, we should do all possible Wick contractions in all possible orders, then divide by $1/P!$ to account for overcounting. When each contraction is distinct this factor drops out, but for repeated contractions this becomes non-trivial as certain permutations of contractions become identical. For example, for $(i\Delta_{12})^2\Delta_{23}$ there are only three distinct permutations,
\begin{equation}
    \frac{1}{3!}((i\Delta_{12})(i\Delta_{12})(i\Delta_{23})+ (i\Delta_{12})(i\Delta_{23})(i\Delta_{12})+ (i\Delta_{23})(i\Delta_{12})(i\Delta_{12})) = \frac{1}{2}(i\Delta_{12})^2(i\Delta_{23})\, ,
\end{equation}
where the order of the $\Delta_{ij}$ in each product corresponds to the order in which we perform the contractions, and therefore it ends up with a factor of $1/2$. In general, this factor can be expressed as
\begin{equation}\label{eq:PropagatorCombinatorics}
    f_P=\frac{1}{P!}\binom{P}{c_1}\binom{P-c_1}{c_2}\cdots\binom{P-c_1-\cdots - c_{k-1}}{c_k} = \frac{1}{c_1! c_2!\cdots c_k!}\, .
\end{equation}
The binomial factors count how many inequivalent permutations there are: first pick all $c_1$ contractions out of $P$, then $c_2$ out of the remaining and so on. Overall, the coefficient associated to the contractions is
\begin{equation}
    N = \frac{1}{2^L}\left(\prod_{1\leq v\leq n}\frac{1}{(m-d_v)!}\right)\frac{1}{c_1! c_2!\cdots c_k!}\, .
\end{equation}

\section{Symmetry factors between the \texorpdfstring{$N$}{N} operator and matrix elements}\label{app: SymmFactors}
Symmetry factors of Feynman graphs are a notoriously confusing topic. In this appendix we will try to demystify them so that our definition of multi-loop Murua coefficients is clear. In addition we will show how to go from the $N$-operator expansion \eqref{eq:Nsum} to that of its matrix elements \eqref{eq:MuruaFormulaMatrixElement}.

Consider a graph $\tau$  appearing in the $N$ operator (note that there are no external legs in the $N$-operator expansion). The exact diagram we pick will not change the general argument laid out here, but for concreteness one can consider the graph
\begin{equation}\label{eq: example diagram}
    \tau=\begin{tikzpicture}[baseline={([yshift=-0.5ex]current bounding box.center)},thick] 
    \draw[Anew] (0,0) -- (0.75,0);
    \draw[Rnew] (0.75,0) -- (0.75,1);
    \draw[Rnew] (0.75,0) -- (1.25,-0.8);
    \draw[Rnew] (1.75,0) -- (1.25,-0.8);
    \draw[Rnew] (1.75,0) -- (1.75,1);
    \draw[Rnew] (1.75,0) -- (2.5,0);
    \draw[out=-45,in=-135,cut] (0.75,1) to (1.75,1);
    \draw[out=135,in=45,cut]   (1.75,1) to (0.75,1);
    \path (2,0) -- (2,-0.5); 
    \draw[fill=white] (0,0) circle(3pt);
    \draw[fill=white] (0.75,0) circle(3pt);
    \draw[fill=white] (0.75,1) circle(3pt);
    \draw[fill=white] (1.75,0) circle(3pt);
    \draw[fill=white] (1.75,1) circle(3pt);
    \draw[fill=white] (2.5,0) circle(3pt);
    \draw[fill=white] (2.5,0) circle(3pt);
    \draw[fill=white] (1.25,-0.8) circle(3pt);
    \end{tikzpicture}\,.
\end{equation}
The coefficient appearing in front of such a graph in the Magnus expansion is given by the ratio of a Murua coefficient $\omega(\tau)$ and a symmetry factor $\sigma(\tau)$, see \eqref{eq:Nsum}. By direct computation, from the Magnus expansion we find this coefficient to be
\begin{equation}\label{eq: exampleGraphCoeff}
    \frac{\omega\left(\begin{tikzpicture}[baseline={([yshift=-0.5ex]current bounding box.center)},scale=0.7,thick] 
    \draw[Anew] (0,0) -- (0.75,0);
    \draw[Rnew] (0.75,0) -- (0.75,1);
    \draw[Rnew] (0.75,0) -- (1.25,-0.8);
    \draw[Rnew] (1.75,0) -- (1.25,-0.8);
    \draw[Rnew] (1.75,0) -- (1.75,1);
    \draw[Rnew] (1.75,0) -- (2.5,0);
    \draw[out=-45,in=-135,cut] (0.75,1) to (1.75,1);
    \draw[out=135,in=45,cut]   (1.75,1) to (0.75,1);
    \path (2,0) -- (2,-0.5); 
    \draw[fill=white] (0,0) circle(3pt);
    \draw[fill=white] (0.75,0) circle(3pt);
    \draw[fill=white] (0.75,1) circle(3pt);
    \draw[fill=white] (1.75,0) circle(3pt);
    \draw[fill=white] (1.75,1) circle(3pt);
    \draw[fill=white] (2.5,0) circle(3pt);
    \draw[fill=white] (2.5,0) circle(3pt);
    \draw[fill=white] (1.25,-0.8) circle(3pt);
    \end{tikzpicture}\right)}{\sigma\left(\begin{tikzpicture}[baseline={([yshift=-0.5ex]current bounding box.center)},scale=0.7,thick] 
    \draw[Anew] (0,0) -- (0.75,0);
    \draw[Rnew] (0.75,0) -- (0.75,1);
    \draw[Rnew] (0.75,0) -- (1.25,-0.8);
    \draw[Rnew] (1.75,0) -- (1.25,-0.8);
    \draw[Rnew] (1.75,0) -- (1.75,1);
    \draw[Rnew] (1.75,0) -- (2.5,0);
    \draw[out=-45,in=-135,cut] (0.75,1) to (1.75,1);
    \draw[out=135,in=45,cut]   (1.75,1) to (0.75,1);
    \path (2,0) -- (2,-0.5); 
    \draw[fill=white] (0,0) circle(3pt);
    \draw[fill=white] (0.75,0) circle(3pt);
    \draw[fill=white] (0.75,1) circle(3pt);
    \draw[fill=white] (1.75,0) circle(3pt);
    \draw[fill=white] (1.75,1) circle(3pt);
    \draw[fill=white] (2.5,0) circle(3pt);
    \draw[fill=white] (2.5,0) circle(3pt);
    \draw[fill=white] (1.25,-0.8) circle(3pt);
    \end{tikzpicture}\right)}=-\frac{1}{3360}\, .
\end{equation}
Since Murua coefficients were originally only defined for tree diagrams \cite{Murua_2006}, how we define $\omega(\tau)$ depends on how we have defined the symmetry factor $\sigma(\tau)$.

The logic we will follow here is to define the symmetry factor of a graph in the same way as the Dyson expansion (while preserving propagator prescriptions). This has the benefit that the Murua coefficients of cut graphs are related to those with fewer cuts by a simple factor, which can be seen in relations \eqref{eq: MuruaRelation} and \eqref{eq: MuruaRelationBest}. 

A detailed derivation of Feynman symmetry factors for $\phi^n$ theories can be found in, for example, \cite{Palmer:2001vq, Saemann:2020oyz}. Here we will simply state the practical rules that allow you to compute the symmetry factor for any graph.

There are two types of symmetry of a generic multiloop graph without external legs $\tau$:\footnote{There are additional factors if we consider tadpoles, see \cite{Palmer:2001vq}, which we omit in this paper.}
\begin{enumerate}
    \item  Permutations of the vertices (along with their attached edges) that leave $\tau$  invariant and respect the propagator prescriptions of the edges.
    \item Any permutation of a duplicated propagator (edge) between the same two vertices. For $n$ duplicated edges there will be $n!$ permutations. In the Magnus expansion these can be duplicated cut or retarded/advanced propagators (which will appear as pieces of larger graphs), e.g. 
    \begin{equation}
        \begin{tikzpicture}[baseline={([yshift=-0.5ex]current bounding box.center)},thick] 
    \draw[out=-75,in=-105,cut] (0,0) to (1.5,0);
    \draw[out=105,in=75,cut]   (1.5,0) to (0,0)node[xshift=0.75cm]{$\vdots$};
    \draw[out=145,in=35,cut]   (1.5,0) to (0,0);
    \path (2,0) -- (2,-0.5); 
    \draw[fill=white] (0,0) circle(3pt);
    \draw[fill=white] (1.5,0) circle(3pt);
    \end{tikzpicture}\,, \quad 
     \begin{tikzpicture}[baseline={([yshift=-0.5ex]current bounding box.center)},thick] 
    \draw[out=-75,in=-105,Rnew] (0,0) to (1.5,0);
    \draw[out=105,in=75,Anew]   (1.5,0) to (0,0)node[xshift=0.75cm]{$\vdots$};
    \draw[out=145,in=35,Anew]   (1.5,0) to (0,0);
    \path (2,0) -- (2,-0.5); 
    \draw[fill=white] (0,0) circle(3pt);
    \draw[fill=white] (1.5,0) circle(3pt);
    \end{tikzpicture}
    \end{equation}
\end{enumerate}
We will denote the symmetry group generated by the above permutations of vertices and edges as $G(\tau)$.
The symmetry factor is then just the order of this symmetry group $\sigma(\tau)=|G(\tau)|$ . For the example in \eqref{eq: example diagram} we have 
\begin{equation}
   \sigma\left(\begin{tikzpicture}[baseline={([yshift=-0.5ex]current bounding box.center)},scale=0.7,thick] 
    \draw[Anew] (0,0) -- (0.75,0);
    \draw[Rnew] (0.75,0) -- (0.75,1);
    \draw[Rnew] (0.75,0) -- (1.25,-0.8);
    \draw[Rnew] (1.75,0) -- (1.25,-0.8);
    \draw[Rnew] (1.75,0) -- (1.75,1);
    \draw[Rnew] (1.75,0) -- (2.5,0);
    \draw[out=-45,in=-135,cut] (0.75,1) to (1.75,1);
    \draw[out=135,in=45,cut]   (1.75,1) to (0.75,1);
    \path (2,0) -- (2,-0.5); 
    \draw[fill=white] (0,0) circle(3pt);
    \draw[fill=white] (0.75,0) circle(3pt);
    \draw[fill=white] (0.75,1) circle(3pt);
    \draw[fill=white] (1.75,0) circle(3pt);
    \draw[fill=white] (1.75,1) circle(3pt);
    \draw[fill=white] (2.5,0) circle(3pt);
    \draw[fill=white] (2.5,0) circle(3pt);
    \draw[fill=white] (1.25,-0.8) circle(3pt);
    \end{tikzpicture}\right)=2\times(2!)=4\, ,
\end{equation}
where one factor of 2 came from flipping the diagram left to right and the factor of $2!$ came from the permutations of the two cut propagators. With this symmetry factor we can derive the Murua coefficient from the overall graph coefficient \eqref{eq: exampleGraphCoeff}
\begin{equation}
    \omega\left(\begin{tikzpicture}[baseline={([yshift=-0.5ex]current bounding box.center)},scale=0.7,thick] 
    \draw[Anew] (0,0) -- (0.75,0);
    \draw[Rnew] (0.75,0) -- (0.75,1);
    \draw[Rnew] (0.75,0) -- (1.25,-0.8);
    \draw[Rnew] (1.75,0) -- (1.25,-0.8);
    \draw[Rnew] (1.75,0) -- (1.75,1);
    \draw[Rnew] (1.75,0) -- (2.5,0);
    \draw[out=-45,in=-135,cut] (0.75,1) to (1.75,1);
    \draw[out=135,in=45,cut]   (1.75,1) to (0.75,1);
    \path (2,0) -- (2,-0.5); 
    \draw[fill=white] (0,0) circle(3pt);
    \draw[fill=white] (0.75,0) circle(3pt);
    \draw[fill=white] (0.75,1) circle(3pt);
    \draw[fill=white] (1.75,0) circle(3pt);
    \draw[fill=white] (1.75,1) circle(3pt);
    \draw[fill=white] (2.5,0) circle(3pt);
    \draw[fill=white] (2.5,0) circle(3pt);
    \draw[fill=white] (1.25,-0.8) circle(3pt);
    \end{tikzpicture}\right)= -\frac{1}{840}= \frac{1}{2^2}\,\omega\left(\begin{tikzpicture}[baseline={([yshift=-0.5ex]current bounding box.center)},scale=0.7,thick] 
    \draw[Anew] (0,0) -- (0.75,0);
    \draw[Rnew] (0.75,0) -- (0.75,1);
    \draw[Rnew] (0.75,0) -- (1.25,-0.8);
    \draw[Rnew] (1.75,0) -- (1.25,-0.8);
    \draw[Rnew] (1.75,0) -- (1.75,1);
    \draw[Rnew] (1.75,0) -- (2.5,0);
    \path (2,0) -- (2,-0.5); 
    \draw[fill=white] (0,0) circle(3pt);
    \draw[fill=white] (0.75,0) circle(3pt);
    \draw[fill=white] (0.75,1) circle(3pt);
    \draw[fill=white] (1.75,0) circle(3pt);
    \draw[fill=white] (1.75,1) circle(3pt);
    \draw[fill=white] (2.5,0) circle(3pt);
    \draw[fill=white] (2.5,0) circle(3pt);
    \draw[fill=white] (1.25,-0.8) circle(3pt);
    \end{tikzpicture}\right)\,,
\end{equation}
which satisfies the relation \eqref{eq: MuruaRelation} as promised.

Once we contract some external states into a graph, the symmetry factor will change. The rules for calculating the symmetry factor remain as above, but now the symmetry \textit{must fix the external legs}. We can see how this works explicitly for the diagram \eqref{eq: example diagram}. For the $N$-operator, this diagram contributes a factor
\begin{equation}
     iN^{(7)}\supset(-i)^7 \! \int \prod_{i=1}^7 d^4x_i\,\,\frac{1}{4}\omega\left(
    \begin{tikzpicture}[baseline={([yshift=-0.5ex]current bounding box.center)},scale=0.7,thick] 
    \draw[Anew] (0,0) -- (0.75,0);
    \draw[Rnew] (0.75,0) -- (0.75,1);
    \draw[Rnew] (0.75,0) -- (1.25,-0.8);
    \draw[Rnew] (1.75,0) -- (1.25,-0.8);
    \draw[Rnew] (1.75,0) -- (1.75,1);
    \draw[Rnew] (1.75,0) -- (2.5,0);
    \draw[out=-45,in=-135,cut] (0.75,1) to (1.75,1);
    \draw[out=135,in=45,cut]   (1.75,1) to (0.75,1);
    \path (2,0) -- (2,-0.5); 
    \draw[fill=white] (0,0) circle(3pt);
    \draw[fill=white] (0.75,0) circle(3pt);
    \draw[fill=white] (0.75,1) circle(3pt);
    \draw[fill=white] (1.75,0) circle(3pt);
    \draw[fill=white] (1.75,1) circle(3pt);
    \draw[fill=white] (2.5,0) circle(3pt);
    \draw[fill=white] (2.5,0) circle(3pt);
    \draw[fill=white] (1.25,-0.8) circle(3pt);
    \end{tikzpicture}\right)
    \begin{tikzpicture}[baseline={([yshift=-0.5ex]current bounding box.center)},scale=0.7,thick,font=\footnotesize] 
    \draw[Anew] (0,0)node[yshift=0.3cm]{1} -- (0.75,0);
    \draw[Rnew] (0.75,0) -- (0.75,1);
    \draw[Rnew] (0.75,0) -- (1.25,-0.8)node[yshift=-0.3cm]{2};
    \draw[Rnew] (1.75,0) -- (1.25,-0.8);
    \draw[Rnew] (1.75,0) -- (1.75,1);
    \draw[Rnew] (1.75,0) -- (2.5,0)node[yshift=0.3cm]{3};
    \draw[out=-45,in=-135,cut] (0.75,1) to (1.75,1);
    \draw[out=135,in=45,cut]   (1.75,1) to (0.75,1);
    \path (2,0) -- (2,-0.5); 
    \draw[fill=white] (0,0) circle(3pt);
    \draw[fill=white] (0.75,0) circle(3pt);
    \draw[fill=white] (0.75,1) circle(3pt);
    \draw[fill=white] (1.75,0) circle(3pt);
    \draw[fill=white] (1.75,1) circle(3pt);
    \draw[fill=white] (2.5,0) circle(3pt);
    \draw[fill=white] (2.5,0) circle(3pt);
    \draw[fill=white] (1.25,-0.8) circle(3pt);
    \end{tikzpicture}\frac{:\phi_1^2\phi_2\phi_3^2:}{(2!)^2}\,.
\end{equation}
We will call \textit{external vertices} the vertices of a graph $\tau$ which can be contracted to external particles. In the example above these external vertices are labelled $1,2,3$.
After contracting with external states, the one diagram above will generate many diagrams which contribute to the five-point $N$-matrix element
\begin{equation}
    \langle0|iN^{(7)}|\phi(p_a)\phi(p_b)\phi(p_c)\phi(p_d)\phi(p_e)\rangle\,,
\end{equation}
with the external legs contracted onto the vertices $1,2,3$ in all possible ways. There are, in principle, $5!$ such contractions, however many of these will be equivalent. There are two ways contractions can be equivalent:
\begin{enumerate}
    \item When the same $n$ external legs contract with one external vertex they can do so $n!$ equivalent ways
    \begin{equation}
        \begin{tikzpicture}[baseline={([yshift=-0.5ex]current bounding box.center)},thick]
    \draw[pattern=north east lines] (0,0) circle(0.5cm);
    \draw[massive] (0.5,0)node[xshift=0.5cm,yshift=0.1cm]{$\vdots$}--(1,0.5);
    \draw[massive] (0.5,0)--(1,0.35);
    \draw[massive] (0.5,0)--(1,-0.5);
    \draw[massive] (0.5,0)--(1,-0.35);
    \draw[fill=white] (0.5,0) circle(3pt);
\end{tikzpicture}
\Bigg\}\, n
    \end{equation}
    This cancels the same numerical factor in $\phi_i^n/(n!)$ for external vertices.
    \item If a symmetry of the underlying uncontracted graph relates two contracted graphs, then they are equivalent. This will happen whenever there exists a symmetry that performs a non-trivial permutation of the external vertices. 
\end{enumerate}
To make the second point precise, let us define $\tilde{G}(\tau)$ as the subgroup of $G(\tau)$ that acts trivially on all external vertices (i.e.~the intersection of the stabilisers of the external vertices). Then, by Lagrange's theorem
\begin{equation}
    \frac{|G(\tau)|}{|\tilde{G}(\tau)|}= |\tilde{G}(\tau): G(\tau)|\,,
\end{equation}
where $\tilde{G}(\tau): G(\tau)$ is the set of left cosets of $\tilde{G}(\tau)$ in $G(\tau)$. Intuitively $|\tilde{G}(\tau):G(\tau)|$ is number of ways of exchanging external vertices using the symmetries $G(\tau)$. In other words $|\tilde{G}(\tau):G(\tau)|$ is the number of symmetries which are inequivalent if you only compare them through their action on external vertices. This is the exactly the number of equivalent contractions of the graph $\tau$ described in point two above. For example, the following two contractions are equivalent and will combine into a single term
\begin{equation}
    \begin{tikzpicture}[baseline={([yshift=-0.5ex]current bounding box.center)},thick] 
    \draw[Anew] (0,0) -- (0.75,0);
    \draw[Rnew] (0.75,0) -- (0.75,1);
    \draw[Rnew] (0.75,0) -- (1.25,-0.8);
    \draw[Rnew] (1.75,0) -- (1.25,-0.8);
    \draw[Rnew] (1.75,0) -- (1.75,1);
    \draw[Rnew] (1.75,0) -- (2.5,0);
    \draw[out=-45,in=-135,cut] (0.75,1) to (1.75,1);
    \draw[out=135,in=45,cut]   (1.75,1) to (0.75,1);
    \path (2,0) -- (2,-0.5); 
    \draw[massive] (0,0) -- (-0.5,0.5)node[xshift=-0.3cm]{$p_a$};
    \draw[massive] (0,0) -- (-0.5,-0.5)node[xshift=-0.3cm]{$p_b$};
    \draw[massive] (2.5,0) -- (3.0,0.5)node[xshift=0.3cm]{$p_d$};
    \draw[massive] (2.5,0) -- (3.0,-0.5)node[xshift=0.3cm]{$p_e$};
    \draw[massive] (1.25,-0.8) -- (1.25,-1.3)node[yshift=-0.3cm]{$p_c$};
    \end{tikzpicture}\,,\quad 
    \begin{tikzpicture}[baseline={([yshift=-0.5ex]current bounding box.center)},thick] 
    \draw[Anew] (0,0) -- (0.75,0);
    \draw[Rnew] (0.75,0) -- (0.75,1);
    \draw[Rnew] (0.75,0) -- (1.25,-0.8);
    \draw[Rnew] (1.75,0) -- (1.25,-0.8);
    \draw[Rnew] (1.75,0) -- (1.75,1);
    \draw[Rnew] (1.75,0) -- (2.5,0);
    \draw[out=-45,in=-135,cut] (0.75,1) to (1.75,1);
    \draw[out=135,in=45,cut]   (1.75,1) to (0.75,1);
    \path (2,0) -- (2,-0.5); 
    \draw[massive] (0,0) -- (-0.5,0.5)node[xshift=-0.3cm]{$p_d$};
    \draw[massive] (0,0) -- (-0.5,-0.5)node[xshift=-0.3cm]{$p_e$};
    \draw[massive] (2.5,0) -- (3.0,0.5)node[xshift=0.3cm]{$p_a$};
    \draw[massive] (2.5,0) -- (3.0,-0.5)node[xshift=0.3cm]{$p_b$};
    \draw[massive] (1.25,-0.8) -- (1.25,-1.3)node[yshift=-0.3cm]{$p_c$};
    \end{tikzpicture}
\end{equation}
which is immediate due to the symmetry that flips the graph left to right. However, the symmetry that swaps the two cut propagators will never relate two different external leg contractions.
Combining the initial symmetry factor of the uncontracted graph with the number of equivalent contractions to external legs we have
\begin{equation}
    \frac{1}{\sigma(\tau)}|\tilde{G}(\tau): G(\tau)|= \frac{1}{\sigma(\tau)}\frac{|G(\tau)|}{|\tilde{G}(\tau)|}= \frac{\sigma(\tau)}{\sigma(\tau)}\frac{1}{|\tilde{G}(\tau)|}=\frac{1}{|\tilde{G}(\tau)|}\,.
\end{equation}
In the main text to reduce notation we defined $\sigma(g)=|\tilde{G}(\tau)|$ where $g$ is the graph $\tau$ including the contracted external legs. For our example above we have
\begin{equation}
    \sigma\left(\begin{tikzpicture}[baseline={([yshift=-0.5ex]current bounding box.center)},scale=0.7,thick] 
    \draw[Anew] (0,0) -- (0.75,0);
    \draw[Rnew] (0.75,0) -- (0.75,1);
    \draw[Rnew] (0.75,0) -- (1.25,-0.8);
    \draw[Rnew] (1.75,0) -- (1.25,-0.8);
    \draw[Rnew] (1.75,0) -- (1.75,1);
    \draw[Rnew] (1.75,0) -- (2.5,0);
    \draw[out=-45,in=-135,cut] (0.75,1) to (1.75,1);
    \draw[out=135,in=45,cut]   (1.75,1) to (0.75,1);
    \path (2,0) -- (2,-0.5); 
    \draw[massive] (0,0) -- (-0.5,0.5)node[xshift=-0.3cm]{$p_a$};
    \draw[massive] (0,0) -- (-0.5,-0.5)node[xshift=-0.3cm]{$p_b$};
    \draw[massive] (2.5,0) -- (3.0,0.5)node[xshift=0.3cm]{$p_d$};
    \draw[massive] (2.5,0) -- (3.0,-0.5)node[xshift=0.3cm]{$p_e$};
    \draw[massive] (1.25,-0.8) -- (1.25,-1.3)node[yshift=-0.3cm]{$p_c$};
    \end{tikzpicture}\right)=2!\,.
\end{equation}
The manipulations in this appendix take us from symmetry factors of $N$-operator graphs in \eqref{eq:Nsum} to symmetry factors of $N$-matrix elements in \eqref{eq:MuruaFormulaMatrixElement}. This process of relating symmetry factors of contracted and uncontracted graphs applies equally well to the Dyson expansion as the Magnus expansion. We note that the Murua coefficient $\omega(\tau)$ is simply a spectator when contracting external legs, and so it trivially only depends on the graph without external legs.

\newpage

\bibliographystyle{JHEP}
\bibliography{ScatEq.bib}

\end{document}